\documentclass{jfm}

\usepackage{newtxtext}
\usepackage{newtxmath}
\usepackage{natbib}
\usepackage{hyperref}
\hypersetup{
  colorlinks=true,
  linkcolor=blue,
  citecolor=blue,
  urlcolor=blue
}
\usepackage{amsmath,amssymb,graphicx,subcaption,float}
\usepackage{booktabs}
\usepackage{multirow}
\usepackage{hyperref}
\usepackage{comment}
\usepackage{xcolor}
\usepackage{booktabs} 
\usepackage{bm}
\usepackage{mathrsfs}

\hypersetup{
    colorlinks = true,
    urlcolor   = blue,
    citecolor  = black,
}

\newcommand{\RomanNumeralCaps}[1]

\title{Turbulence Closure in RANS and Flow Inference around a Cylinder using PINNs and Sparse Experimental Data}

\author{
  Zhen Zhang\aff{1}\footnotemark[1],
  Khemraj Shukla\aff{1}\footnotemark[1],
  Zhicheng Wang\aff{1}\footnotemark[1], 
  Anthony Morales\aff{2}\footnotemark[1],
  Theo Käufer\aff{3}\footnotemark[1],
  Sheikh Salauddin\aff{2},
  Nathan Walters\aff{2},
  David Barrett\aff{3},
  Kareem Ahmed\aff{2},
  Michael S. Triantafyllou\aff{3},
  George Em Karniadakis\aff{1}\footnotemark[2]
}
\affiliation{\aff{1}Division of Applied Mathematics, Brown University
\aff{2}Department of Mechanical and Aerospace Engineering, University of Central Florida
\aff{3}Department of Mechanical Engineering, Massachusetts Institute of Technology
}

\begin{document}
\maketitle

\begingroup
\renewcommand\thefootnote{\fnsymbol{footnote}} 
\footnotetext[1]{These authors contributed equally to this work.}
\footnotetext[2]{Email address for correspondence: george\_karniadakis@brown.edu}
\endgroup

\begin{abstract}
Traditional Reynolds-averaged Navier-Stokes (RANS) closures, based on the Boussinesq eddy viscosity hypothesis and calibrated on canonical flows, often yield inaccurate predictions of both mean flow and turbulence statistics. Here, we consider flow past a circular cylinder over a range of Reynolds numbers ($3,900$-$100,000$) and Mach numbers ($0$-$0.3$), encompassing incompressible and weakly compressible regimes, with the goal of improving predictions of mean velocity and Reynolds forces. To this end, we assemble a cross-validated dataset comprising hydrodynamic particle image velocimetry (PIV) in a towing tank, aerodynamic PIV in a wind tunnel, and high-fidelity spectral element DNS and LES. Analysis of these data reveals a universal distribution of Reynolds stresses across the parameter space, which provides the foundation for a data-driven closure. We employ physics-informed neural networks (PINNs), trained with the unclosed RANS equations, to infer the velocity field and Reynolds-stress forcing from boundary information alone. The resulting closure, embedded in a forward PINN solver \textcolor{black}{and the numerical solver OpenFOAM}, significantly improves RANS predictions of both mean flow and turbulence statistics relative to conventional models.

\end{abstract}


\section{Introduction}

Flow past a circular cylinder is both a rich physical phenomenon and a canonical benchmark for turbulence modeling. Across $Re \approx 300$–$300{,}000$, the wake is fully turbulent while the boundary layer remains laminar. Classical RANS closures—constrained by the assumptions used to close the Reynolds-stress term—struggle to predict the mean field and, more critically, the Reynolds stresses. By contrast, DNS resolves all dynamically relevant scales at low to moderate $Re$, while at higher $Re$ PIV provides spatially resolved measurements of mean velocity and Reynolds stresses, albeit over limited windows. These observations motivate two questions: (i) how can we infer full-field, physically consistent flow states from sparse, patchwise PIV data; and (ii) how can we overcome the structural limitations of classical RANS closures so that both the mean flow as well as the turbulent contribution are predicted accurately?

\textbf{Drawbacks of RANS closure:}
The Boussinesq assumption collapses the full Reynolds-stress tensor to an isotropic eddy-viscosity response, $\tau_{ij} = 2\nu_t S_{ij} - \frac{2}{3}k\,\delta_{ij}$. In flows with strong streamline curvature, rotation, or rapid distortion, the true stress anisotropy departs markedly from this linear form, as shown by~\cite{schmitt2007boussinesq}, leading to discrepancies in separation location, pressure distribution, and global forces. These errors persist even when $\nu_t$ is calibrated to match a single target, because the structural restriction is the dominant error source. 
Non-linear turbulence models, \textcolor{black}{such as those introduced by~\cite{spalart2000strategies} and ~\cite{cambon1999linear}}, extend the Boussinesq hypothesis by allowing the Reynolds-stress anisotropy to depend on non-linear combinations of the mean strain and rotation rate tensors, so they can represent effects like curvature, rotation, and secondary flows that linear eddy-viscosity models miss. Coefficients may come from asymptotic analysis and calibration or from data-driven fitting. Limitations include sensitivity to coefficient choices and potential stiffness or instability if not embedded in conservative solver formulations with appropriate limiters. Nonlinear turbulence closures are still far from accurately matching both mean flow and Reynolds stress.
Instead of representing the full Reynolds-stress tensor, one can model the Reynolds forcing vector, as shown by~\cite{amarloo2022frozen,amarloo2023data}, the net turbulent contribution that appears as a source term in the mean momentum equations. This targets what the solver directly requires, and integrates cleanly into RANS/URANS by adding a learned body force at each control volume. It avoids tensor realizability constraints and is straightforward to supervise from DNS/LES and PIV. This approach also supports a residual-learning variant, where the forcing vector is decomposed into a baseline Boussinesq part plus a learned correction, achieving both numerical stability and accuracy in the Reynolds forcing. 

\textbf{Data-driven turbulence closure:}
Recent data-driven closures promise richer stress representations, including augmentation of the classical RANS models, \textcolor{black}{such as~\cite{Duraisamy1} and~\cite{Yufei1}, direct prediction of the eddy viscosity, such as~\cite{weiwei1}, prediction of the anisotropic part of Reynolds stresses, see~\cite{ling1} and~\cite{xiao1}, discrepancy of Reynolds stresses, see~\cite{xiao2} and~\cite{yan1}, and the Reynolds forcing vector, see~\cite{amarloo2023data} and~\cite{CRUZ2019104258}}. 
There are many issues of the data-driven turbulence closure models. \textcolor{black}{It is now well recognized that purely \emph{a priori} data fitting of Reynolds stresses or modeled forces, in which machine--learning models are trained to match high-fidelity data in an offline manner, often fail to deliver accurate predictions of quantities of interest (QoIs) once embedded in a CFD solver. Despite achieving low regression error against DNS or LES targets, such models may lead to large discrepancies in integrated or flow-level metrics, including mean velocities, separation characteristics, and force coefficients. This inconsistency arises because the learned closures are optimized independently of the governing equations and the numerical discretization in which they are ultimately deployed. In contrast, \emph{a posteriori} training strategies explicitly couple the learning process with the flow solver and typically optimize the model parameters with respect to QoIs directly, rather than enforcing pointwise agreement with Reynolds stresses or forcing terms. As a result, \emph{a posteriori} approaches generally do not aim to reproduce the true underlying stress or force fields, but instead seek closures that compensate for both modeling deficiencies and structural errors inherent in the chosen model form, leading to improved predictive performance at the level of QoIs. Moreover, while the circular cylinder is one of the most extensively studied benchmark problems in fluid mechanics, data-driven turbulence closure studies on cylinder flows remain comparatively scarce, highlighting an important gap between canonical CFD validation cases and current machine-learning-based modeling efforts.} Therefore, our focus is to pursue a \emph{closure discovery} that is accurate both in mean flow and turbulence statistics for cylinder flows.
Eddy-viscosity-based methods, including direct prediction and augmentation of RANS models, are not enough to fully represent the Reynolds stresses or force vector. Among the Reynolds stresses and the forcing vector, we choose to approximate the forcing vector directly because the turbulent contribution to the governing equation will be directly modeled without the divergence operator, and there are less degrees of freedom than in the Reynolds stresses. 

\textbf{PINNs and flow inference:}
\textcolor{black}{Physics-informed neural networks (PINNs) provide a flexible way to fuse sparse and noisy measurements with the governing equations to recover hidden or unmeasured features of a flow, as shown by~\cite{toscano2025aivt},~\cite{shukla2020physics},~\cite{shukla2025neurosem},~\cite{kiyani2023framework},~\cite{kiyani2024characterization}, and~\cite{karniadakis2021physics}.} In PINNs, a neural network represents one or more fields (e.g., velocity, pressure, and possibly auxiliary closure terms), and minimizes a composite loss that balances data mismatch at sensor locations, residuals of the PDE operator (e.g., momentum and continuity), and boundary/initial-condition penalties. Because the governing physics enters as soft constraints, PINNs naturally enable: 
\emph{flow inference} (estimating latent quantities such as pressure or Reynolds stresses from velocity-only PIV data); 
\emph{flow reconstruction} (extending partially observed subdomains to solver-ready, domain-wide fields); and 
\emph{flow super-resolution} (lifting coarse spatial or temporal measurements to finer resolutions while respecting physical conservation laws). 
\textcolor{black}{For example, \cite{cai2021flow} reconstructed three-dimensional velocity and pressure fields from tomographic background-oriented Schlieren measurements and validated the inferred fields against independent PIV experiments; \cite{patel2024turbulence} recovered the mean flow velocity for flow past a cylinder at $Re=200$; \cite{eivazi2022physics} solved the RANS equations using sparse in-domain flow measurements; \cite{hanrahan2023studying} demonstrated that embedding the RANS equations within PINNs enables accurate reconstruction of turbulent mean flows from measurement densities representative of experimental campaigns; and \cite{Eivazi_2024} showed that physics-informed deep-learning models can reconstruct and super-resolve flow fields from experimental data, including time-resolved hot-wire anemometry and PIV measurements.}

\textbf{Contributions:}
This paper is organized as follows: After the introduction and problem setup in \autoref{sec:setup}, we build a diverse dataset for flows past a cylinder using PIV and DNS/LES and introduce a physics-informed data correction method in \autoref{sec:data}. We use PINNs with the unclosed RANS equation to infer the full-field flow states and Reynolds forcing based on only the boundary information in ~\autoref{sec:infer}. We then build a data-driven turbulence closure model based on the universality of Reynolds stress distribution, and integrate it with a forward PINN solver \textcolor{black}{and the numerical solver OpenFOAM} in ~\autoref{sec:closure}. Finally, we draw the conclusion in \autoref{sec:conclusion}. 

In summary, our contributions are:
\begin{itemize}
    \item \textbf{Diverse dataset.} We assemble a multi-regime cylinder dataset combining incompressible and compressible DNS/LES and PIV, covering a broad Reynolds number range.
    \item \textbf{Flow inference with sparse data.} We use PINNs to infer the full-field mean velocity and Reynolds forcing from boundary information, solving a numerically under-determined problem.
    \item \textbf{Universality of Reynolds stresses.} We find that the distribution of Reynolds stresses/forcing is similar along the Reynolds numbers for both incompressible and weakly compressible flows past a cylinder.
    \item \textbf{Turbulence model discovery.} We propose and evaluate data-driven closures that target \emph{simultaneous} accuracy in Reynolds forcing and mean velocity.
    \item \textbf{Turbulence model–solver integration.} We integrate the data-driven closure model with \textcolor{black}{both PINNs and OpenFOAM and find that the proposed data-driven closure model can improve the accuracy in both mean velocity and the Reynolds stresses.}
\end{itemize}

\section{Problem Setup}\label{sec:setup}

The flow past a cylinder is the focus of this study. The flow equations are normalized by the cylinder diameter $D$ and freestream velocity $U_\infty$. In all setups, the origin of the 2D coordinate system is at the center of the cylinder. Hydrodynamic and aerodynamic PIVs are used to measure the mean flow fields and turbulence statistics in the wake region. High-fidelity spectral-element-based DNS and LES are also conducted in a wide range of Reynolds numbers for both incompressible and weakly compressible regimes.

\begin{figure}
    \centering
    \includegraphics[width=.9\linewidth]{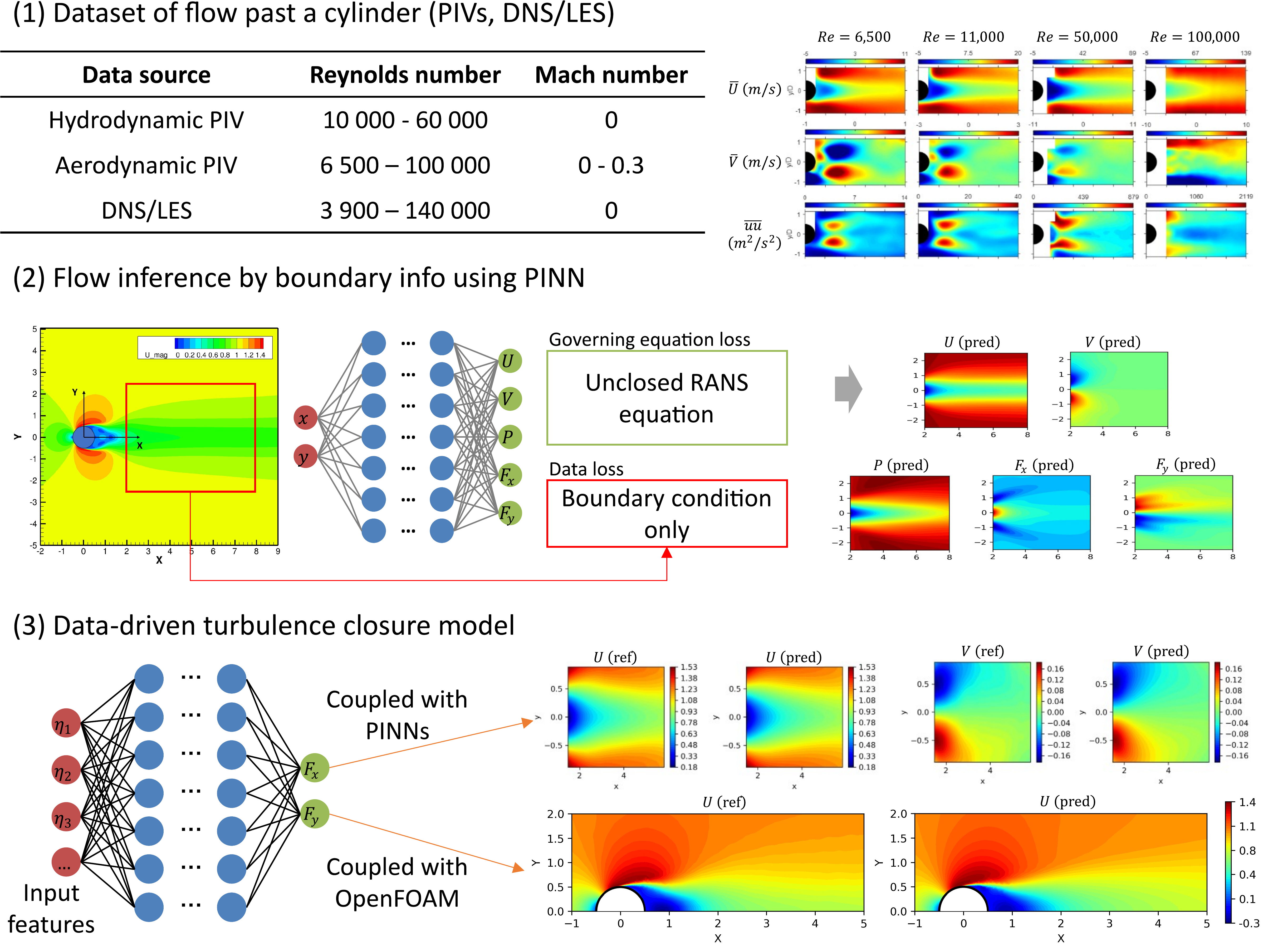}
    \caption{\textcolor{black}{Overview of the paper. (1) PIV and DNS/LES are used to establish a dataset of flow past a cylinder. The range of the key parameters, $Re$ and $Ma$, is listed. (2) PINNs are used to infer the flow fields within a domain $\Omega$ based on the unclosed RANS equation and the boundary conditions at $\partial \Omega$ for both incompressible and weakly compressible regimes. (3) Data-driven turbulence closure model is built and integrated with the forward PINN solver and the numerical solver OpenFOAM, investigating the accuracy of both velocity and Reynolds forcing fields.}}
    \label{fig:sketch}
\end{figure}

\autoref{fig:sketch} shows a sketch of the overview of our work. There are three primary aims in this paper: (i) Build a diverse dataset for flows past a cylinder using PIV and DNS/LES. (ii) Infer the flow field and the Reynolds forcing using limited measurements and PINNs. (iii) Build data-driven turbulence closure models and integrate them with PINNs and numerical solvers.

One of the key ideas in this paper is to use the unclosed data-driven RANS equation without any assumptions on the turbulence closure and represent the Reynolds forcing directly. In this case, we can simultaneously pursue the accuracy of both velocity and Reynolds forcing fields. The steady 2D unclosed RANS equation with Reynolds forcing for incompressible flows is:
\begin{align}\label{eq:RANS_Euqtaion}
\begin{aligned}
\frac{\partial U}{\partial x}+\frac{\partial V}{\partial y}=0,\\
U \frac{\partial U}{\partial x}+V \frac{\partial U}{\partial y}+\frac{1}{\rho} \frac{\partial P}{\partial x}-\nu\left(\frac{\partial^2 U}{\partial x^2}+\frac{\partial^2 U}{\partial y^2}\right)-F_x=0, \\
U \frac{\partial V}{\partial x}+V \frac{\partial V}{\partial y}+\frac{1}{\rho} \frac{\partial P}{\partial y}-\nu\left(\frac{\partial^2 V}{\partial x^2}+\frac{\partial^2 V}{\partial y^2}\right)-F_y=0, 
\end{aligned}
\end{align}
where $U,V,P$ are time-averaged state variable, $\nu$ is the molecular viscosity, and $\rho=1$ is the density. $F_x,F_y$ are the Reynolds forcing vector, defined by $\mathbf{F} = \nabla \cdot \tau_{Re}$, where $[\tau_{Re}]_{ij}=-\overline{u'_iu'_j}$ is the Reynolds stress tensor.

\section{Data Generation and Cross Validation}\label{sec:data}
\subsection{Hydrodynamic Particle Image Velocimetry}

The hydrodynamic PIV experiments
were performed using the robotic towing tank facility at the Massachusetts Institute of Technology. A detailed description of the facility can be found in~\cite{fan2019robotic}. 
The towing tank has a length of 10 meters in the main towing direction and a cross section of 1 m $\times$ 1 m.

A vertically mounted cylinder with a diameter $D=50.8$ mm is towed by a gantry robot that is mounted on a rail system at the top of the towing tank. The cylinder is made from aluminum and has a smooth surface, which is anodized in black to reduce light reflection on the cylinder's surface. Endplates are attached to mitigate the effects arising from the finite extent of the cylinder and the free surface, resulting in an effective length between the endplates of $L\approx0.6$ m, which corresponds to an aspect ratio $L/D=12$. The camera (Optronis Cyclone 2000) and the lens (Zeiss Dimension 2/35) are mounted behind the cylinder in a waterproof housing. The camera images a horizontal light sheet perpendicular to the cylinder's main axis, which is generated by a 40W pulsed CW laser (Optolutions LD-PS/40), fixed onto the gantry robot but positioned outside the towing tank. Thus, the relative position between the camera, cylinder and laser is fixed.  To visualize the flow, polyamide particles of size 5 $\mu$m were used, which over the duration of the experiment showed negligible floating or sedimentation. A camera calibration was performed to obtain the pixel scaling factor and compensate for optical distortions. The camera calibration resulted in a pixel scaling factor of 5.744 pixels/mm.
The images were recorded as time series, and the time step $\Delta t$ between the frames was adapted to the Reynolds number and is shown in \autoref{tab:MIT_parameters}.  
Due to the struts of the towing tank, the laser light is occasionally blocked. These events were discarded during the preprocessing.
To provide a variety of data, runs at several different Reynolds numbers were performed. 

\begin{figure}
    \centering
    \includegraphics[width=\linewidth]{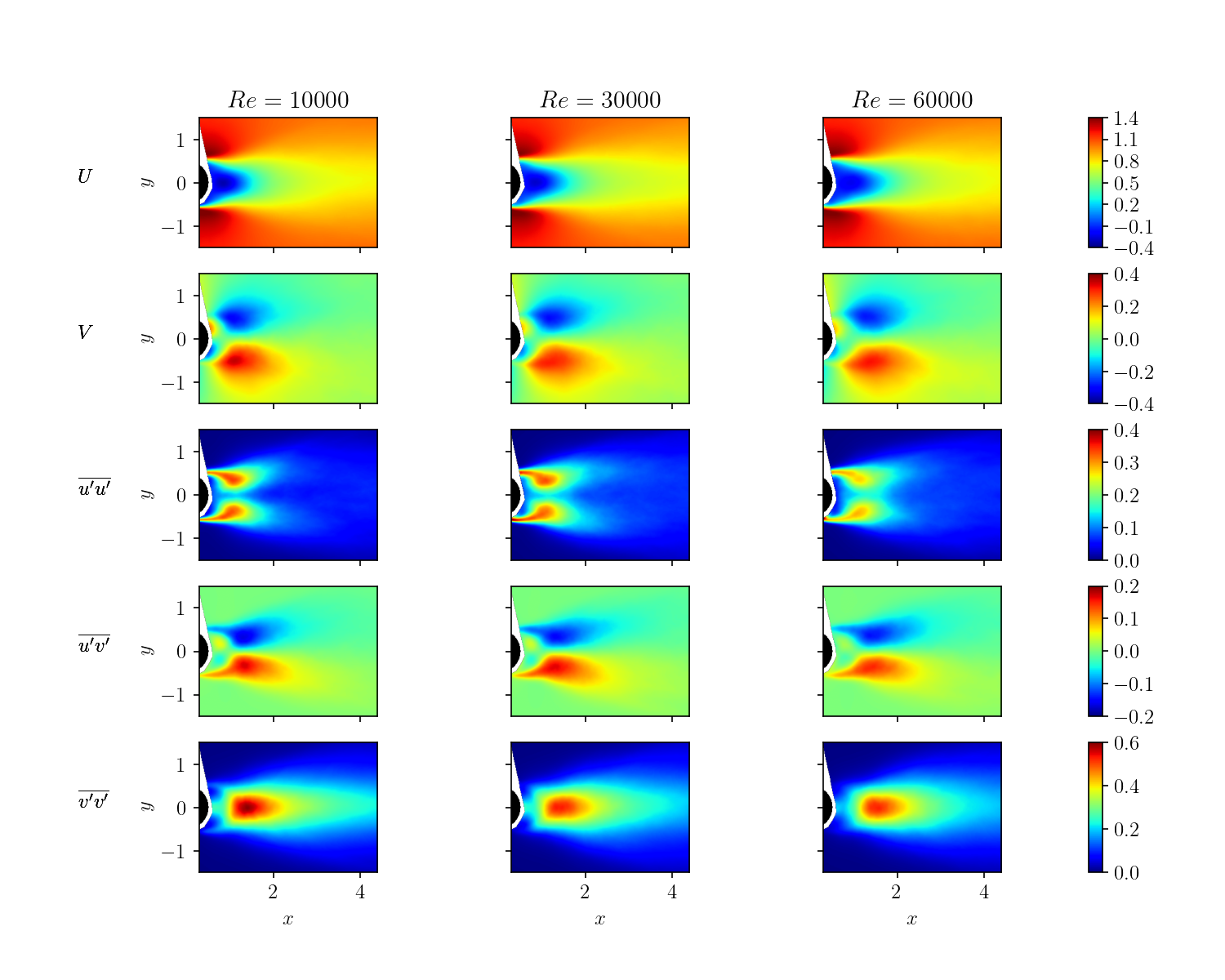}
    \caption{Hydrodynamics PIV: Overview of the mean velocity components $U$ and $V$, and the Reynolds stress components $\overline{u'u'}$, $\overline{u'v'}$, $\overline{v'v'}$ for $Re = 10,000$, $Re = 20,000$, $Re = 30,000$ (left to right).The white region was masked during the processing. The black region depicts the cylinder.}
    \label{fig:MIT_field}
\end{figure}

An overview of important parameters like the towing velocity $U_\mathrm{tow}$, final interrogation window size, overlap, field size, the spatial resolution as ratio of interrogation window edge length $l_\mathrm{win}$ divided by the Kolmogorov length $l_\mathrm{k}=D*Re^{-3/4}$ and number of snapshots, the time between the snapshots $\Delta t$ are shown in \autoref{tab:MIT_parameters}.
\begin{table}
    \centering
    \caption{Parameters of the hydrodynamic PIV measurements at $Re$= 10,000, 30,000 and 60,000. $U_{\mathrm{tow}}$ is the towing velocity, win. size the size of the interrogation windows in pixels, overlap the overlap percentage of the interrogation windows, field size is the size of the vector field in terms of vectors,  $l_\mathrm{win}$ / $l_\mathrm{k}$ denotes the relative spatial resolution, no. snapshots the number of snapshots, and $\Delta t$ the time between the snapshots.}
    \label{tab:MIT_parameters}
    \begin{tabular}{cccccccc}
        \toprule
        Re & $U_{\text{tow}}$ in m/s& win. size& overlap & field size & $l_\mathrm{win}$ / $l_\mathrm{k}$  & no. snapshots & $\Delta t$ in ms \\
        \midrule
        10,000  & 0.188 &$16\times 16$& $75 \%$& 301$\times$220& 55 & 30,025 & 4\\ 
        30,000  & 0.563 &$16\times 16$& $75 \%$  & 301$\times$220& 125 &  20,125 & 2  \\
        60,000 & 1.125 &$16\times 16$& $75 \%$ & 301$\times$220& 210 & 19,900  & 1  \\
        
        \bottomrule
    \end{tabular}
\end{table}
For each Reynolds number, five runs of data were recorded, insufficiently illuminated regions and reflections masked out, and processed {\color{black}with the LaVision DAVIS 11 flow visualization software using} multipass PIV with a final interrogation window size of $16\times16$ with an overlap of 75$\%$, which corresponds to about $2.8$ mm $\times$ 2.8 mm.  The data was post-processed using the universal outlier detection~\cite{westerweel2005universal} and gaps in the data filled by interpolation. The flow field after processing covered about 200 mm $\times$ 150 mm.
Subsequently, the data were non-dimensionalized by dividing length scales by the cylinder diameter $D$ and velocities by the towing velocity $U_\mathrm{tow}$. Per $Re$, the data from all runs were combined, {\color{black} resulting in a total of close to 90 vortex shedding cycles per $Re$}, to obtain converged mean statistics.
\autoref{fig:MIT_field} shows mean fields of the dimensionless velocities $U$, $V$, and Reynolds stresses $\overline{u'u'}$, $\overline{u'v'}$, $\overline{v'v'}$ for the various Reynolds numbers. All quantities show a structural similarity across all Reynolds numbers; however, differences remain, i.e, the recirculation bubble visible in the $U$ field tends to grow slightly with $Re$. Additionally, the Reynolds stress appears more concentrated at lower $Re$.

\subsection{Aerodynamic Particle Image Velocimetry}
Aerodynamic particle image velocimetry (PIV) experiments were conducted in a high-speed wind tunnel facility at the University of Central Florida (UCF). Details of the wind tunnel geometry, rig instrumentation, and flow metering methods are described in previous work \citep{morales2019mechanisms,morales2022role}. PIV measurements were performed in an optically accessible test section with a rectangular cross section measuring 127 mm × 45 mm (depth × height). For this study, the test section was fitted with a smooth circular cylinder to generate a turbulent wake. The cylinder has a diameter $D$ = 15 mm and spanwise length $L$ = 127 mm to cover the full depth of the test section. The cylinder was mounted in the vertical center of the channel using a side plate and screw. This configuration corresponds to a blockage ratio of 33\% (ratio of cylinder diameter to channel height) and an aspect ratio $L/D = 8.47$.


The PIV system consists of a seeder with tracer particles, a high-speed laser, sheet forming optics, and a camera mounted perpendicular to the facility. 0.5 $\mu$m aluminum oxide (Al$_2$O$_3$) tracer particles are injected into the main flow path upstream of the wind tunnel using a pressure-driven seeder. Airflow through the seeder was metered using a shop air source (102 psi) and choke orifice. The tracer particles were illuminated in the test section using a dual-cavity, solid state, 532 nm, Nd:YAG laser (LDP-200MQG Dual). The laser beam was formed into a thin sheet using a 1000 mm focusing optic and and a -25.4 mm focal length cylindrical lens. The laser sheet was then directed into the test section using a 45-degree  mirror with the sheet centered in the spanwise direction of the test section. Tracer particle motion was captured with a high-speed CMOS camera (Photron SA-Z 2100K) with a $f/2.8$, 24-85 mm focal length lens. The laser and camera systems were synchronized with a BNC model pulse/delay generator to collect sequential PIV images at a 40,000 Hz  over a \textcolor{black}{total sample duration of 0.27 seconds, corresponding to approximately 11,000 samples.} The camera captures a domain of approximately 90 mm in the streamwise direction and the full height of the channel 45 mm. The setup results in time-resolved images with a size of 512$\times$1024 pixels \textcolor{black}{(pix)}, corresponding to a pixel resolution of 91 $\mu$m/\textcolor{black}{pix}. The images were collected on Photron Fastcam software and exported for vector field processing. The PIV data was processed in LaVision DaVis 10 software using two interrogation box sizes for cross-correlations. The initial and final box sizes, including the overlapping window size, were adjusted for the different Reynolds number conditions (detailed below) to provide high-resolution velocity vector fields.

A summary of the experimental conditions is provided in \autoref{tab: UCF incompressible tests}. The conditions are defined based on the Reynolds number $Re = U_0D/\nu$, where $U_0$ is the velocity upstream of the cylinder, $D$ is the cylinder diamater (15 mm), and $\nu$ is the kinematic viscosity of air. The Stokes number is included, calculated using the formulation from \cite{raffel2018particle}. The final window size used for PIV data processing ($W$) is provided for each test condition, as well as the overlapping window size (in \%). The resulting vector grid resolution $\lambda$ is listed for each test case, and also quoted with respect to the Kolmogorov length scales as $\lambda/l_k$. Here, $l_k = l_0Re_T^{-3/4}$, where $l_0$ are the integral scales, and $Re_T = u'l_0/\nu$ is the turbulent Reynolds number \citep{pope2001turbulent}.

\begin{table}
    \centering
    \caption{Experimental conditions for high-speed aerodynamic particle image velocimetry measurements in \textcolor{black}{University of Central Florida} wind tunnel experiments.}
    \label{tab: UCF incompressible tests}
    \begin{tabular}{cccccccc}
        \toprule
        Re$_D$ & U$_0$ (m/s) & $St_k$ & W (pixels) & overlap (\%) & $\lambda$ ($\mu$m) & $l_k$ ($\mu$m) & $\lambda/l_k$ \\
        \midrule
        6,500 & 6.5 & 0.001 & $12\times12$ pix & 50\%   & 546 & 117 & 4.7 \\
        11,000 & 11 & 0.002 & $12\times12$ pix & 50\%   & 546 & 79  & 7.0 \\
        50,000 & 50 & 0.010 & $24\times24$ pix & 87.5\% & 273 & 25  & 10.5 \\
        100,000 & 100 & 0.021 & $32\times32$ pix & 87.5\% & 364 & 15  & 24.3 \\
        \bottomrule
    \end{tabular}
\end{table}

\begin{figure}
\centering
\includegraphics[width=1\textwidth]{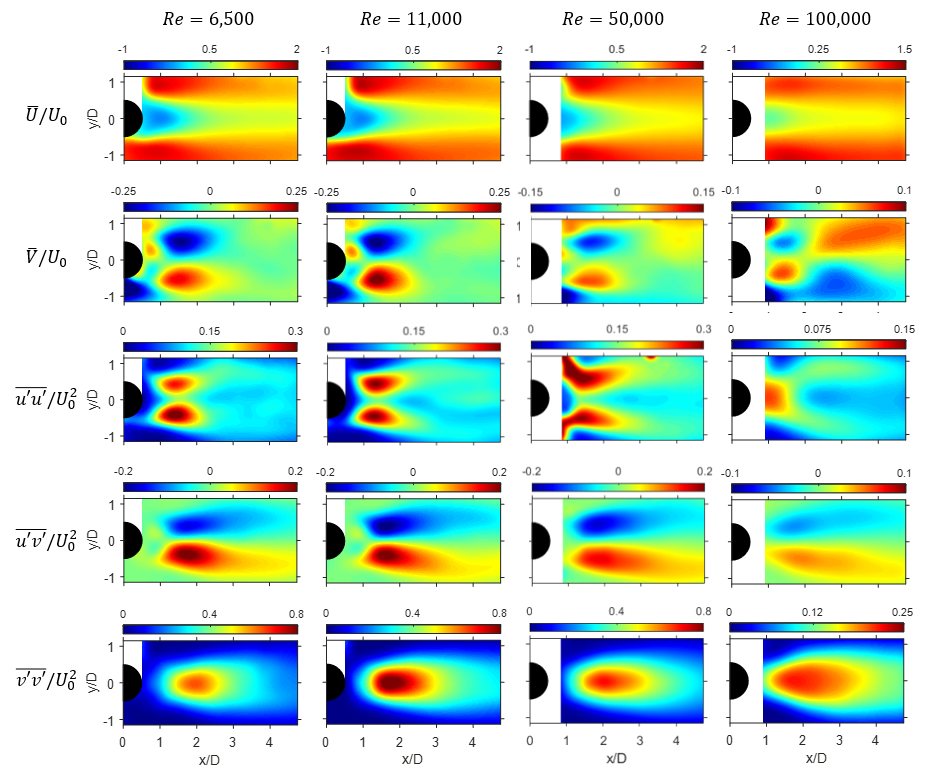}
\caption{Aerodynamic PIV: Time-averaged velocity components and Reynolds stresses measured  from Re = 6,500 - 100,000. White regions indicate areas where vector data could not be resolved, and the black portion marks the location of the cylinder.}
\label{fig: ucf_flow}
\end{figure}

Time-averaged streamwise and cross-stream velocity fields, together with Reynolds stress components obtained from aerodynamic PIV measurements, are shown in \autoref{fig: ucf_flow}. \textcolor{black}{Statistical convergence was verified with all reported Reynolds stresses within +/- 4\% of their asymptotic values}. Despite the varying Reynolds number, the Reynolds stress fields exhibit structural similarity across cases. For instance, the Reynolds shear stress ($\overline{u'v'}$) consistently displays an antisymmetric distribution, with negative values concentrated in the upper shear layer and positive values in the lower shear layer. The normal stress components ($\overline{u'u'}$, $\overline{v'v'}$) preserve similar spatial organizations, with regions of elevated intensity aligned with the shear layers and wake centerline. These observations indicate a degree of self-similarity in the RANS terms; although the recirculation length tends to decrease with elevated Reynolds numbers, the underlying distribution of Reynolds stresses retains an invariant structural pattern. This persistence suggests that the turbulence production and distribution mechanisms in the near-wake remain fundamentally the same across Reynolds numbers.

\subsection{Numerical Simulation}

\autoref{tab:param_inc} shows the parameter ranges of incompressible cylinder flows simulated using the open source spectral element code nekRS \textcolor{black}{proposed by}~\cite{NEKRS}. In particular, the direct numerical simulation (DNS) is used in the simulations of $Re<60,000$, and the large eddy simulation is used in the simulation for flows at $Re=60,000$ and $Re=140,000$, where the entropy viscosity method (EVM) \textcolor{black}{demonstrated by}~\cite{wang2019entropy} is employed to model the unclosed subgrid scale turbulence. Note that the EVM has been validated by simulation of incompressible flow past a cylinder at $Re=140,000$, \textcolor{black}{as shown by}~\cite{Wang_RL_JFM_2023}. More validations and the mesh independence study on the implementation of the EVM on nekRS can be found in \autoref{fig:evm_vali1} \autoref{fig:evm_vali2}, and \autoref{fig:evm_vali3} in \autoref{app:validation}. 

\begin{table}
    \caption{Parameters used in simulations of incompressible flow past a cylinder. $D$ is the diameter of the cylinder, $L_z/D$ is the aspect ratio, $N_{xy}$ is the number of elements in $x-y$ plane, $N_c$ is the number of elements along circumference of the cylinder, $N_z$ is the number of elements along the axis of the cylinder, $L_r/D$ is the thickness of the first layer elements around the cylinder. \textcolor{black}{The dimensions of the computational domain is $[-7.5D, 25D] \times [-10D, 10D]$ for the streamwise($x$) and crossflow ($y$) directions, and the cylinder center is located at $x=0, \, y=0$.}} 
    \centering
    \begin{tabular}{cccccccccc}
    \toprule
         Case ID&Re&Type&SEM Order&$L_z/D$&$N_{xy}$&$N_c$&$N_z$&$L_r/D$&DOFs (millions) \\
    \midrule
         1&3,900&DNS& 7 &6.4&1607&96&64&0.002& 35.3\\
         2&5,000&DNS&  7&9.6&1607&96&64&0.002& 35.3 \\
         3&11,000&DNS& 7 & 6.4&2220&96&72&0.001&54.8\\
         4&30,000&DNS& 7 &3.2&2200&96&72&0.001&54.8\\
         5&60,000&LES& 6 & 4&2101&96&72&0.001&21.8\\
         6&140,000&LES& 7 &4&2101&96&48&0.0012&34.7\\
    \bottomrule
    \end{tabular}
    \label{tab:param_inc}
\end{table}



\subsection{Cross Validation}

\begin{figure}
  \centering
  \begin{subfigure}[t]{0.48\textwidth}
    \includegraphics[width=\linewidth]{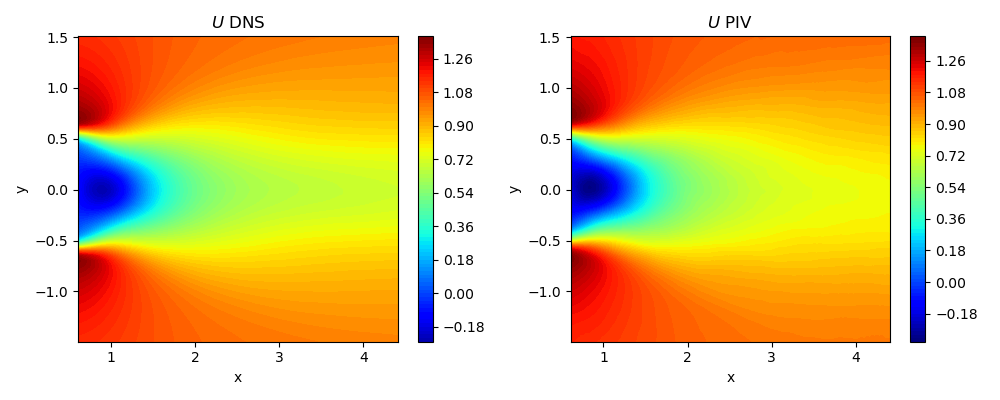}
    \caption{Comparison of $U$}
    \label{fig:re11k-u}
  \end{subfigure}\hfill
  \begin{subfigure}[t]{0.48\textwidth}
    \includegraphics[width=\linewidth]{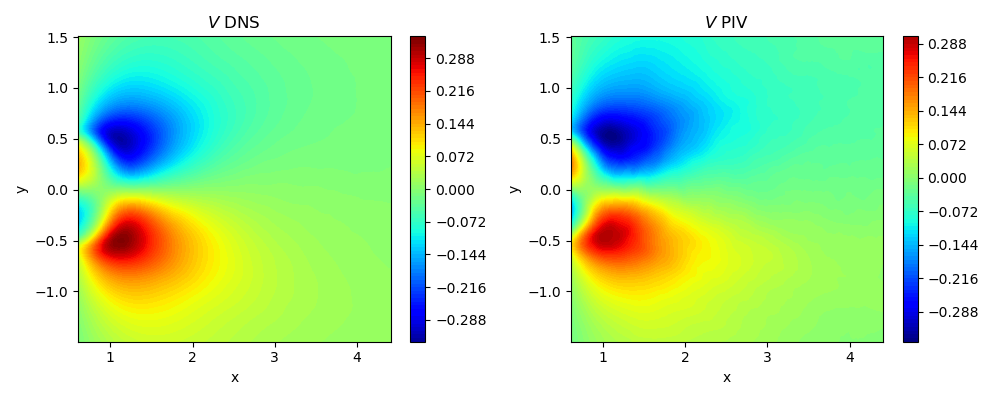}
    \caption{Comparison of $V$}
    \label{fig:re11k-v}
  \end{subfigure}
  \begin{subfigure}[t]{0.48\textwidth}
    \includegraphics[width=\linewidth]{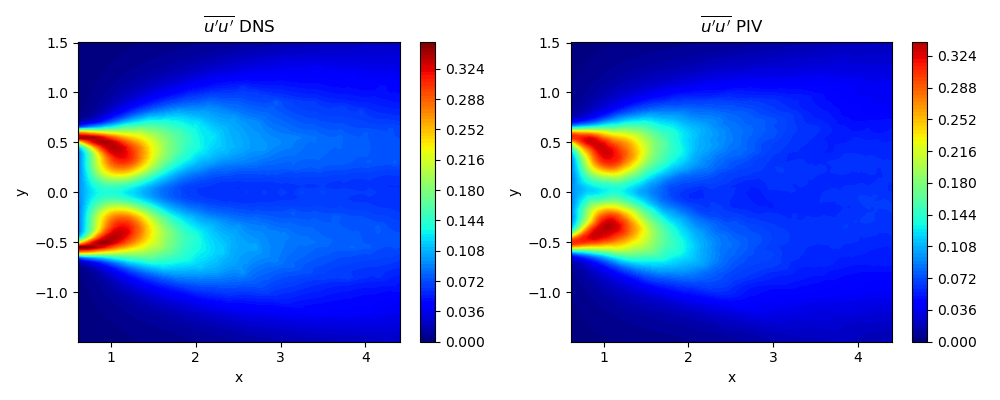}
    \caption{Comparison of $\overline{u'u'}$}
    \label{fig:re11k-p}
  \end{subfigure}\hfill
  \begin{subfigure}[t]{0.48\textwidth}
    \includegraphics[width=\linewidth]{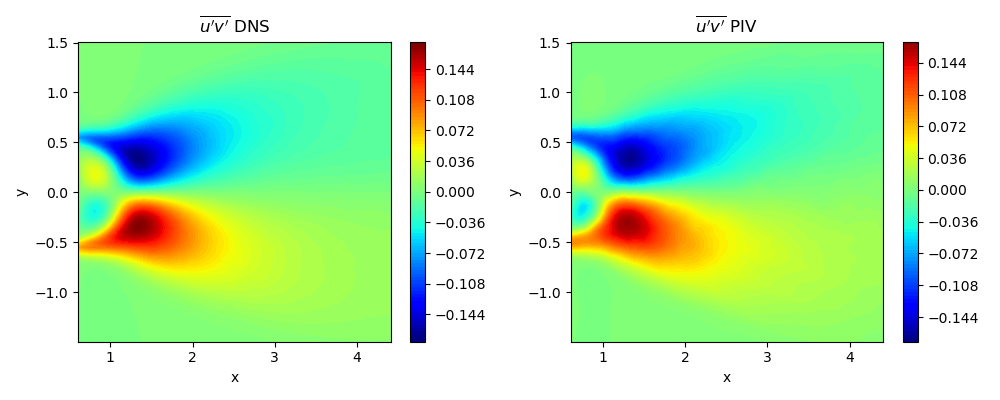}
    \caption{Comparison of $\overline{u'v'}$}
    \label{fig:re11k-fx}
  \end{subfigure}
  \begin{subfigure}[t]{0.48\textwidth}
    \includegraphics[width=\linewidth]{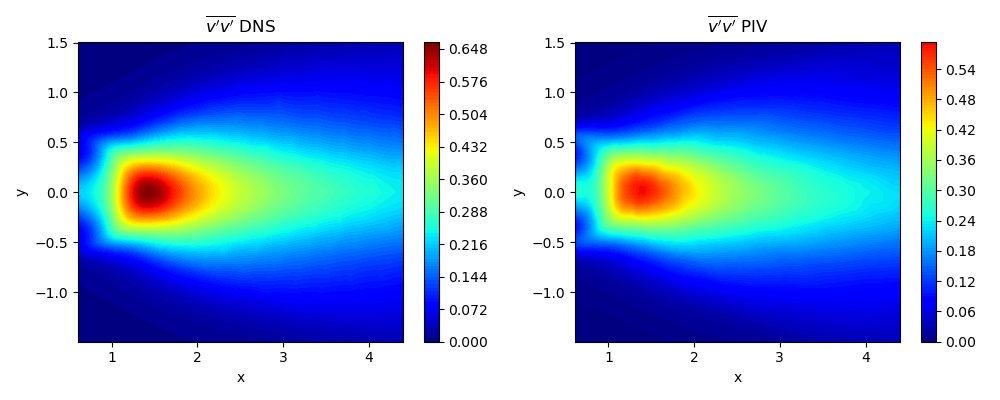}
    \caption{Comparison of $\overline{v'v'}$}
  \end{subfigure}
  \caption{Comparison between time-averaged DNS and hydrodynamic PIV fields. The mean velocity $U$ and $V$ and the Reynolds stresses $\overline{u'u'}$, $\overline{u'v'}$, and $\overline{v'v'}$ are shown. The Reynolds number for DNS is $Re_{DNS}=11,000$, while the Reynolds number for PIV is $Re_{PIV}=10,000$.}
  \label{fig:compare-field}
\end{figure}

\begin{figure}
  \centering
  \begin{subfigure}{0.32\textwidth}
  \centering
    \includegraphics[width=\linewidth]{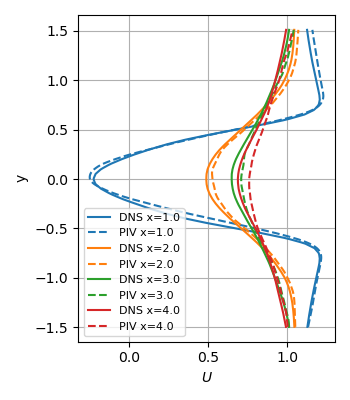}
    \caption{Comparison of $U$}
  \end{subfigure}
  \centering
  \begin{subfigure}{0.32\textwidth}
  \centering
    \includegraphics[width=\linewidth]{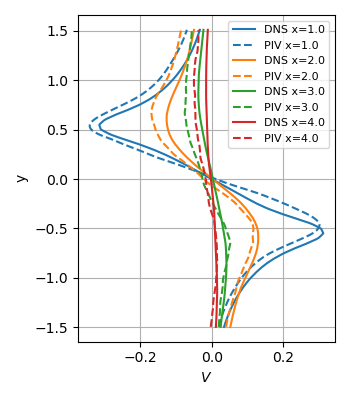}
    \caption{Comparison of $V$}
  \end{subfigure}
  \centering
  \begin{subfigure}{0.32\textwidth}
  \centering
    \includegraphics[width=\linewidth]{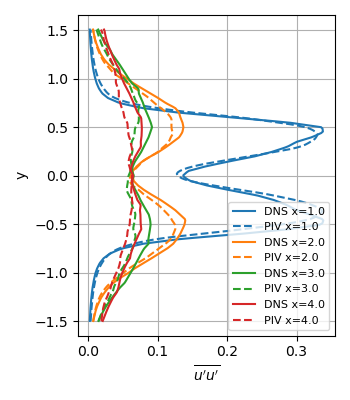}
    \caption{Comparison of $\overline{u'u'}$}
  \end{subfigure}
  \centering
  \begin{subfigure}{0.32\textwidth}
  \centering
    \includegraphics[width=\linewidth]{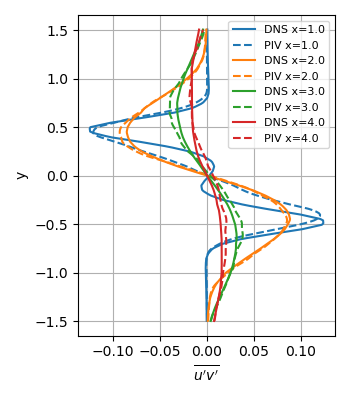}
    \caption{Comparison of $\overline{u'v'}$}
  \end{subfigure}
  \centering
  \begin{subfigure}{0.32\textwidth}
  \centering
    \includegraphics[width=\linewidth]{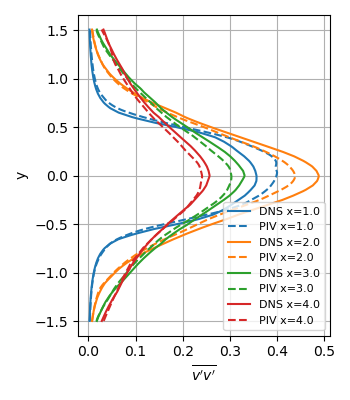}
    \caption{Comparison of $\overline{v'v'}$}
  \end{subfigure}
  \caption{Comparison between time-averaged DNS and hydrodynamic PIV results at four streamwise positions $x=1,2,3,4$. The mean velocity $U$ and $V$ and the Reynolds stresses $\overline{u'u'}$, $\overline{u'v'}$, and $\overline{v'v'}$ are shown. The Reynolds number for DNS is $Re_{DNS}=11,000$, while the Reynolds number for PIV is $Re_{PIV}=10,000$.}
  \label{fig:compare-lines}
\end{figure}

\autoref{fig:compare-field} and \autoref{fig:compare-lines} cross-validate DNS and hydrodynamic PIV. The Reynolds number for DNS is $Re=11,000$, while for hydrodynamic PIV is $Re=10,000$. Both assume an unconfined setup and use $U_\infty$ normalization, enabling a one-to-one comparison. The data were co-registered and time-averaged for mean fields and Reynolds stresses.
Field comparisons (\autoref{fig:compare-field}) show \textcolor{black}{qualitative} agreement in $U$, $V$, and $\overline{u'u'}$, $\overline{v'v'}$, $\overline{u'v'}$: the $U$-deficit and wake spread are consistent; $V$ exhibits the expected centerline symmetry; stress peaks align across shear layers; and far-wake decay rates match. Line profiles (\autoref{fig:compare-lines}) confirm similar velocity deficits, half-widths, and stress-peak magnitudes/locations.
\textcolor{black}{Minor discrepancies are observed in the $\overline{v'v'}$ component, which can be attributed to spatial filtering and signal-to-noise ratio limitations in the PIV measurements, as well as differences in spatial resolution and averaging procedures in the DNS. In addition, noticeable deviations in $\overline{u'v'}$ at $x=1$ are observed. These differences may be associated with experimental uncertainties in the recirculation bubble region, where the mean streamwise velocity becomes negative near the centerline, as well as possible effects of cylinder vibration in the towing tank. Overall, the DNS and PIV datasets show good quantitative agreement in both mean flow and second-order statistics, supporting their joint use for model development and benchmarking.}

\begin{figure}
  \centering
  \begin{subfigure}{0.8\textwidth}
    \includegraphics[width=\linewidth]{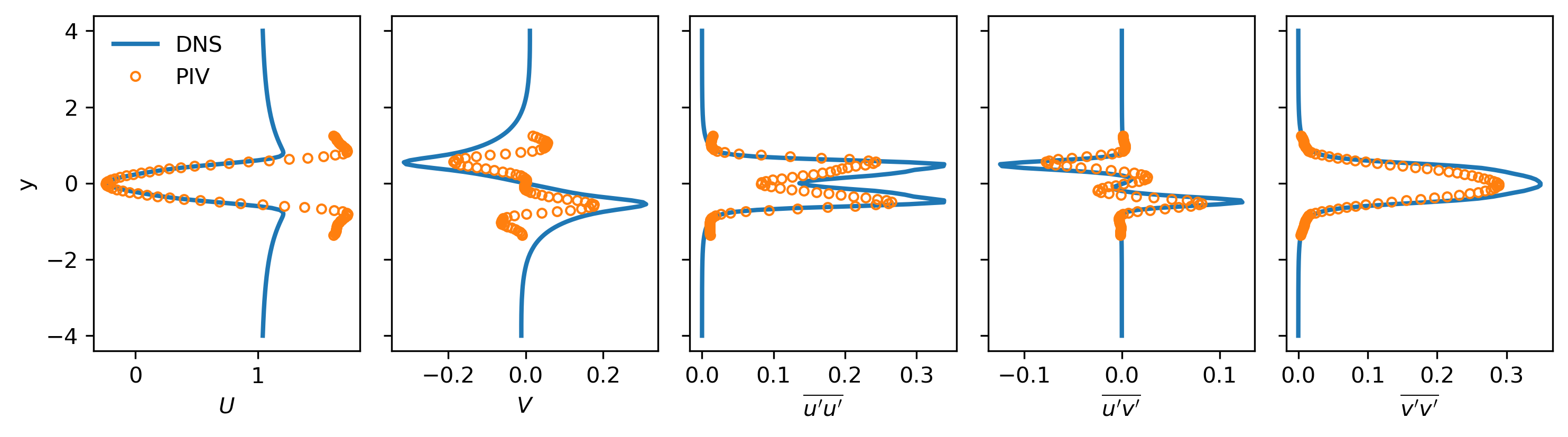}
    \caption{Comparison at $x=1$}
 \end{subfigure}
 \centering
 \begin{subfigure}{0.8\textwidth}
    \includegraphics[width=\linewidth]{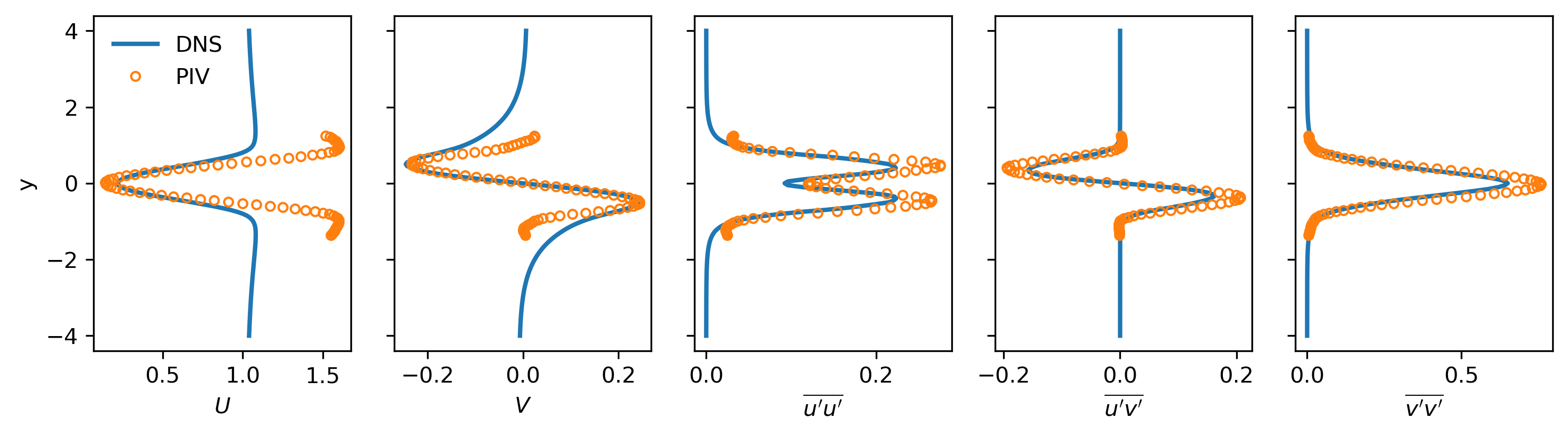}
    \caption{Comparison at $x=1.5$}
  \end{subfigure}
  \caption{Comparison between time-averaged DNS and aerodynamic PIV results at two streamwise positions $x=1,1.5$. The aerodynamic PIV is conducted in a confined wind tunnel with a blockage ratio of $1/3$, as shown in \autoref{fig: ucf_flow}, so only fields near the cylinder are compared here. The mean velocity $U$ and $V$ and the Reynolds stresses $\overline{u'u'}$, $\overline{u'v'}$, and $\overline{v'v'}$ are shown. The Reynolds number for both DNS and PIV is $Re_{DNS}=11,000$.}
  \label{fig:compare-line-UCF}
\end{figure}

\autoref{fig:compare-line-UCF} compares DNS with aerodynamic PIV at $Re=11,000$. The PIV was acquired in a confined wind tunnel whose test-section height is $3D$, giving a nominal blockage ratio of $1/3$. Because this configuration differs from the unconfined DNS, a full-field one-to-one comparison is not meaningful; instead we focus on near-wake line profiles where wall-confinement effects are less dominant. Profiles are extracted at two streamwise locations, $x=1$ and $x=1.5$, after co-registration and normalization by $U_\infty$.
The mean streamwise velocity exhibits a good velocity deficit at both stations, with comparable wake half-widths and centerline values. The cross-stream velocity at $x=1.5$ also agrees well, showing the expected antisymmetric structure and near-zero centerline value. The Reynolds stresses—$\overline{u'u'}$, $\overline{v'v'}$, and $\overline{u'v'}$—match in pattern: peaks for $\overline{u'u'}$, $\overline{v'v'}$, and $\overline{u'v'}$ appear at similar transverse locations. Magnitude differences are more apparent in $V$ and in the stress levels, consistent with tunnel confinement (reduced lateral spreading, modified pressure recovery, and altered turbulence production) and with known PIV limitations near walls (spatial filtering, out-of-plane motion, seeding nonuniformity). Overall, the near-wake agreement is strong in shape and feature placement, while the observed amplitude deviations provide a quantitative envelope for confinement and measurement effects. Thus, \autoref{fig:compare-line-UCF} supports using the aerodynamic PIV to validate DNS in the immediate wake, while cautioning against far-wake or full-field equivalence under confined conditions.

\subsection{Physics-Informed Data Postprocessing}\label{sec:correction}

After obtaining the time-averaged DNS and PIV data, we introduced a physics-informed data postprocessing method to correct the data, ensuring that the continuity and momentum equations are satisfied.

We consider the two‐dimensional incompressible \textcolor{blue}{RANS} equations on a doubly‐periodic domain $\Omega = [0,L_x]\times[0,L_y]$. For a non-periodic domain, we can linearly pad a buffer layer around its boundary, where the outer boundary of the buffer layer is uniform, so that the padded domain satisfies the periodic condition. Let the kinematic viscosity be $\nu$ and assume unit density ($\rho=1$) for simplicity. The RANS momentum equation in steady form, with the Reynolds‐stress contribution modeled as an external forcing $\mathbf{F}(x,y)$, reads:
\begin{equation}
  (\mathbf{U}\cdot\nabla)\,\mathbf{U}
  \;=\;-\,\nabla P \;+\;\nu\,\nabla^{2}\mathbf{U} \;+\;\mathbf{F}\, \quad\text{in }\Omega,
  \label{eq:momentum}
\end{equation}
together with the incompressibility constraint
\begin{equation}
  \nabla\cdot\mathbf{U} \;=\; 0 \quad\text{in }\Omega\,.
  \label{eq:incompressible}
\end{equation}

Suppose we have discrete measurements of the velocity field $\mathbf{U}_{\mathrm{m}}(x,y) = \bigl(u_{\mathrm{m}}(x,y),\,v_{\mathrm{m}}(x,y)\bigr)$ and of the forcing $\mathbf{F}_{\mathrm{m}}(x,y)$ on a uniform grid in $\Omega$.  In general, $\mathbf{U}_{\mathrm{m}}$ does not exactly satisfy $\nabla\cdot\mathbf{U}_{\mathrm{m}}=0$ due to measurement error.  Our goal is to:

\begin{itemize}
  \item Enforce a divergence‐free velocity field $\mathbf{U}$ by applying a Helmholtz decomposition to $\mathbf{U}_{\mathrm{m}}$ and solving a Poisson equation for a scalar potential $\phi$ via a spectral (Fourier) method.
  \item Compute the pressure field $p(x,y)$ from the corrected divergence‐free velocity $\mathbf{U}$ and the measured forcing $\mathbf{F}_{\mathrm{m}}$, ensuring consistency with the steady momentum equation \autoref{eq:momentum}.
  \item Reconstruct a corrected forcing $\mathbf{F}_{\mathrm{new}}(x,y)$ such that the momentum equation \autoref{eq:momentum} holds exactly when $\mathbf{U}$ and $p$ from steps (1)–(2) are used.
\end{itemize}

\subsubsection{Helmholtz Decomposition and Divergence Correction}

\begin{figure}
    \centering
    \includegraphics[width=\linewidth]{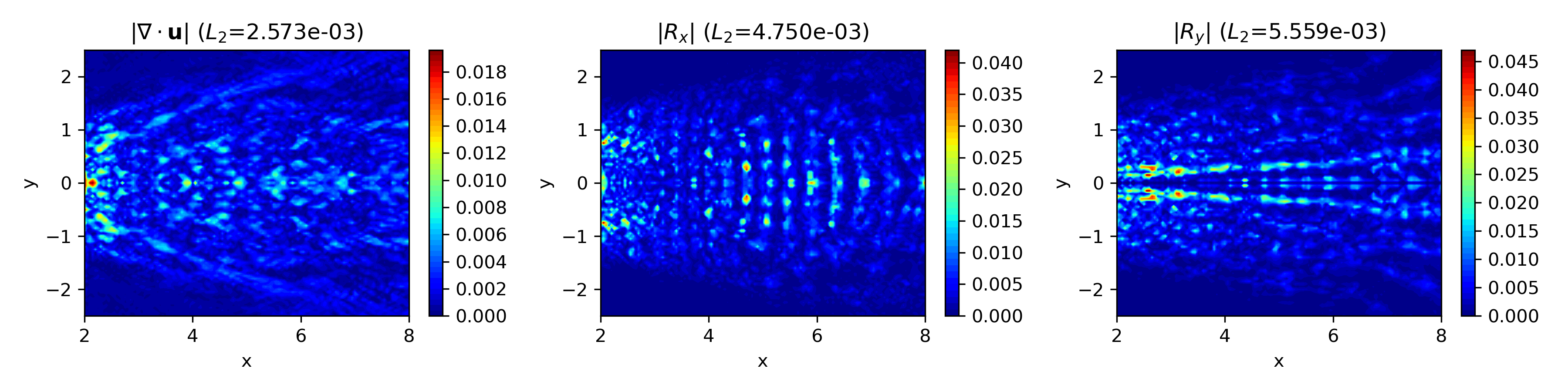}
    \caption{\textcolor{black}{Residuals of the divergence and momentum equations before correction. After the correction, all residuals become zero, and the equations are satisfied. The dataset is incompressible time-averaged DNS data at $Re=11,000$.}}
    \label{fig:data_correction_res}
\end{figure}

\begin{figure}
    \centering
    \begin{subfigure}{0.7\textwidth}
        \centering
        \includegraphics[width=\linewidth]{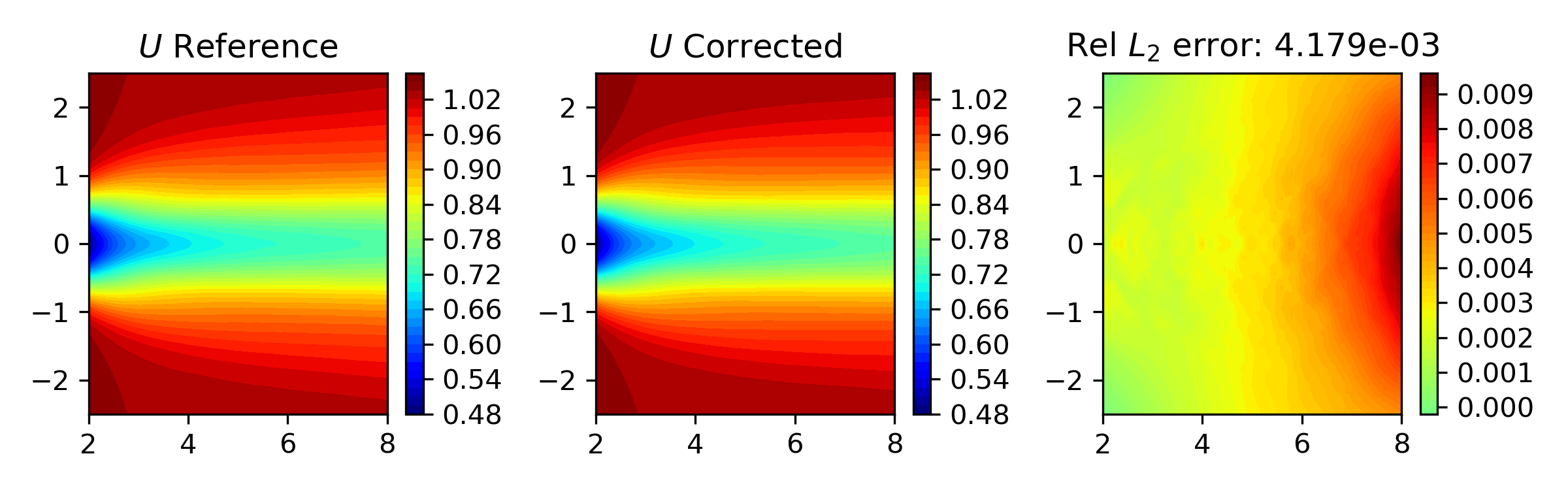}
    \end{subfigure}
    \begin{subfigure}{0.7\textwidth}
        \centering
        \includegraphics[width=\linewidth]{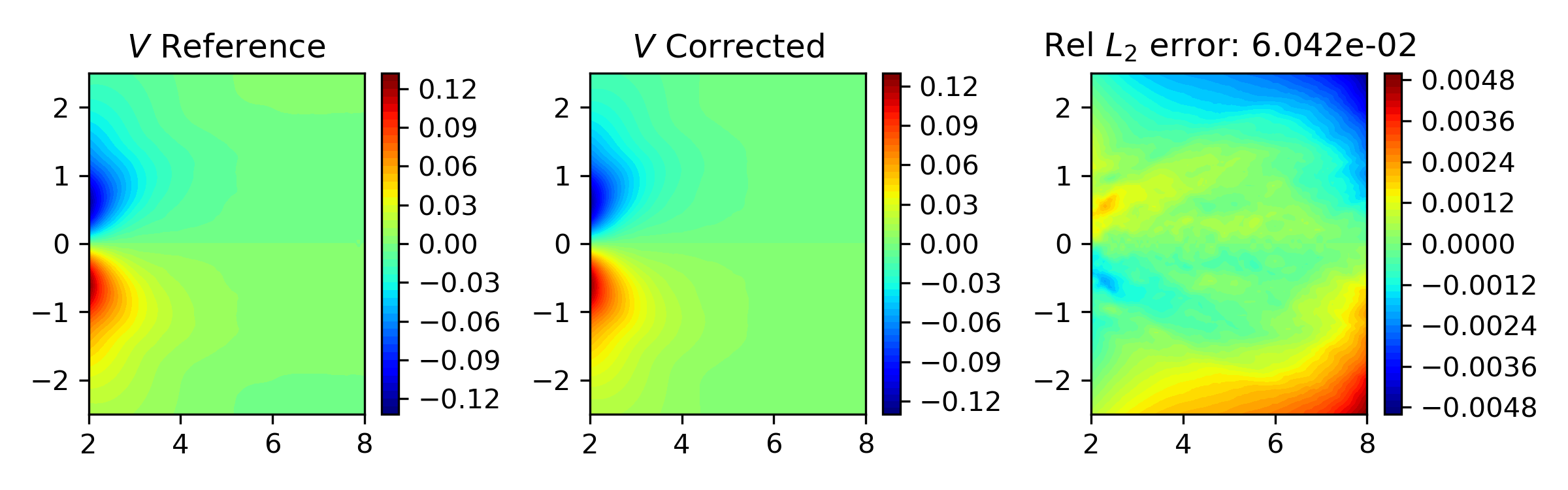}
    \end{subfigure}
    \caption{Velocity correction results based on the Helmholtz decomposition. The reference is incompressible time-averaged DNS data at $Re=11,000$. The corrected velocity components, differences, and relative $L_2$ errors are shown.}
    \label{fig:data_correction_DNS_Re11k_UV}
\end{figure}

Any sufficiently smooth vector field $\mathbf{v}(x,y)$ on a periodic domain can be decomposed uniquely into a solenoidal divergence‐free \textcolor{black}{(sol)} part and a gradient of a scalar potential:
\begin{equation}
\textcolor{black}{
  \mathbf{v}(x,y) \;=\; \mathbf{v}_{\mathrm{sol}}(x,y)\;+\;\nabla\phi(x,y)\,, 
  \qquad \nabla\cdot\mathbf{v}_{\mathrm{sol}} = 0\,.}
  \label{eq:helmholtz_general}
\end{equation}
Apply \autoref{eq:helmholtz_general} to the measured velocity $\mathbf{U}_{\mathrm{m}}(x,y)$:
\begin{equation}
  \mathbf{U}_{\mathrm{m}}(x,y)
  \;=\;\mathbf{U}(x,y)\;+\;\nabla\phi(x,y)\,, 
  \quad\text{with}\;\nabla\cdot\mathbf{U}=0.
  \label{eq:helmholtz_um}
\end{equation}
Taking the divergence of \autoref{eq:helmholtz_um} gives
\begin{equation}
  \nabla\cdot\mathbf{U}_{\mathrm{m}}
  \;=\;\nabla\cdot\bigl(\mathbf{U} + \nabla\phi\bigr)
  \;=\;0 \;+\;\nabla^{2}\phi,
  \label{eq:divergence_phi}
\end{equation}
so that $\phi$ satisfies the Poisson equation
\begin{equation}
  \nabla^{2}\phi(x,y) \;=\;\nabla\cdot\,\mathbf{U}_{\mathrm{m}}(x,y).
  \label{eq:poisson_phi}
\end{equation}
This Poisson equation can be solved by a spectral method, \textcolor{black}{and the periodic boundary condition can be satisfied by the linear padding discussed before.} Having found $\phi(x,y)$, define the corrected velocity field
\begin{equation}
  \mathbf{U}(x,y)
  \;=\;\mathbf{U}_{\mathrm{m}}(x,y)\;-\;\nabla\phi(x,y).
  \label{eq:u_corrected}
\end{equation}
Thus, $\mathbf{U}$ is divergence‐free on the discrete grid (up to spectral‐aliasing errors). \textcolor{black}{\autoref{fig:data_correction_res} shows the divergence residual before the correction, and \autoref{fig:data_correction_DNS_Re11k_UV} compares the velocity before and after the correction. Results show that the correction is small in magnitude but sufficient to satisfy the equation.}

\subsubsection{Pressure Computation from $\mathbf{U}$ and $\mathbf{F}_{\mathrm{m}}$}

For the DNS dataset, time-averaged pressure is available, so there is no need to compute pressure. However, for the PIV dataset, there is no pressure measurement, so one can obtain pressure information based on the incompressible condition.

With the divergence‐free velocity $\mathbf{U}(x,y)$ known, we now reconstruct the pressure $P(x,y)$ so that the steady momentum equation \autoref{eq:momentum} is satisfied using the measured forcing $\mathbf{F}_{\mathrm{m}}(x,y)$.  For simplicity, set $\rho=1$.

Since $\nabla\cdot\mathbf{U}=0$, the momentum \autoref{eq:momentum} can be rearranged to express the pressure gradient:
\begin{equation}
  \nabla P 
  \;=\; -\,(\mathbf{U}\cdot\nabla)\,\mathbf{U}
        \;+\;\nu\,\nabla^{2}\mathbf{U}
        \;+\;\mathbf{F}_{\mathrm{m}}(x,y)\,.
  \label{eq:pressure_gradient}
\end{equation}
To determine $P(x,y)$ (up to an additive constant), we take the divergence of \autoref{eq:pressure_gradient}, yielding a Poisson equation for the scalar field $P$:
\begin{equation}
  \nabla\cdot\bigl(\nabla P\bigr)
  \;=\;\nabla^{2}P
  \;=\; -\,\nabla\cdot\bigl[(\mathbf{U}\cdot\nabla)\,\mathbf{U}\bigr]
         \;+\;\nabla\cdot\mathbf{F}_{\mathrm{m}}.
  \label{eq:poisson_pressure}
\end{equation}
After \textcolor{black}{linearly padding to satisfy the periodic boundary condition and} solving the Poisson equation by a spectral method, we can get the pressure field $P$.

\subsubsection{Reconstruction of the Corrected Forcing $\mathbf{F}_{\mathrm{new}}$}

\begin{figure}
    \centering
    \begin{subfigure}{0.7\textwidth}
        \centering
        \includegraphics[width=\linewidth]{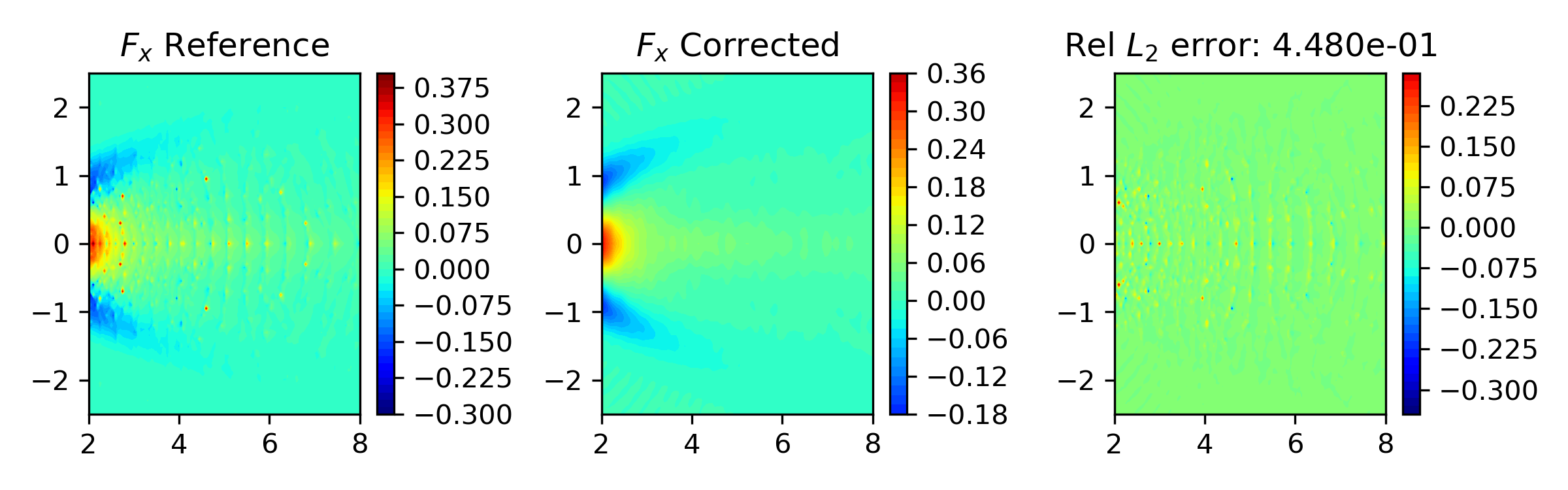}
    \end{subfigure}
    \begin{subfigure}{0.7\textwidth}
        \centering
        \includegraphics[width=\linewidth]{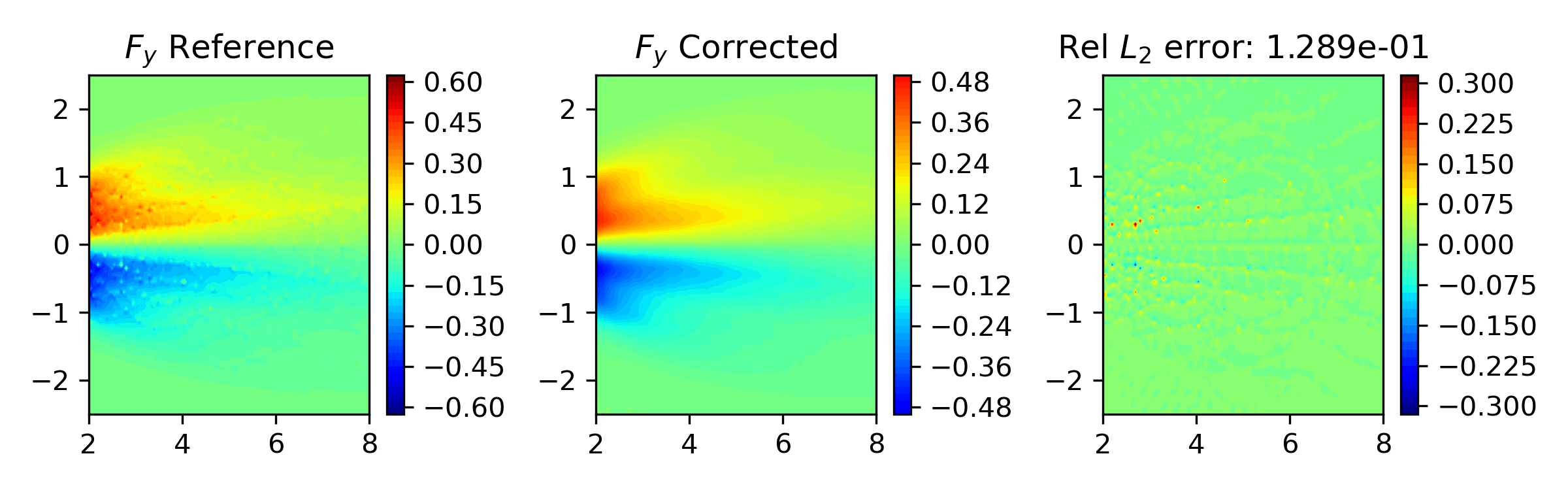}
    \end{subfigure}
    \caption{Forcing term correction results. The reference is incompressible time-averaged DNS data at $Re=11,000$. The corrected forcing term components, differences, and relative $L_2$ errors are shown.}
    \label{fig:data_correction_DNS_Re11k_F}
\end{figure}

Even though $\mathbf{F}_{\mathrm{m}}$ was used to compute $P$, the momentum equation~(\autoref{eq:momentum}) may not hold for $\mathbf{U},P,\mathbf{F}_m$, so we now define a new forcing $\mathbf{F}_{\mathrm{new}}(x,y)$ that exactly enforces \autoref{eq:momentum} for the divergence‐free $\mathbf{U}$ and the computed pressure $P$.  Rearranging \autoref{eq:momentum} yields
\begin{equation}
  \mathbf{F}_{\mathrm{new}}(x,y)
  \;=\; (\mathbf{U}\cdot\nabla)\,\mathbf{U}
         \;+\;\nabla P \;-\;\nu\,\nabla^{2}\mathbf{U}.
  \label{eq:f_new}
\end{equation}
Therefore, by construction, $\mathbf{U}$, $P$, and $\mathbf{F}_{\mathrm{new}}$ satisfy
\[
  (\mathbf{U}\cdot\nabla)\,\mathbf{U} \;=\; -\,\nabla P \;+\;\nu\,\nabla^{2}\mathbf{U} \;+\;\mathbf{F}_{\mathrm{new}}
  \quad\text{and}\quad
  \nabla\cdot\mathbf{U} = 0,
\]
so that the corrected data set $(\mathbf{U},\,p,\,\mathbf{F}_{\mathrm{new}})$ is \emph{self‐consistent} with the steady incompressible RANS momentum equation. \textcolor{black}{\autoref{fig:data_correction_res} shows the residual of momentum equations before the correction, and \autoref{fig:data_correction_DNS_Re11k_F} compares the Reynolds forces before and after the correction. Again, the correction keeps the pattern of Reynolds forces, reduces the noise coming from non-smoothnesses at the cell boundaries in the spectral element method in DNS data, and is sufficient to satisfy the governing equations.} A similar process can be used to force an unsteady dataset satisfying the unsteady incompressible NS equation, such as the phase-averaged URANS and LES datasets.

\section{Flow Inference}\label{sec:infer}

In PIV measurements, there is a trade-off between field of view and spatial resolution. One can only have high spatial resolution in some limited domains. We aim to use PINNs to infer the flow field in the whole domain based on limited measurements.

\subsection{Flow inference with unclosed RANS equation}
In this section, we formulate a problem where we assume that only measurable quantities on the domain boundary are known. We use PINNs with the unclosed RANS equation with the Reynolds force vector to infer all quantities within that domain. 

We select a domain in the wake behind the cylinder. The size of the domain $\Omega$ is $x\in[2,8], y\in[-2.5,2.5]$. The time-averaged velocity $U, V$ and Reynolds force vector $F_x, F_y$ at the domain boundary are used to calculate the data loss $L_{data}$. The number of the boundary points is $N_{BC}=440$; $N_{PDE}=50,000$ randomly sampled interior points are used to calculate the PDE loss $L_{PDE}$. The total loss is defined as 
\begin{equation}
    L_{total} = L_{data} + \lambda_{PDE}L_{PDE},
\end{equation}
where the weight $\lambda_{PDE}$ is used to control the contribution of the PDE loss and is changed adaptively. 

\begin{figure}
    \centering
    \includegraphics[width=0.6\linewidth]{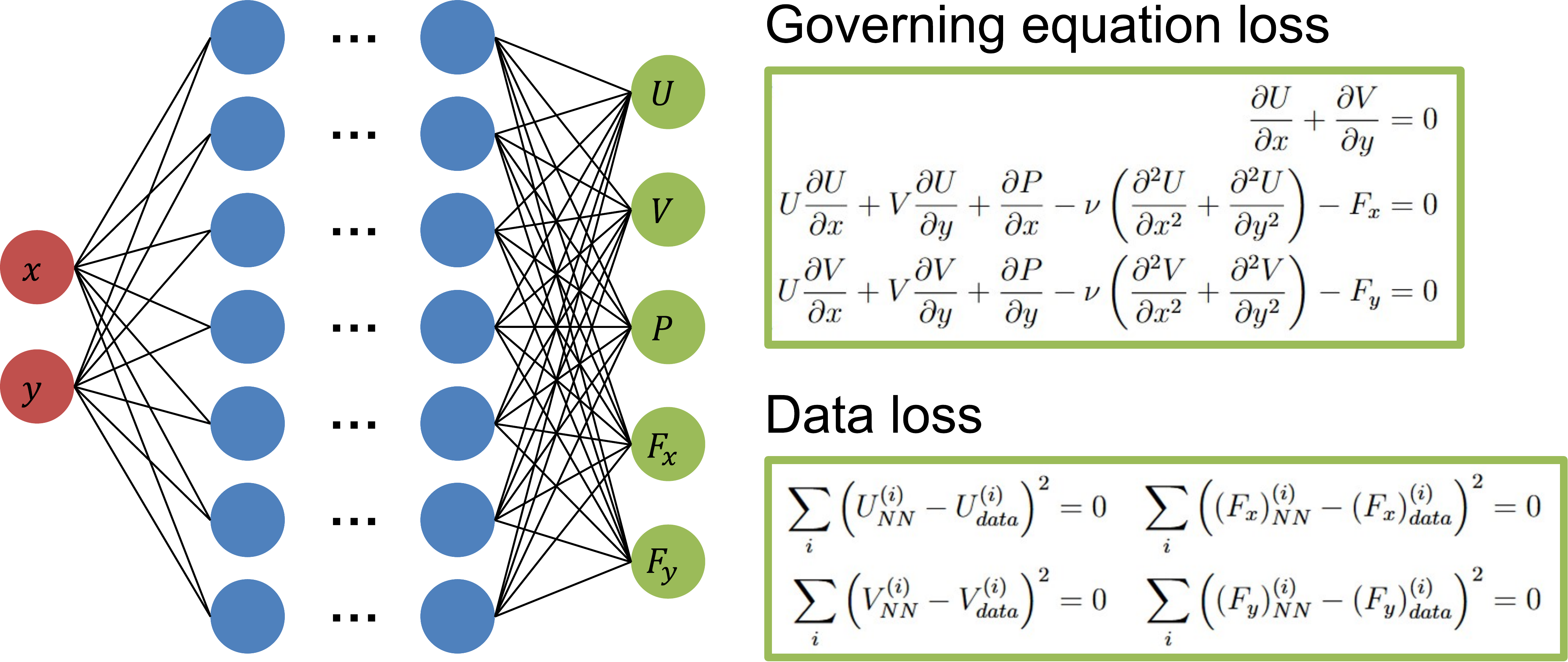}
    \caption{PINNs architecture for flow inference. An MLP is used to approximate flow fields. The 2D incompressible Navier-Stokes equation with the unclosed Reynolds force is used as physical regularization. Measurable variables $U$, $V$, $F_x$, and $F_y$ at the boundary are used for data loss.}
    \label{fig:PINN_incom}
\end{figure}

\autoref{fig:PINN_incom} illustrates the PINN architecture employed in this study. A multilayer perceptron (MLP) is used to predict the state variables, consisting of four hidden layers with 64 neurons per layer and \(\tanh\) activation functions. The network is trained using the Adam optimizer for \(N_{\text{epochs}} = 100{,}000\) full-batch iterations. \textcolor{black}{The initial learning rate is set to \(1 \times 10^{-3}\), and a step-decay scheduler is adopted, reducing the learning rate by a factor of 0.3 every 10{,}000 iterations for a total of five decay steps. In parallel, the PDE loss weight \(\lambda_{\mathrm{PDE}}\) is gradually increased from 0.01 to 1 during training. An example evolution of the learning rate and \(\lambda_{\mathrm{PDE}}\), together with the training loss and testing error, is provided in the Appendix (\autoref{fig:flow_inference_DNS_Re11k_history}).}
To further accelerate convergence, the residual-based attention (RBA) strategy proposed by \citet{rba} is employed. The RBA parameters in Eq.~(12) of their work are set to \(\gamma = 0.99\) and \(\eta = 0.1\), which allow the spatially varying weights at interior and boundary points with large local residuals to be amplified by up to a factor of 10.

\begin{figure}
    \centering
    \begin{subfigure}{0.9\textwidth}
        \centering
        \includegraphics[width=\linewidth]{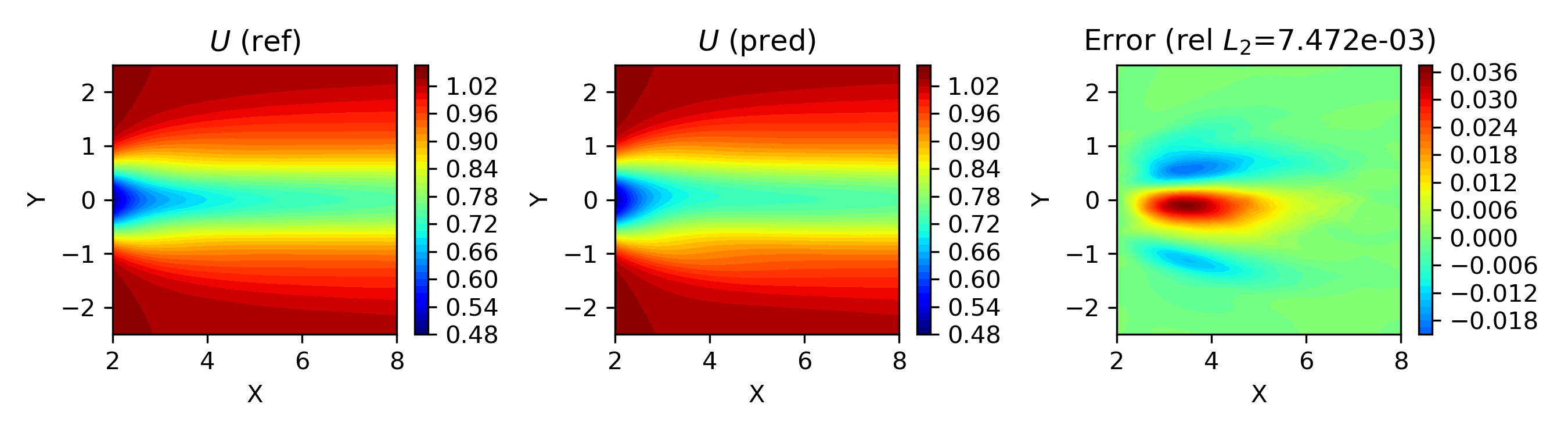}
    \end{subfigure}
    \begin{subfigure}{0.9\textwidth}
        \centering
        \includegraphics[width=\linewidth]{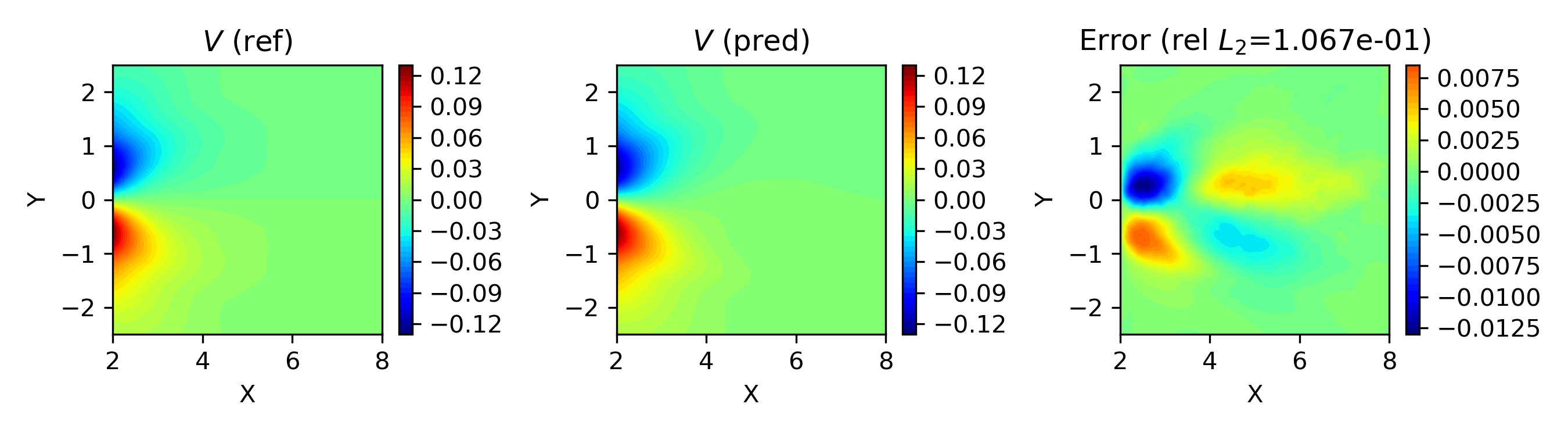}
    \end{subfigure}
    \begin{subfigure}{0.9\textwidth}
        \centering
        \includegraphics[width=\linewidth]{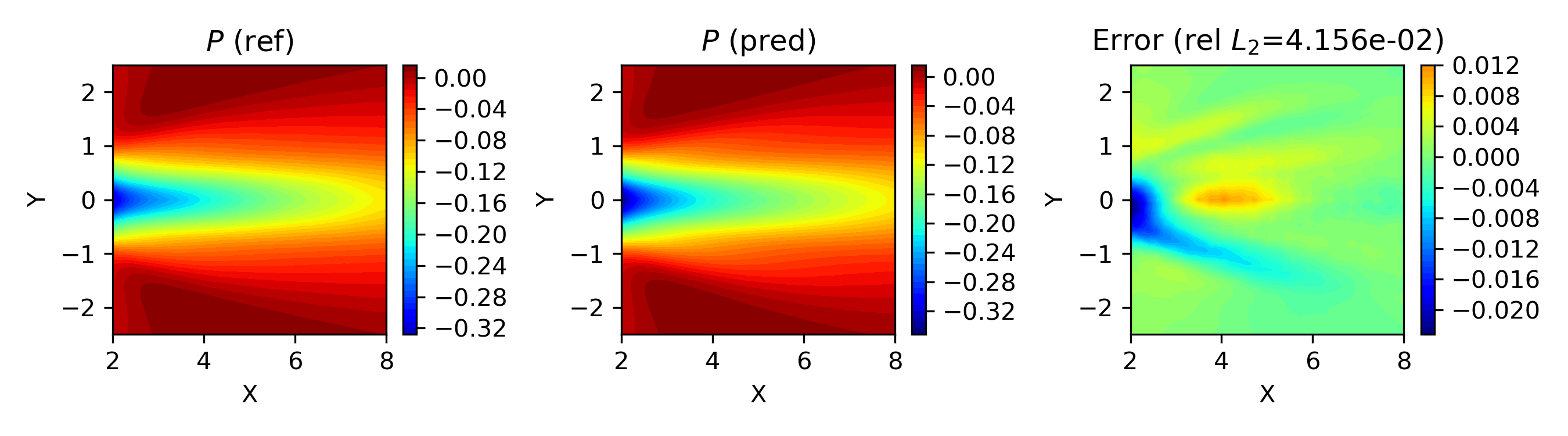}
    \end{subfigure}
    \begin{subfigure}{0.9\textwidth}
        \centering
        \includegraphics[width=\linewidth]{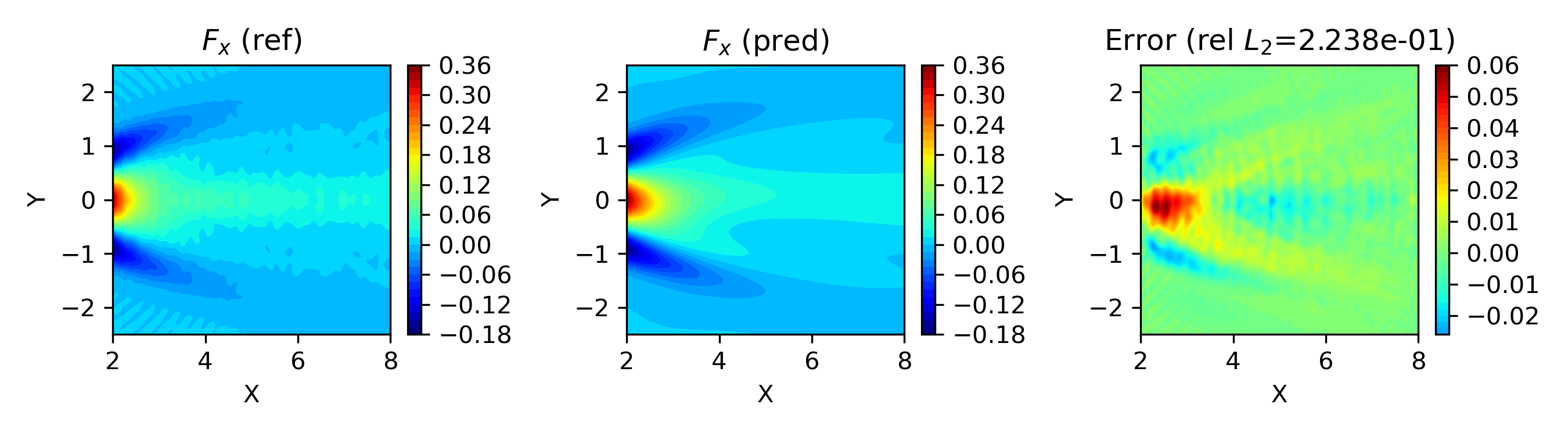}
    \end{subfigure}
    \begin{subfigure}{0.9\textwidth}
        \centering
        \includegraphics[width=\linewidth]{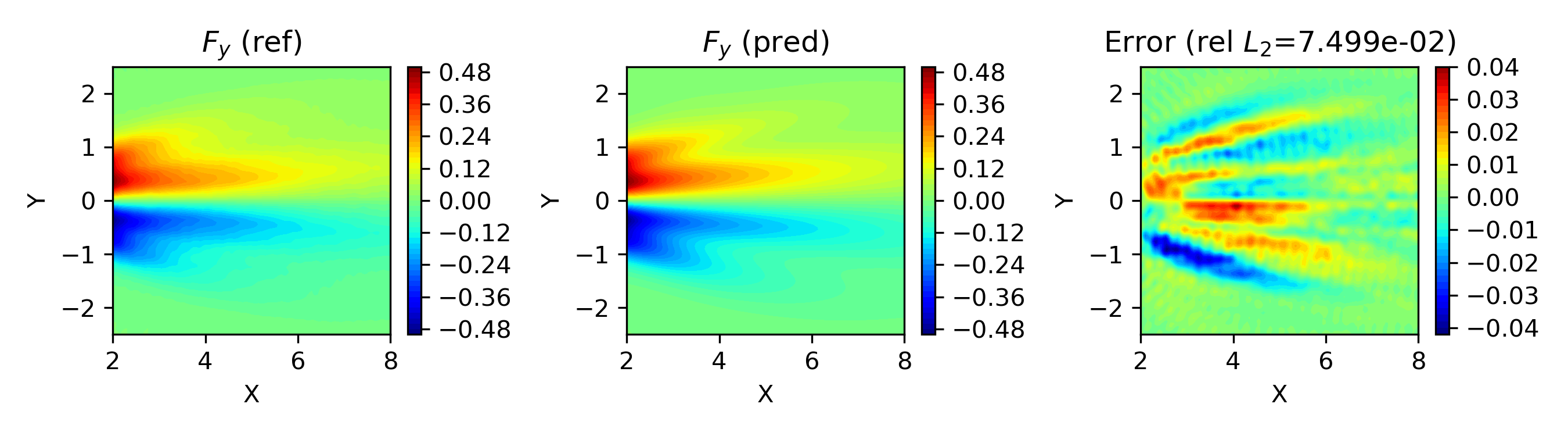}
    \end{subfigure}
    \caption{Flow inference for incompressible cylinder flow at $Re=11,000$. The first column is reference values, the second column is PINNs reconstruction, and the last column is the pointwise error. Reference data is corrected time-averaged DNS. Prediction is obtained by PINNs only using the measurable data of $U,V,F_x,F_y$ at the domain boundary. Error distributions and the relative $L_2$ errors are shown in the third column.}
    \label{fig:flow_inference_DNS_Re11k_compare}
\end{figure}

\autoref{fig:flow_inference_DNS_Re11k_compare} compares the prediction of PINNs against the reference DNS data. The relative $l_2$ error is also shown, which is defined as
\begin{equation}
    E=\frac{||U_{pred}-U_{ref}||_{l_2(\Omega)}}{||U_{ref}||_{l_2(\Omega)}}.
\end{equation}
All flow fields are satisfactorily inferred. In particular, no pressure information is available for PINNs, but PINNs can infer a pressure field with around $5-10\%$ error. The forcing terms are also well inferred, and the prediction is smoother than the reference, where there is some noise due to the nature of the weak solution that is obtained from the spectral element method.

\autoref{tab:error_summary} summarizes the $L_2$ errors in the flow inference problem at different Reynolds numbers with and without the physics-informed correction. We see that the physics-informed correction can distinctly enhance the accuracy in PINNs flow inference. This implies that when data itself satisfies the equation, the physics regularization in PINNs will better determine the solution. Key flow inference results based on corrected data are shown in \autoref{app:flow_inference}. Three cases are based on corrected aerodynamic PIV measurements. In these cases, the inlet Mach number is less than $0.15$, and the maximal value of the Mach number in the field is less than $0.3$. Thus, these data can be treated as incompressible.
\begin{table}
  \caption{Relative $L_2$ errors of $U, V, P, F_x, F_y$ of the flow inference based on different datasets. Results obtained from corrected (default) and uncorrected data are compared. Data loss is calculated only at the domain boundary.}
  \centering
  \begin{tabular}{lccccc}
    \toprule
    Case & $e_U [\%]$& $e_V [\%]$& $e_P [\%]$& $e_{F_x} [\%]$& $e_{F_y} [\%]$\\
    \midrule
    DNS, $Re=3,900$   & 0.884& 5.91& 9.95& 21.1& 16.2\\
    DNS, $Re=5,000$   & 0.568& 5.39& 6.76& 16.5& 14.3\\
    DNS, $Re=11,000$ &0.747& 10.7& 4.16& 22.4& 7.50\\
    LES, $Re=60,000$ & 0.829& 21.9& 4.78& 29.6& 12.0\\
    LES, $Re=140,000$ & 0.575& 12.3& 3.73& 24.5& 9.46\\
    Aerodynamic PIV, $Re=6,500$ & 0.672& 3.00& 6.00& 17.6& 12.3\\
    Aerodynamic PIV, $Re=11,000$ & 0.660& 2.63& 7.14& 20.0& 16.7\\
    Aerodynamic PIV, $Re=50,000$ & 1.23& 5.35& 4.13& 15.1& 10.1\\
    Hydrodynamic PIV, $Re=10,000$ & 0.379& 4.68& 1.73& 21.2& 10.4\\
    Hydrodynamic PIV, $Re=30,000$ & 0.813& 8.40& 5.89& 26.0& 17.3\\
    Hydrodynamic PIV, $Re=60,000$ & 0.781& 5.69& 3.06& 15.0& 15.2\\
    DNS, $Re=3,900$ (uncorrected)& 0.906& 5.39& 15.6& 32.6& 43.7\\
    DNS, $Re=5,000$ (uncorrected)& 1.79& 10.8& 16.8& 56.3& 42.0\\
    DNS, $Re=11,000$ (uncorrected)          & 0.796& 12.5& 11.3& 56.7& 31.1\\
    LES, $Re=60,000$ (uncorrected)& 1.49& 37.3& 19.1& 114& 68.9\\
    LES, $Re=140,000$ (uncorrected)& 1.24& 28.0& 13.5& 94.8& 47.8\\
    \bottomrule
  \end{tabular}
  \label{tab:error_summary}
\end{table}

\autoref{tab:internal_data} shows the flow inference where the data loss is calculated using more data than the domain boundary data only. Concretely, uniformly distributed (for instance, $5\times5$) internal points are added to the data loss. Like boundary information, only measurable variables $U, V, F_x, F_y$ are used at the internal points. We see that more internal data points can more accurately determine the solution, even for the one-point case. Key results shown in this table are plotted in \autoref{app:flow_inference}.
\begin{table}
\caption{Relative $L_2$ errors of $U, V, P, F_x, F_y$ of the flow inference using the corrected time-averaged DNS data at $Re=11,000$. The data loss is calculated not only at the boundary, but also using internal points, which are uniformly distributed within the domain. Measurable $U, V, F_x, F_y$ are given at the boundary and at those internal points.}
\centering
\begin{tabular}{lccccc}
\toprule
Case & $e_U [\%]$& $e_V [\%]$& $e_P [\%]$& $e_{F_x} [\%]$& $e_{F_y} [\%]$\\
\midrule
BC only &0.747& 10.7& 4.16& 22.4& 7.50\\
BC + 1 $\times$ 1   & 0.279& 6.65& 7.29& 14.6& 9.03\\
BC + 3 $\times$ 3   & 0.176& 2.46& 4.28& 10.7& 9.53\\
BC + 5 $\times$ 5   & 0.100& 2.03& 2.81& 10.8& 4.51\\
\bottomrule
\end{tabular}
\label{tab:internal_data}
\end{table}

\subsection{Flow inference with Helmholtz and Turbulence augmented approach using boundary data for Re=3,900}
In this study, we consider the divergence of the Reynolds stress tensors is treated as a forcing vector $[F_x, F_y]^{\top}$ (the Reynolds forcing vector), thereby reducing the closure problem from six individual Reynolds stresses to three forcing terms (or to two terms when there is no spanwise mean flow). However, inferring all these flowfield becomes an ill-posed problem as we have three equations and  five unknowns. To reduce this seemingly ill-posednes of the problem further and hence gain inference accuracy, we apply a Helmholtz decomposition on Reynolds forcing vector, following the approaches of \cite{foures2014data} and \cite{sliwinski2023mean}. In this decomposition, the forcing is decomposed into a scalar (potential) component, 
denoted by $\phi$, and a divergence-free (solenoidal) vector component, denoted by $F_{s,i}$. The divergence-free condition of the latter provides an additional equation in addition to \autoref{eq:RANS_Euqtaion}. This decomposition is expressed as
\begin{align}\label{eq:hz_decomp}
\bm{F}=[F_x, F_y]^\top \equiv \left[\frac{1}{\rho} \frac{\partial \phi}{\partial x}+F_{s, x}, \frac{1}{\rho} \frac{\partial \phi}{\partial y}+F_{s, y}\right]^\top,
\end{align}
where $F_{s, x}$ and $F_{s, y}$ represent the solenoidal components of forces in $x$ and $y$ directions.
Substituting the \autoref{eq:hz_decomp} into the \autoref{eq:RANS_Euqtaion} yields a system of four partial differential equations as shown in \autoref{eq:RANS_HZ}. In \cite{patel2024turbulence}, \autoref{eq:RANS_HZ} is employed to reconstruct the mean flow field using a PINN for laminar flow past a cylinder at $Re = 200$. In the present work, we infer both the Reynolds forcing and the mean flow field for turbulent flow past a cylinder at $Re = 3,900$.
\begin{figure}
    \centering
   \includegraphics[width=0.8\textwidth]{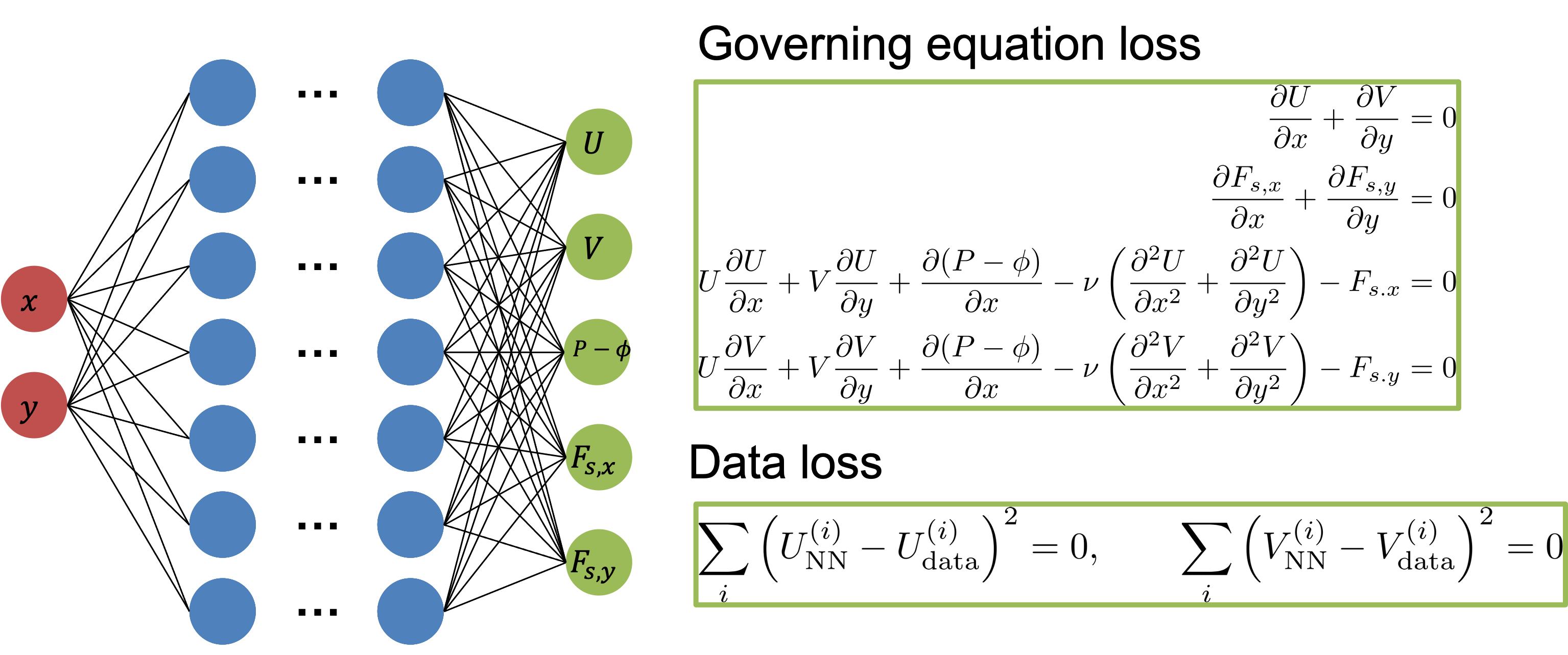}
    \caption{Architecture of the PINN and the data distribution used for its training. A deep neural network, parameterized by a set of weights and biases, approximates a continuous mapping from spatial coordinates to flow variables. The network is trained using high-fidelity coarse measurements and the RANS equation modified by the Helmholtz decomposition, as shown in \autoref{eq:RANS_HZ}.}
\label{fig:pinn_hd_setup}
\end{figure}

\subsubsection{PINNs architecture, hyper-parameters and results on flow inference}
The overall architecture of the Physics-Informed Neural Network (PINN) employed to infer the flow field is illustrated in \autoref{fig:pinn_hd_setup}. In \autoref{fig:pinn_hd_setup}, a deep neural network, parameterized by a set of trainable weights and biases, is used to approximate a continuous mapping from the spatial coordinates to the flow variables of interest. This network is trained using  (i) high-fidelity but coarse measurements prescribed along the boundary, and (ii) the Reynolds–Averaged Navier–Stokes (RANS) equations modified via the Helmholtz decomposition, as formulated in \autoref{eq:RANS_HZ}. For training, we choose a domain for PINN training ranging between $[1.5,7.5] \times [-2, 2]$. We employ a fully connected neural network with 10 hidden layers, each consisting of 32 neurons. The optimization is performed in two stages: first, the network is trained for 15,000 iterations using the first-order Adam optimizer to achieve a robust initial fit, and then the optimization switches to a quasi-Newton BFGS algorithm with backtracking (\cite{kiyani2025optimizing}) line search for an additional 8,000 iterations to refine convergence. The training data consists of 1,000 boundary points and 20,000 residual points, sampled independently and identically distributed (i.i.d.) across the respective domains. 

\begin{figure}
  \begin{subfigure}[b]{\textwidth}
    \centering
     \includegraphics[width=\textwidth, trim=100 80 100 100, clip]{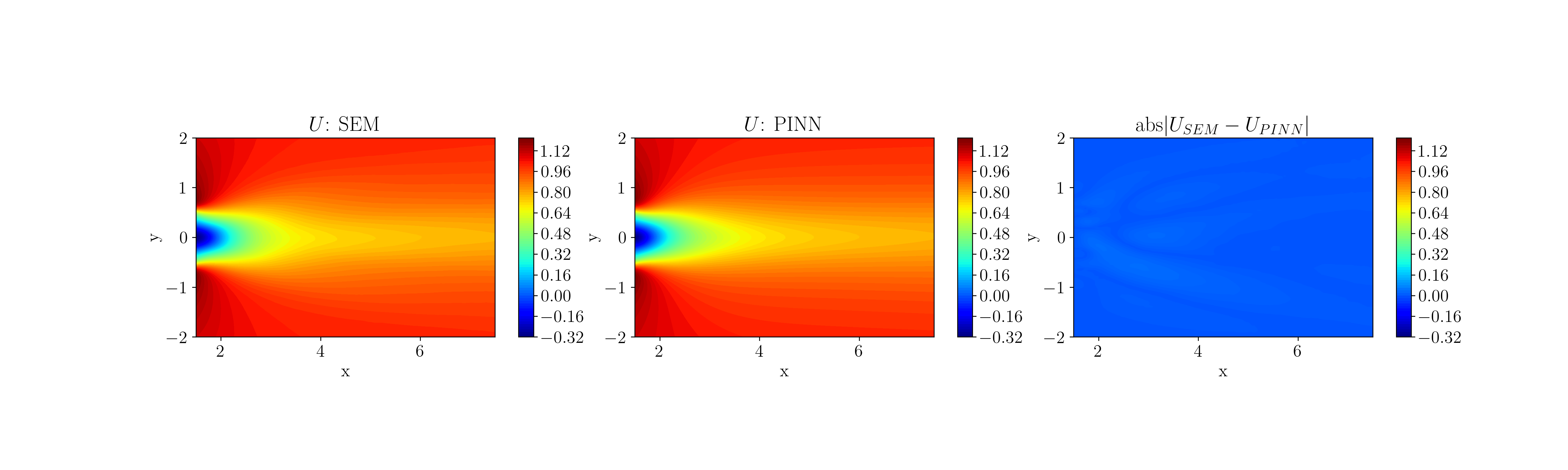}
    \caption{Reference (SEM) vs Inferred (PINN) $(U)$ using the \autoref{eq:RANS_HZ}, Rel. $L_2$ err: 0.98\%}
    \label{fig:pinn_hd_u}
  \end{subfigure}
  \hfill
  \\
  \begin{subfigure}[b]{\textwidth}
    \centering
   \includegraphics[width=\textwidth, trim=100 80 100 100, clip]{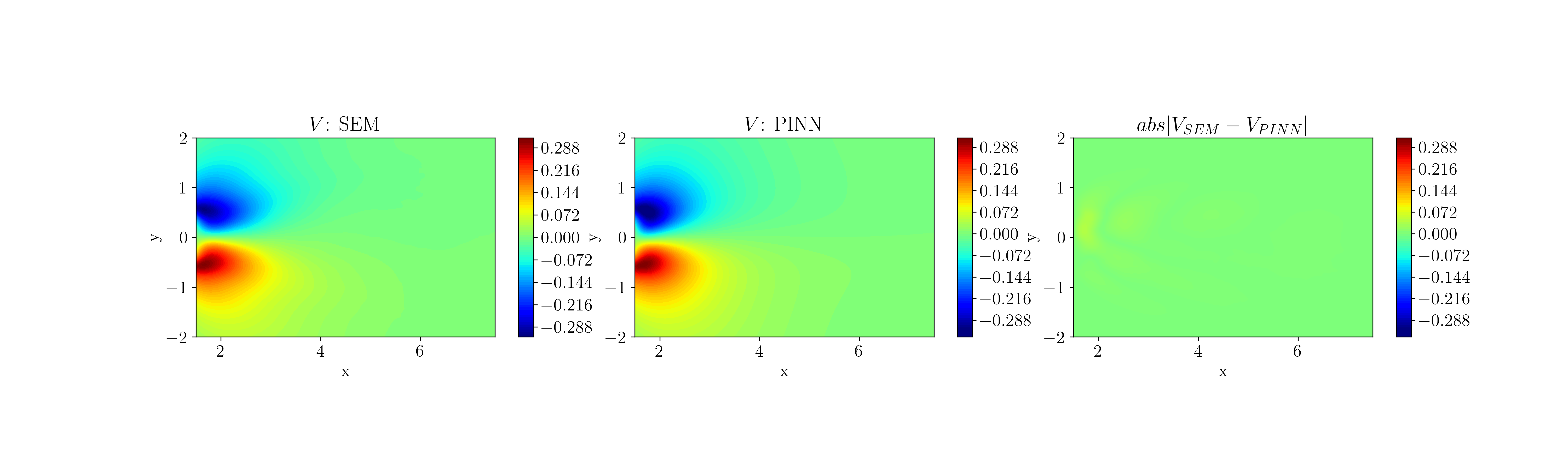}
    \caption{Reference (SEM) vs Inferred (PINN) $(V)$ using the \autoref{eq:RANS_HZ}, Rel. $L_2$ err: 7.13\%}
    \label{fig:pinn_hd_v}
  \end{subfigure}
   \begin{subfigure}[b]{\textwidth}
    \centering
   \includegraphics[width=\textwidth, trim=100 80 100 100, clip]{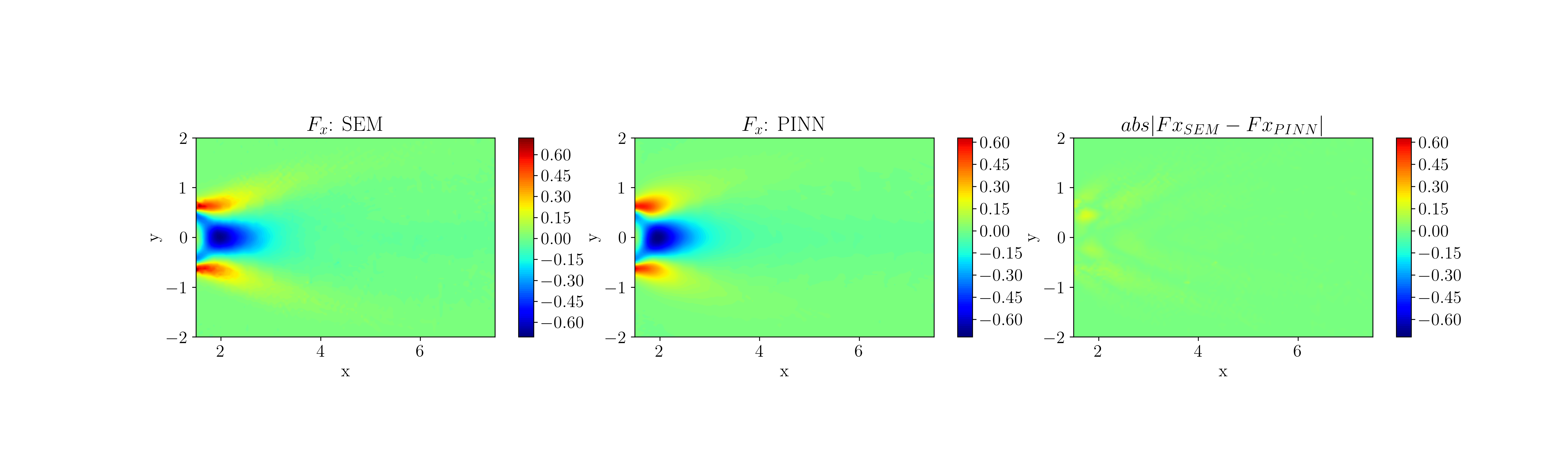}
    \caption{Reference (SEM) vs Inferred (PINN) $(F_x)$ using the \autoref{eq:RANS_HZ}, Rel. $L_2$ err: 16.89\%}
    \label{fig:pinn_hd_fx}
  \end{subfigure}
     \begin{subfigure}[b]{\textwidth}
    \centering
   \includegraphics[width=\textwidth, trim=100 80 100 100, clip]{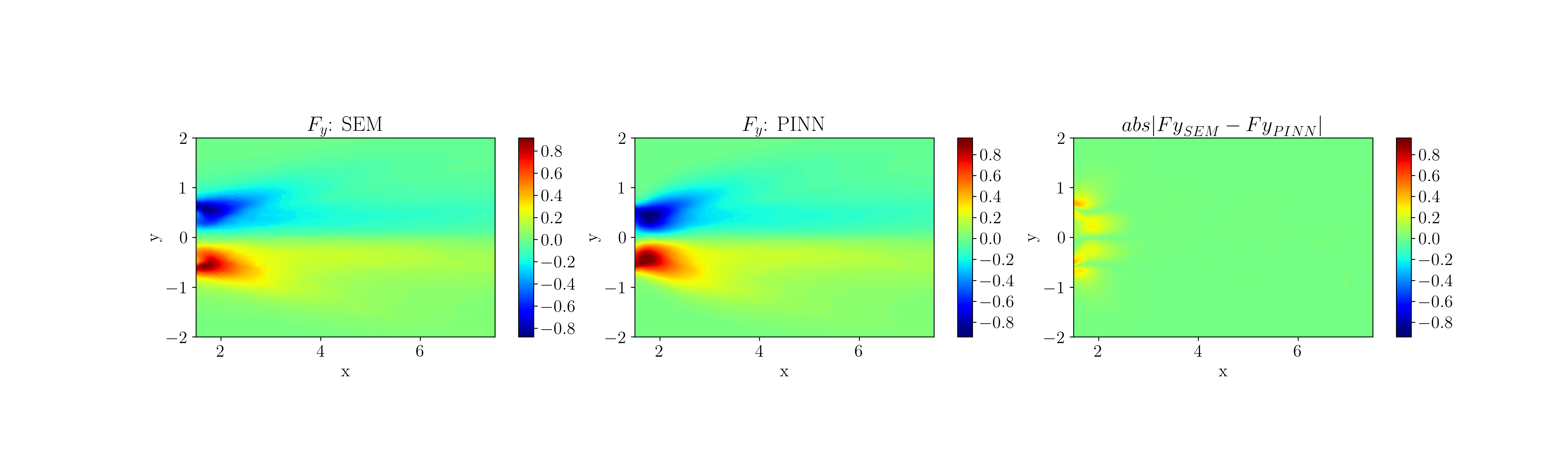}
    \caption{Reference (SEM) vs Inferred (PINN) $(F_y)$ using the \autoref{eq:RANS_HZ}, Rel. $L_2$ err: 26.61\%}
    \label{fig:pinn_hd_fy}
  \end{subfigure}
  \caption{Comparison between the reference and PINN-inferred flow fields:
(a)~$U$, (b)~$V$, (c)~$F_{x}$, and (d)~$F_{y}$ obtained using boundary data with the Helmholtz decomposition \autoref{eq:RANS_HZ}. The left, middle, and right panels in each subfigure display the reference flow field (computed using the spectral element method), the PINN-inferred flow field, and the absolute pointwise error, respectively.
Note that the flow fields reconstructed using the Helmholtz decomposition exhibit higher accuracy than those predicted with the standard RANS equation.}
\label{fig:pinn_hd_all}
\end{figure}

The reconstructed velocity fields $(U, V)$ fields obtained after training the PINN are presented in \autoref{fig:pinn_hd_all}. Panels (a), (b), (c), and (d) display the inferred distributions of $U$, $V$, $F_{x}$, and $F_{y}$, respectively. A quantitative comparison of the relative $L_{2}$ errors between the reference solution and the PINN-predicted flow fields—using both the standard RANS equation \autoref{eq:RANS_Euqtaion} and the RANS equation augmented with the Helmholtz decomposition \autoref{eq:RANS_HZ}—is provided in \autoref{tab:RANS_models}. As shown in \autoref{fig:pinn_hd_all}(a) and summarized in \autoref{tab:RANS_models}, incorporating the Helmholtz decomposition into the PINN framework yields substantially improved predictions over those obtained using the unmodified RANS equation with only boundary data. Specifically, the error in $V$ is reduced by approximately a factor of two, while the accuracy in $U$ improves by nearly a factor of three. The scale of $V$ is very small and it is very hard to recover the solution for $V$. The prediction errors in $F_{x}$ and $F_{y}$ also show marked improvement under the Helmholtz-decomposed formulation. 

\begin{table}
\centering
\caption{Comparison of errors in inferred flows using RANS, RANS with Helmholtz decomposition, and RANS with Helmholtz decomposition combined with a turbulence model.}
\label{tab:RANS_models}
\begin{tabular}{cccc}
\hline
\textbf{Flowfields} & \textbf{RANS model} (\autoref{eq:RANS_Euqtaion})  & \textbf{HD-RANS} (\autoref{eq:RANS_HZ}) & \textbf{with Turbulence model} (\autoref{eq:RANS_HZ_SA}) \\ \hline
$U$ & 3.25\% & 0.98\% & 0.83\% \\ 
$V$ & 14.5\% & 7.13\% & 4.00\% \\ 
$F_x$ & 38.6\% & 16.89\% & 22.66\% \\ 
$F_y$ & 37.5\% & 26.61\% & 23.85\% \\ \hline
\end{tabular}
\end{table}

To further investigate the capabilities of the physics-informed framework for flow-field inference, we augment the governing PINN formulation with a turbulence model tailored for the wake region, as proposed by \cite{spalart1992one}. The resulting augmented system of equations is expressed in \autoref{eq:RANS_HZ_SA}. For training this network, we use the same set of hyperparameters as in the previous case that employed only the Helmholtz decomposition. The architecture of PINN and flow inference results are discussed in \autoref{app:RANS_SA}. A relative $L_2$ error metric is provided in \autoref{tab:RANS_models}, which compares the relative $L_2$ error for all the three variants of RANS equations (\autoref{eq:RANS_Euqtaion}, \autoref{eq:RANS_HZ}, \autoref{eq:RANS_HZ_SA}). It is to be noted that the flow fields reconstructed using the turbulence-augmented model \autoref{eq:RANS_HZ_SA} exhibit higher accuracy than those predicted by both the standard RANS equation and the RANS equation with Helmholtz-decomposed forcing—except for $F_x$. This discrepancy may arise from the turbulence model’s limited accuracy in the wake region.

\section{Turbulence Closure Model}\label{sec:closure}
\subsection{Similarity of Reynolds Stress for different Reynolds Numbers}

\begin{figure}
    \centering
      \begin{subfigure}{0.45\textwidth}
    \includegraphics[width=\textwidth]{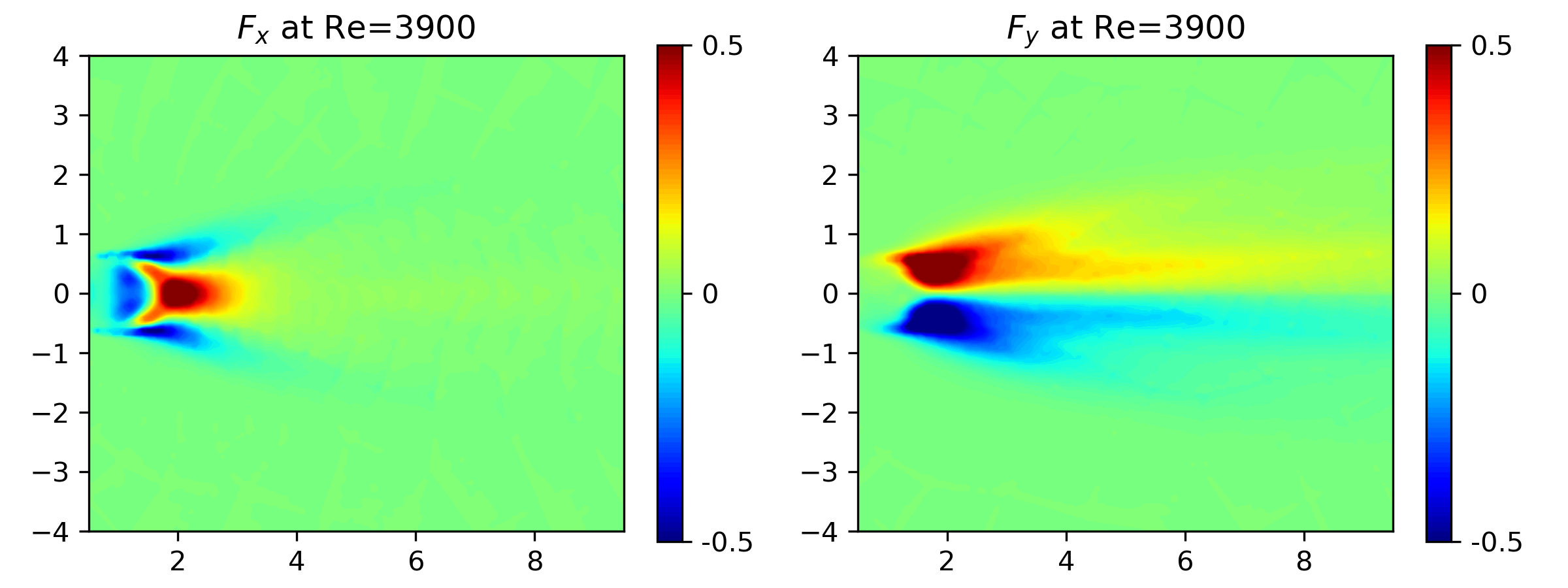}
    \caption{$Re=3,900$}
    \end{subfigure}
      \begin{subfigure}{0.45\textwidth}
    \includegraphics[width=\textwidth]{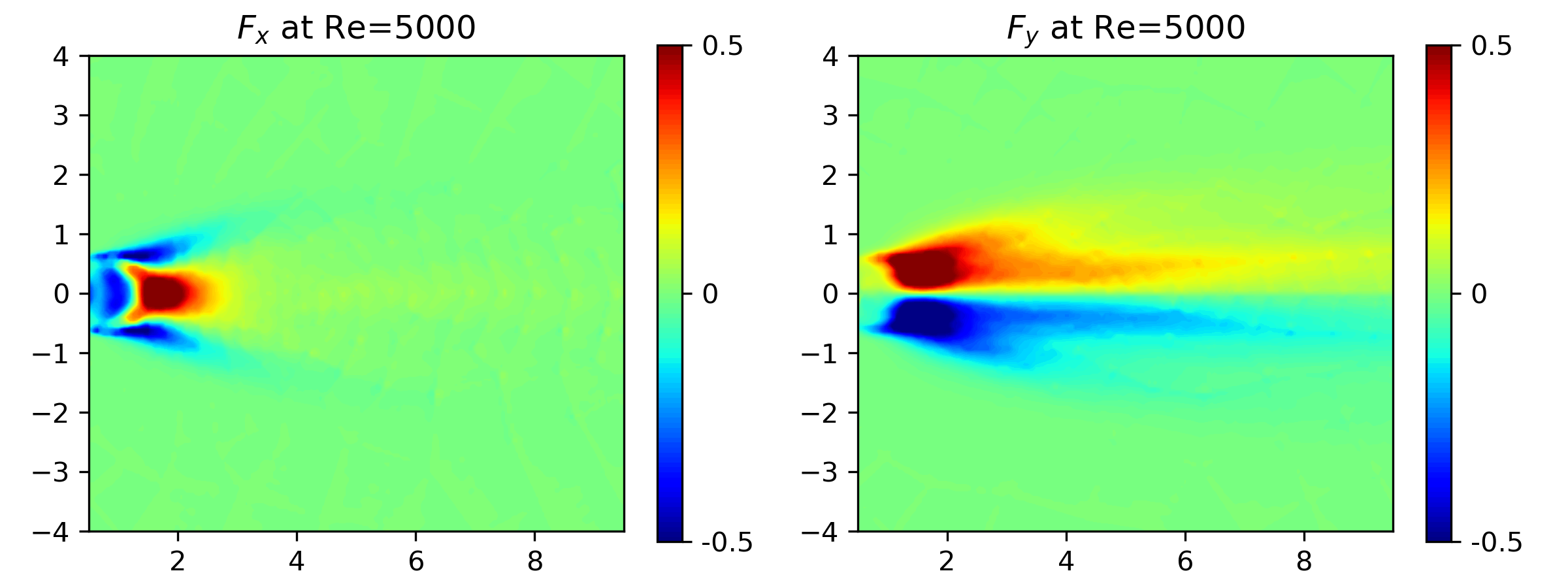}
    \caption{$Re=5,000$}
    \end{subfigure}
     \begin{subfigure}{0.45\textwidth}
        \includegraphics[width=\textwidth]{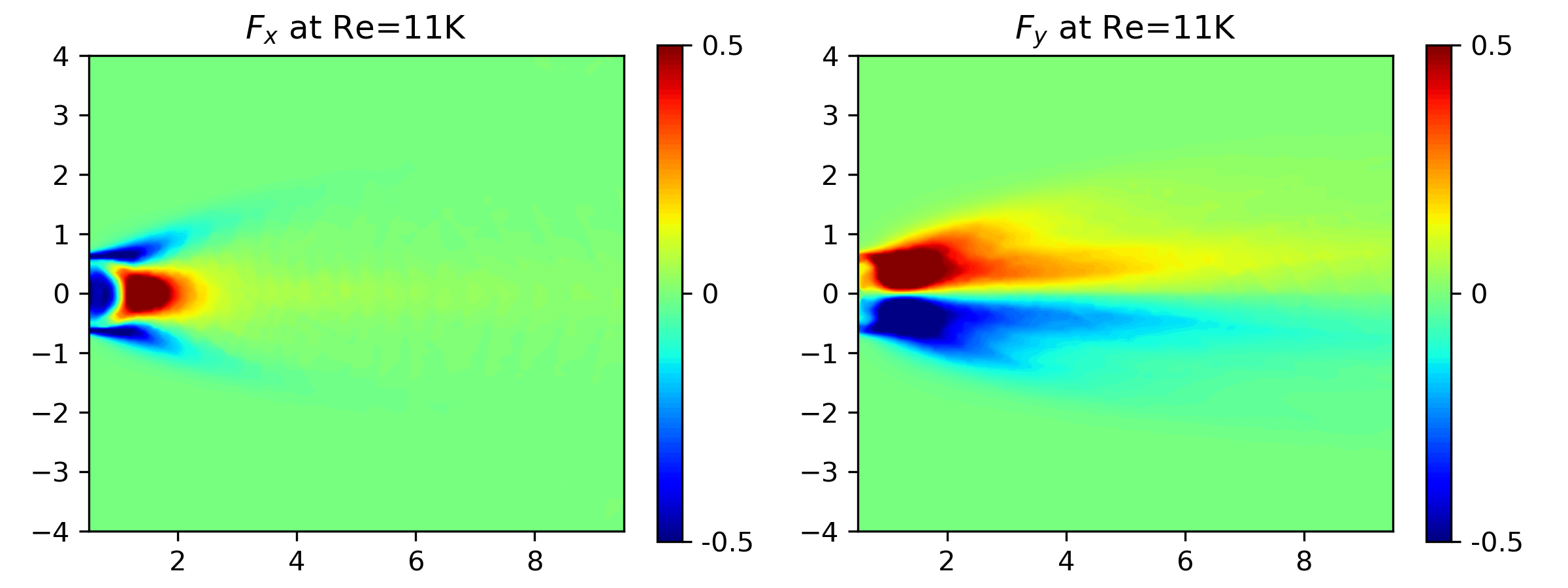}
        \caption{$Re=11,000$}
           \end{subfigure}
        \begin{subfigure}{0.45\textwidth}
        \includegraphics[width=\textwidth]{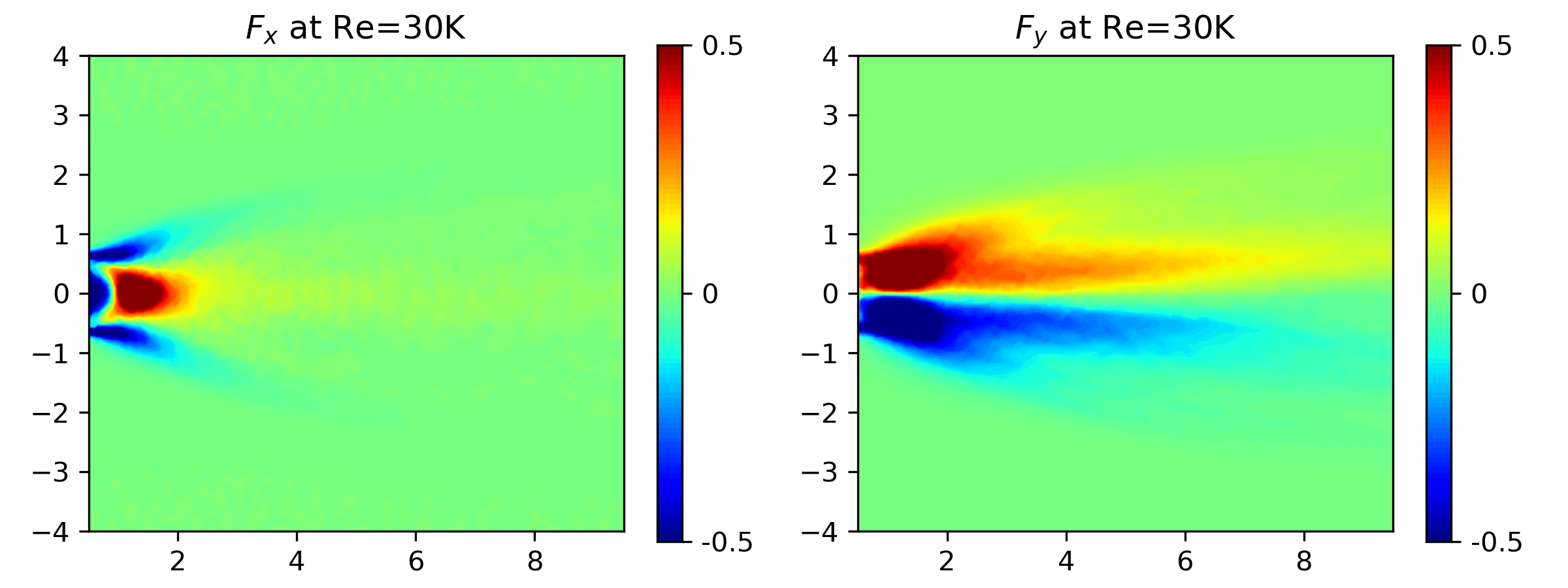}
        \caption{$Re=30,000$}
           \end{subfigure}
        \begin{subfigure}{0.45\textwidth}
        \includegraphics[width=\textwidth]{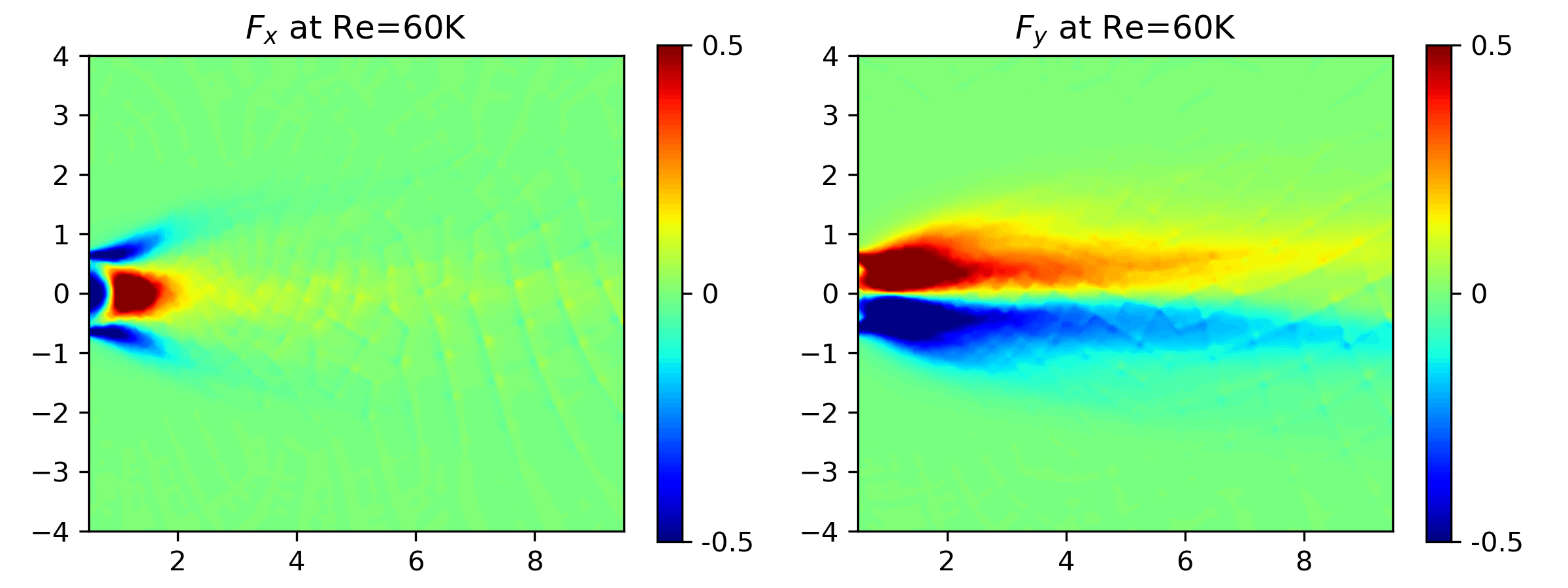}
        \caption{$Re=60,000$}
             \end{subfigure}
        \begin{subfigure}{0.45\textwidth}
        \includegraphics[width=\textwidth]{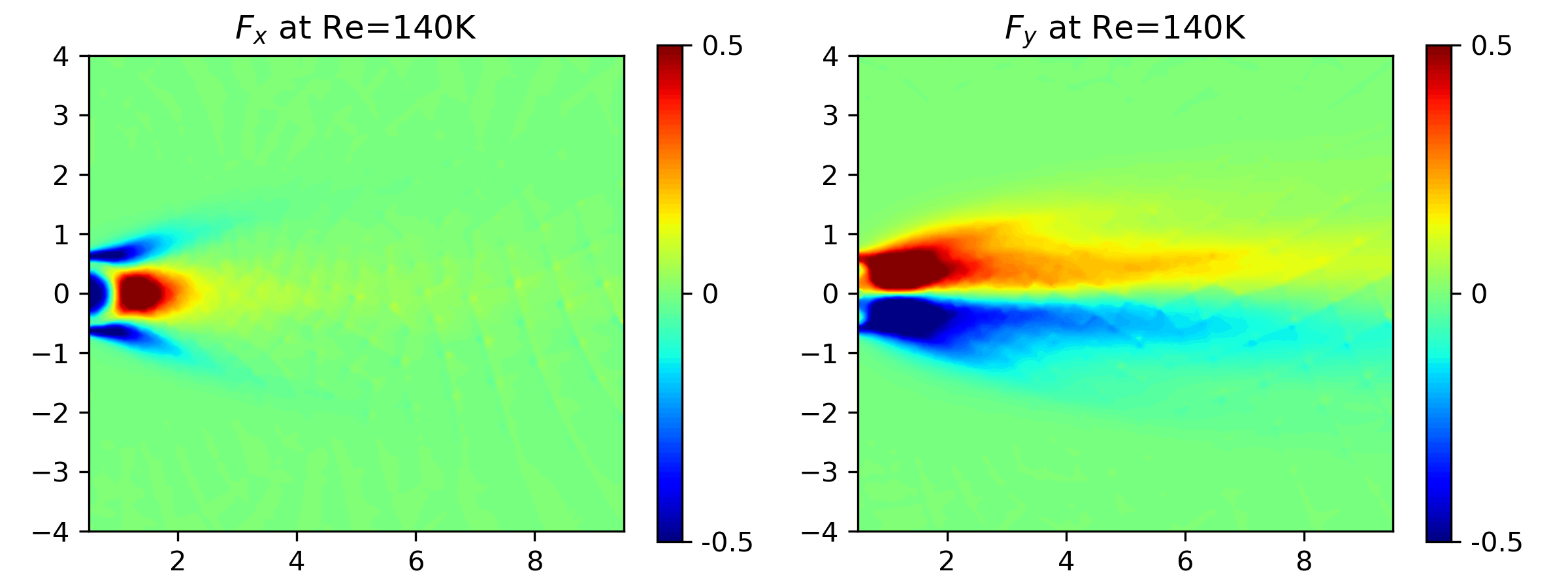}
        \caption{$Re=140,000$}
             \end{subfigure}
             \caption{The similarity of Reynolds force vector $F_x$ and $F_y$ fields at six different Reynolds numbers for incompressible cylinder flows. In each subfigure, $F_x$ is shown left, while $F_y$ is shown right. Results are taken from the time-averaged DNS data.}
    \label{fig:similarity_FxFy}
\end{figure}

\autoref{fig:similarity_FxFy} shows the Reynolds force $F_x$ and $F_y$ in the incompressible RANS equation. Results from six different Reynolds numbers are compared. The force $F_x$ and $F_y$ show similarity among different Reynolds numbers, indicating there is a universal law in incompressible cylinder flows. Apart from simulation data, the same trend can also be found in PIV measurements. \autoref{fig:MIT_field} shows the similarity of both mean velocity and Reynolds force in hydrodynamic PIV of $Re=10,000-60,000$. \autoref{fig: ucf_flow} shows the same similarity in aerodynamic PIV of $Re=6,500-100,000$. Note that when the Mach number reaches $Ma=0.3$, where $Re=100,000$ in the compressible flow, the similarity of mean velocity as well as Reynolds stresses breaks, indicating that the Mach number plays an important role in compressible turbulence closures. A generalizable turbulence closure model for compressible flow is highly interesting and will be a future focus.

\subsection{Neural-Network Turbulence Closure Model}

Based on time-averaged DNS data, we trained a neural network as the closure model. Model inputs are time-averaged velocity components $U$ and $V$, and their derivatives $U_x, U_y, V_x$. Due to the divergence-free condition of the incompressible flow, the term $V_y=-U_x$ is excluded. The model outputs are the Reynolds forcing $F_x$ and $F_y$. A fully-connected network with 3 hidden layers and 128 neurons per layer is used. The activation function is ReLU. We train this neural network using the data of two Reynolds numbers and test it with the other three Reynolds numbers. \autoref{tab:fxfy-errors} summarizes the relative $L_2$ errors of all training and testing datasets. Based on the similarity of Reynolds stresses along the Reynolds number, the forcing term can be well predicted, even in the extrapolation case. The largest error occurs at $Re=3900$, where the length of the recirculation region is longer than that in the other tested Reynolds numbers, as shown in \autoref{fig:similarity_FxFy}. 

\begin{table}
    \centering
    \caption{Errors of the neural network closure model.}
    \label{tab:fxfy-errors}
    \renewcommand{\arraystretch}{1.2}
    \begin{tabular}{cccc}
        \hline
        \textbf{Re} & \textbf{Train/Test} & \textbf{$F_x$ error [\%]}& \textbf{$F_y$ error [\%]} \\
        \hline
        3900  & Test  & 39.6& 45.9\\
        5000  & Train & 10.9& 11.7\\
        11{,}000 & Test  & 15.7& 16.0\\
        30{,}000 & Train & 10.9& 11.1\\
        60{,}000 & Test  & 22.9& 30.2\\
        140{,}000 & Test  & 18.3& 19.5\\
        \hline
    \end{tabular}
\end{table}

\begin{figure}
    \centering
    \includegraphics[width=0.8\linewidth]{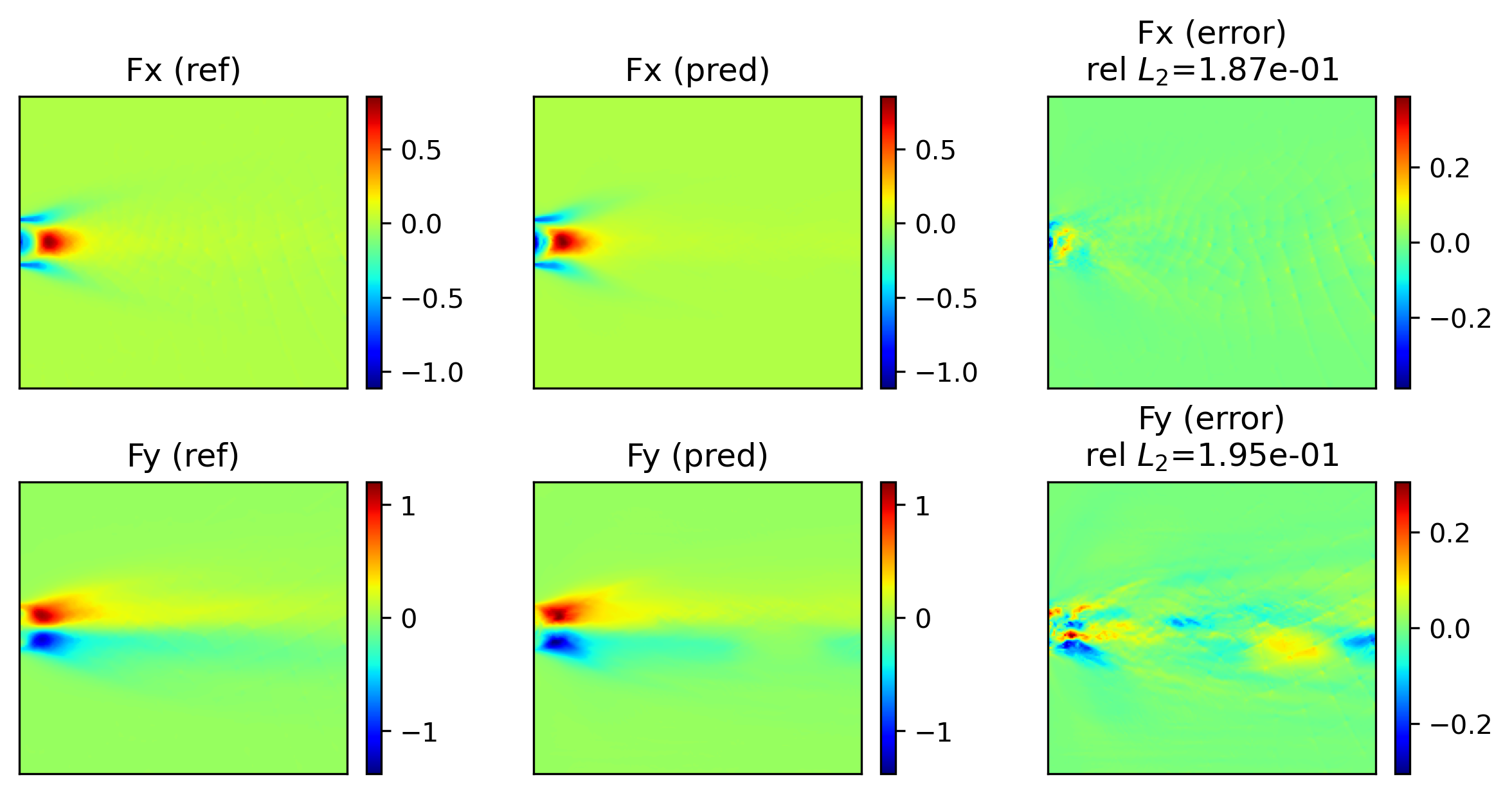}
    \caption{Prediction of the NN closure model at the unseen test Reynolds number $Re=140\,000$. The first column is the reference data from DNS, the second column is the prediction of the NN closure model, and the third column is the pointwise error. The relative $L_2$ error is also shown in the third column.}
    \label{fig:closure_NN_140k}
\end{figure}

In \autoref{fig:closure_NN_140k}, the neural network (NN) turbulence closure model is evaluated at an unseen Reynolds number of $Re=140{,}000$. The predicted force fields $F_x$ and $F_y$ capture the overall structure and spatial distribution of the reference data reasonably well, though some localized discrepancies remain visible.  

\begin{figure}
    \centering
    \includegraphics[width=0.4\linewidth]{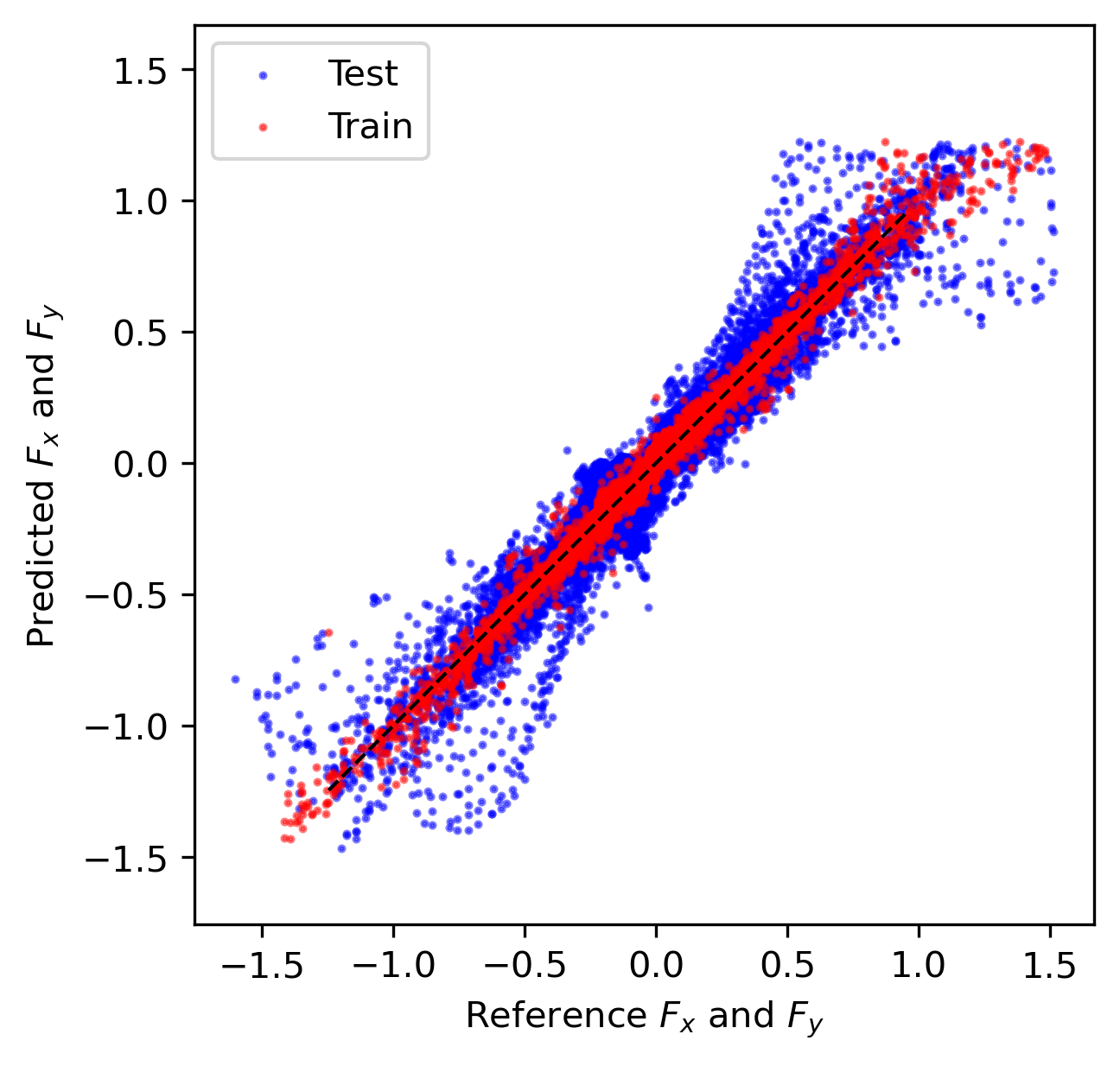}
    \caption{Overall comparison of the NN closure model on training and testing datasets. The horizontal axis is the Reynolds force components $F_x$ and $F_y$ from the DNS data at all Reynolds numbers, while the vertical axis is the closure model's prediction of the corresponding data points. The error at the line of $y=x$ is zero.}
    \label{fig:closure_overall}
\end{figure}

\autoref{fig:closure_overall} shows the overall comparison of the NN closure model. Red dots are training data, while the blue dots are the testing data. The model shows good generalization ability, and the model's performance can be further improved by using more data during training. These results highlight both the promise of the data-driven closure approach and the importance of including a wide range of flow conditions in the training dataset for robust generalization.

\subsection{Integrating the Closure Model with PINNs}

In this section, we integrate the closure model with PINNs to solve a forward problem like in a numerical CFD solver. We have two setups, where different kinds of closure models are used to solve a steady RANS of the flow past a cylinder problem. All of the data used in this section are taken from corrected time-average PIV measurements at $Re=11,000$.

\begin{figure}
    \centering
    \includegraphics[width=0.8\linewidth]{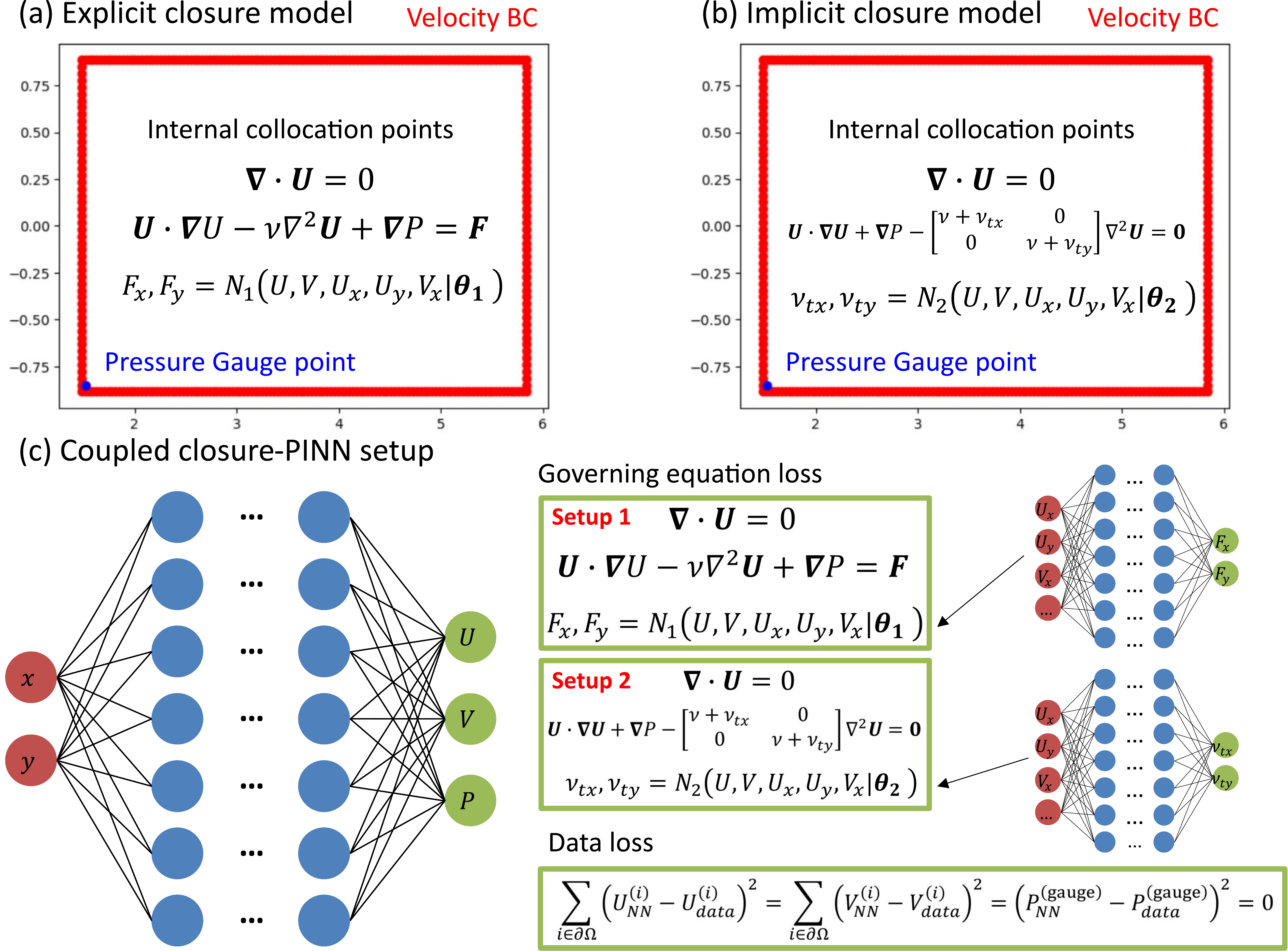}
    \caption{\textcolor{black}{Integrating different closure models with the PINN solver. Velocity BC on \(\partial\Omega\) and single pressure (gauge) are used as data loss. (a) Setup 1, an explicit closure model $N_1$ for Reynolds force on $\Omega$ is given. (b) Setup 2, an implicit closure model $N_2$ for the eddy viscosity matrix on $\Omega$ is given. (c) coupled closure-PINN system. For each closure model, there are two neural networks: one based on coordinates $x,y$ predicts state variables $U,V,P$;  the other is the closure network.}}
    \label{fig:closure_PINN}
\end{figure}

\emph{Setup 1: Explicit closure model.}
\autoref{fig:closure_PINN}(a) shows the first setup. The velocity boundary condition on $\partial \Omega$ and the pressure gauge point are given. A neural network closure model is trained to predict the force term from the mean velocity $U,V$ and their derivatives $U_x,U_y,V_x$. Because of the divergence-free property of the corrected dataset, the $V_y=-U_x$ term is omitted. 

\emph{Setup 2: Implicit closure model.}
\autoref{fig:closure_PINN}(b) shows the second setup. Like in the first setup, the velocity boundary condition on $\partial \Omega$ and the pressure gauge point are given. The force term is treated differently, where a NN closure model is built to predict the eddy viscosity matrix, and this eddy viscosity matrix is used to compute the force term. The input features include the mean velocity $U,V$ and their derivatives $U_x,U_y,V_x$. Again, because of the divergence-free property of the corrected dataset, the $V_y=-U_x$ term is omitted.

The reason why we try this eddy viscosity matrix setup in addition to the previous one is as follows. If we train a NN closure model for the force directly, the force will be an explicit term and will appear on the right-hand side of the linear system in a CFD solver. According to~\cite{illcondition}, the condition number of the linear system will be large for explicit force, and thus the error of the NS equation will be enlarged. To address this issue, we tried to make the force term implicit by representing it using the eddy viscosity matrix. 

The original steady NS equation with force term is 
\begin{equation}\label{eq:original}
  \mathbf{U}\!\cdot\!\nabla \mathbf{U}
  \;+\; 
  \nabla P
  \;-\; 
  \nu\,\nabla^{2}\mathbf{U}
  \;=\; 
  \mathbf{F}(x,y).
\end{equation}

We can choose two eddy‐viscosity fields \(\nu_{tx}(x,y)\) and \(\nu_{ty}(x,y)\) such that
\begin{equation}\label{eq:ft_def}
  \mathbf{F}(x,y)
  \;=\; 
  \begin{pmatrix}
    F_x \\ F_y
  \end{pmatrix}
  \;=\; 
  {\begin{pmatrix}
      \nu_{tx}(x,y) & 0 \\[0.5em]
      0 & \nu_{ty}(x,y)
    \end{pmatrix}}
  \,\nabla^{2}\mathbf{U}(x,y).
\end{equation}
In other words, we model \(\mathbf{F}\) as an \emph{anisotropic}, diagonal eddy‐viscosity acting on \(\nabla^{2}\mathbf{u}\).  
Substituting \autoref{eq:ft_def} into \autoref{eq:original} gives
\begin{align}
  \mathbf{U}\!\cdot\!\nabla \mathbf{U} \;+\; \nabla P 
  \;-\; {\Bigl(\nu\,\mathbf{I} + \mathrm{diag}(\nu_{t1},\,\nu_{t2})\Bigr)}
  \,\nabla^{2}\mathbf{U}
  \;=&\; 
  \mathbf{0}.
  \label{eq:implicit_form}
\end{align}
In this way, we can transfer the explicit force into a purely implicit form without losing accuracy.

\begin{figure}
    \centering
    \includegraphics[width=0.8\linewidth]{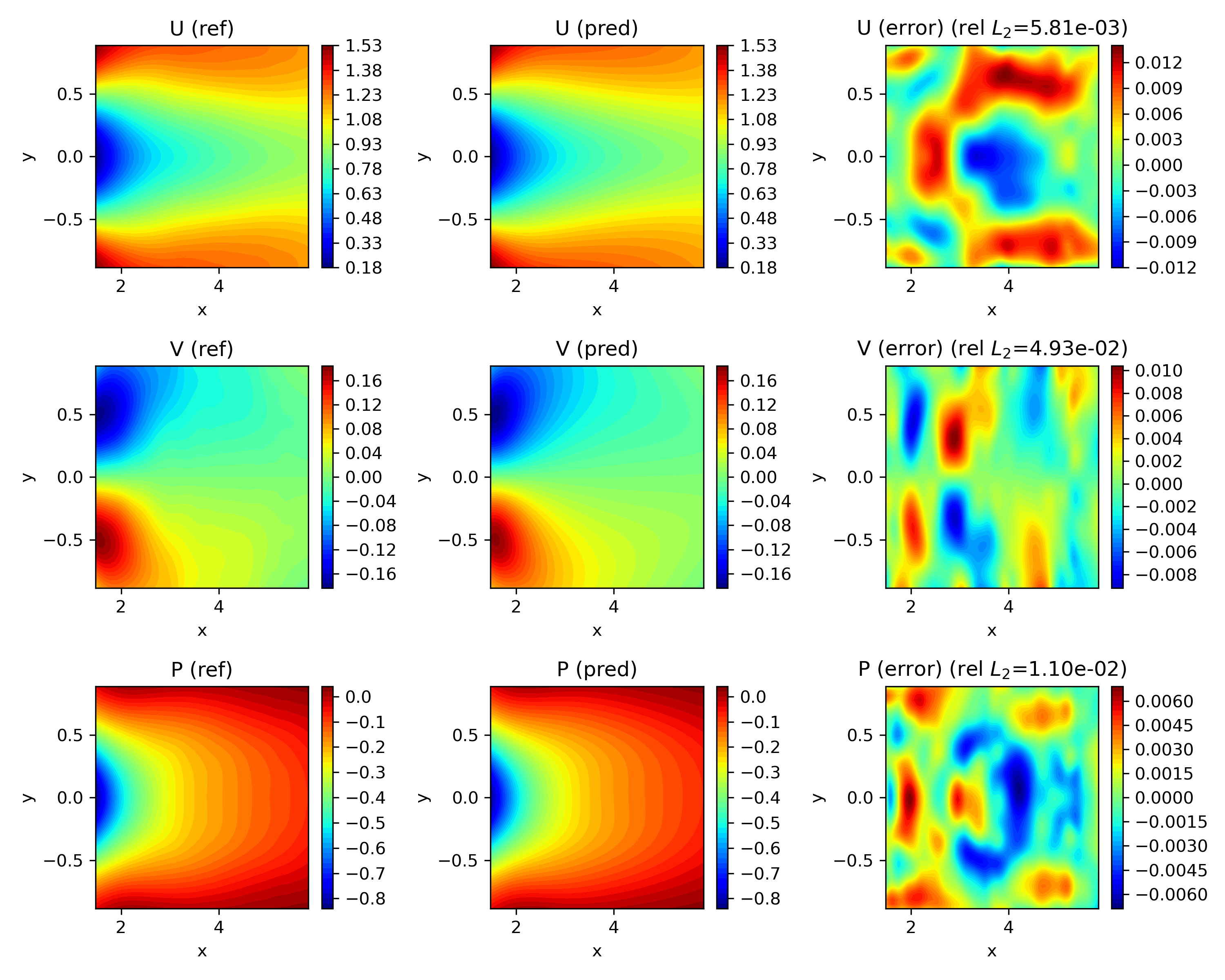}
    \caption{Flow solution of the forward PINN with the explicit closure model. The first column is the reference data from corrected areodynamic PIV at $Re=11\,000$, the second column is the prediction of PINN coupled with the explicit closure model, and the third column is the pointwise error. The relative $L_2$ error is given for each variable.}
    \label{fig:pinn_forward_uvp}
\end{figure}

\autoref{fig:pinn_forward_uvp} shows the result of the forward PINN with the explicit model. Note that the forward problem starts from a randomly initialized flow field, and the final error is small, showing that the coupled flow-turbulence system can converge to the point where the labeled data is. This good convergence indicates the feasibility of the pre-trained NN closure model to be coupled with the CFD solver. This result also indicates that a data-driven turbulence closure model is suitable for PINN-based flow solvers, which opens the door for PINN to solve turbulence problems that arise in real applications.

\begin{figure}
    \centering
    \includegraphics[width=0.8\linewidth]{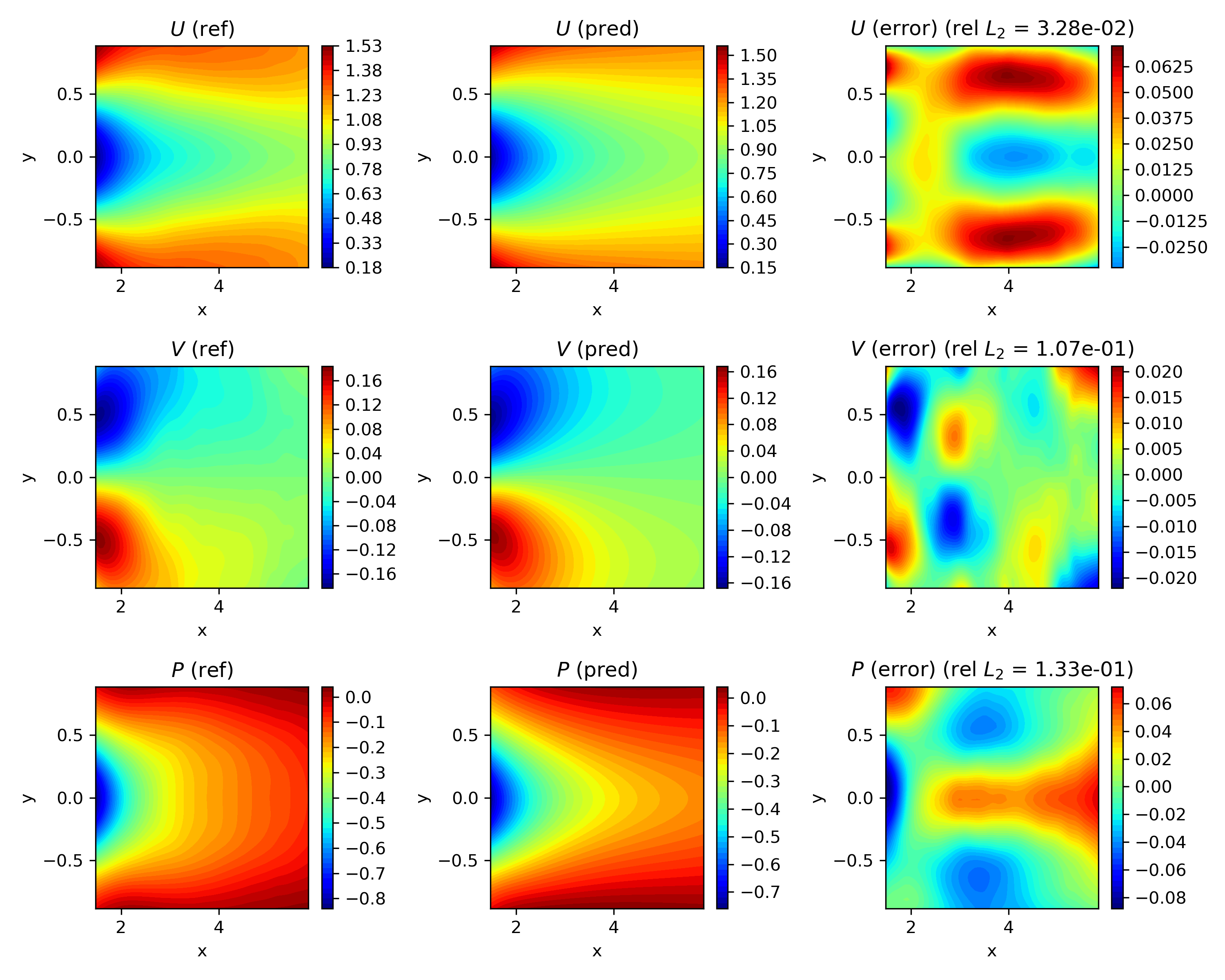}
    \caption{Flow solution of the forward PINN with the implicit closure model. The first column is the reference data from corrected areodynamic PIV at $Re=11\,000$, the second column is the prediction of PINN coupled with the explicit closure model, and the third column is the pointwise error. The relative $L_2$ error is given for each variable.}
    \label{fig:pinn_forward3_uvp}
\end{figure}

\autoref{fig:pinn_forward3_uvp} shows the result of the forward PINN problem with the implicit closure model. Compared with the explicit model, the error in this implicit eddy viscosity closure setup is larger, but it is still comparable. The good accuracy shows that the implicit coupling between flow and turbulence can still converge to the correct point where the labeled data is, which again indicates the feasibility of using the implicit closure model in a CFD solver. Considering the better numerical stability as well as the smaller condition number of the linear system in a CFD solver, this implicit strategy is promising to integrate any data-driven turbulence model into an existing CFD solver. 

\autoref{tab:forward-error} summarizes errors we obtained in both forward problems using different turbulence closures. A posteriori errors of force terms are obtained by testing the closure model together with the PINN solver. The input features of the closure model in these two scenarios are different. Both strategies can integrate data-driven closure models into the forward PINN solver with satisfactory accuracy. The explicit closure model achieves better results with PINNs, where no numerical stability issues occur. The implicit closure model via the eddy viscosity matrix is more suitable for numerical PDE solvers, where numerical stability is important.

\begin{table}
    \centering
    \caption{Summary of errors in PINN's forward solution with different turbulence closure models}
    \label{tab:forward-error}
    \renewcommand{\arraystretch}{1.15}
    \begin{tabular}{lccccccc}
        \hline
        \multirow{2}{*}{\textbf{Rel $L_2$ error [\%]}}
        & \multirow{2}{*}{\textbf{$U$}}
        & \multirow{2}{*}{\textbf{$V$}}
        & \multirow{2}{*}{\textbf{$P$}}
        & \multicolumn{2}{c}{\textit{a priori}}
        & \multicolumn{2}{c}{\textit{a posteriori}} \\
        \cline{5-8}
        & & & & $F_x$ & $F_y$ & $F_x$ & $F_y$ \\
        \hline
        Explicit closure model & 0.581 & 4.93 & 1.10 &  4.96&  4.74&  7.26&  5.55\\
        Implicit closure model & 3.28  & 10.7 & 13.3 &  6.85&  3.50&  17.6&  24.7\\
        \hline
    \end{tabular}
\end{table}

The PINN results demonstrate that both explicit and implicit closure models can be coupled with a flow solver to solve steady RANS problems. When integrated into a traditional CFD solver, however, their behavior differs significantly. The explicit formulation, where the closure model directly predicts the Reynolds force term added to the right-hand side of the equations, offers slightly better accuracy in PINNs but leads to a poorly conditioned linear system, amplifying numerical errors and causing instability at high Reynolds numbers~\cite{illcondition}. In contrast, the implicit formulation represents the force through an eddy-viscosity matrix embedded on the left-hand side, similar to conventional turbulence models. While it produces slightly higher errors in PINNs, this approach greatly improves numerical stability and solver convergence, making it better suited for large-scale CFD applications. Therefore, we propose a dual strategy: use explicit force in the PINN environment for rapid prototyping and accuracy studies, and adopt the implicit eddy-viscosity form for deployment in practical CFD solvers where stability and scalability are essential.

\subsection{\textcolor{black}{Integrating the Closure Model with OpenFOAM}}

\textcolor{black}{
The proposed data-driven turbulence closure is coupled with the steady incompressible RANS solver \texttt{simpleFoam} in OpenFOAM. The computational domain is the top half ($y>=0$) to avoid time-dependent antisymmetric modes over the centerline ($y=0$). To guarantee numerical stability, we used a hybrid implicit/explicit method, where a baseline turbulence model for eddy viscosity and a neural network for residual Reynolds forces are adopted together. The baseline turbulence model is the Spalart--Allmaras (S--A) model, which provides an eddy viscosity $\nu_t$ through the transported variable $\tilde{\nu}$. The steady momentum equations are written as
\begin{equation}
\nabla \cdot (\mathbf{U} \otimes \mathbf{U})
= -\nabla p
+ \nabla \cdot \left[ (\nu + \nu_t)\nabla \mathbf{U} \right]
+ \mathbf{F}_{\mathrm{NN}},
\label{eq:rans_nn}
\end{equation}
where $\mathbf{F}_{\mathrm{NN}}$ denotes the Reynolds-force correction predicted by a neural network. This term represents the residual divergence of the Reynolds stress not captured by the baseline S--A model. The transport equation for $\tilde{\nu}$ is left unchanged, and the NN acts as an additive correction to the momentum equations. 
Since the near-wall region of the cylinder flow remains laminar over the Reynolds number range \(Re \approx 300\text{--}300{,}000\), a predefined wall-damping function is applied to smoothly reduce the neural-network output to zero at the wall. The damping function is chosen as a cubic polynomial,
\[
f(x) =
\begin{cases}
3x^2 - 2x^3, & 0 \le x \le 1, \\
1, & x > 1 .
\end{cases}
\]
where $x = d/d^*$, $d$ denotes the wall-normal distance, and $d^* = 0.1D$.}

\textcolor{black}{
The NN inputs consist of the wall distance $d$, the normalized eddy-viscosity $\tilde{\nu}/200\nu$, and the mean velocity gradients $\partial U/\partial x$, $\partial U/\partial y$, $\partial V/\partial x$, and $\partial V/\partial y$. All input features $\eta$ are normalized using
\begin{equation}
\hat{\eta}= \frac{\eta}{1 + |\eta|},
\end{equation}
to ensure boundedness and numerical robustness. The NN outputs the two components of the Reynolds-force correction $\mathbf{F}_{\mathrm{NN}} = (F_x, F_y)$, which are applied pointwise within the SIMPLE iterations.}

\textcolor{black}{The NN parameters are trained using the discrete-adjoint framework provided by DAFoam, developed by \cite{he2020dafoam}, which differentiates through the fully converged steady RANS system. The objective function penalizes discrepancies in both the mean velocity and total Reynolds force fields within the wake region,
\begin{equation}
\mathcal{J}
= \sum_{i \in \Omega_{\mathrm{1}}}
\left\| \mathbf{U}_i - \mathbf{U}_i^{\mathrm{DNS}} \right\|^2
+ \lambda
\sum_{i \in \Omega_{\mathrm{2}}}
\left\| \mathbf{F}_{\mathrm{NN},i} 
+ \left[\nabla \cdot \left( \nu_t\nabla \mathbf{U} \right)\right]_i
- \mathbf{F}_i^{\mathrm{DNS}} \right\|^2,
\end{equation}
where $\Omega_{\mathrm{1}}$ denotes the whole calculation domain, and $\Omega_{\mathrm{2}}$ denotes the wake observation region ($x\in[1,4], y\in[0,2]$), excluding the near wall region where the flow is laminar. 
The weight is chosen to be $\lambda=0.1$. The discrete adjoint method can calculate the derivatives of the objective with respect to the NN parameters, and the BFGS method is used to optimize the NN parameters.}

\textcolor{black}{The proposed model is compared against the baseline S--A model and a state-of-the-art NN-augmented S--A formulation, as shown by \cite{holland2019field}, in which a neural network predicts a multiplicative correction $\beta(\mathbf{x})$ applied to the production term of the S--A equation and the NN is trained coupled with the solver using the discrete adjoint method.}

\begin{figure}
    \centering
      \begin{subfigure}{0.4\textwidth}
    \includegraphics[width=\textwidth]{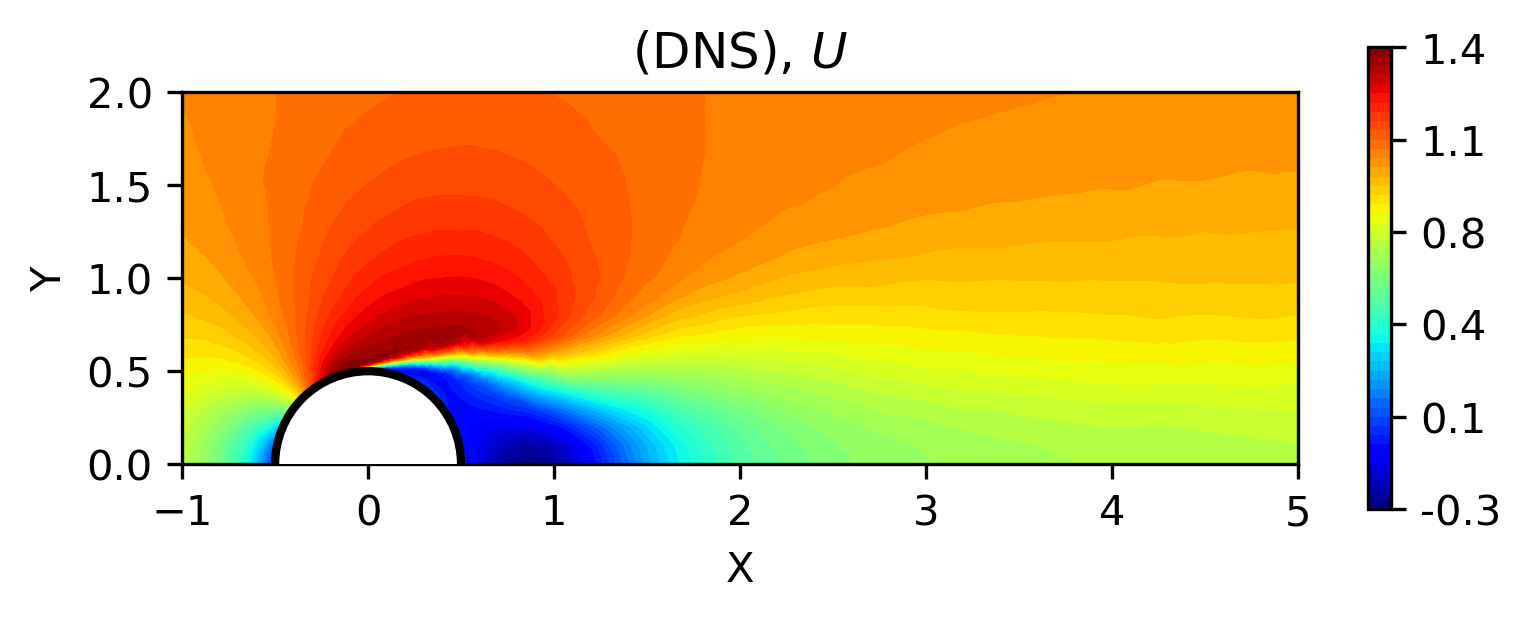}
    \end{subfigure}
      \begin{subfigure}{0.4\textwidth}
    \includegraphics[width=\textwidth]{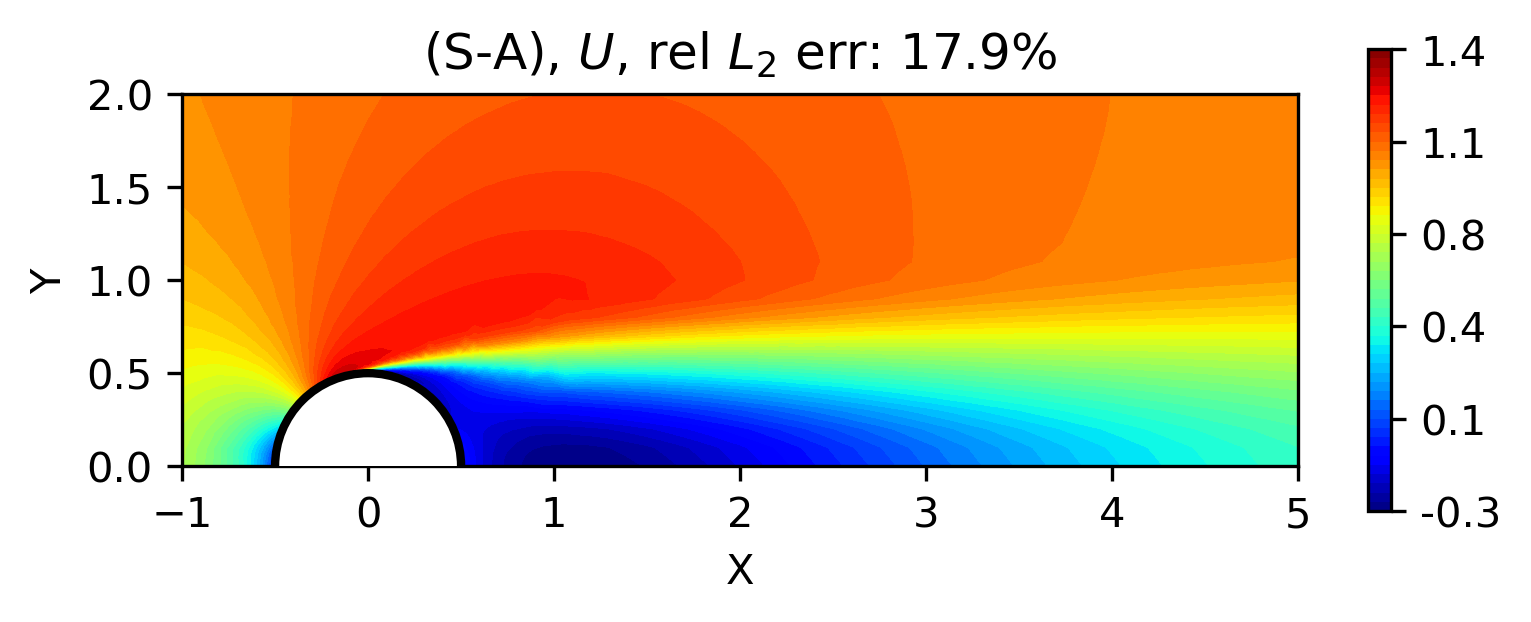}
    \end{subfigure}
      \begin{subfigure}{0.4\textwidth}
    \includegraphics[width=\textwidth]{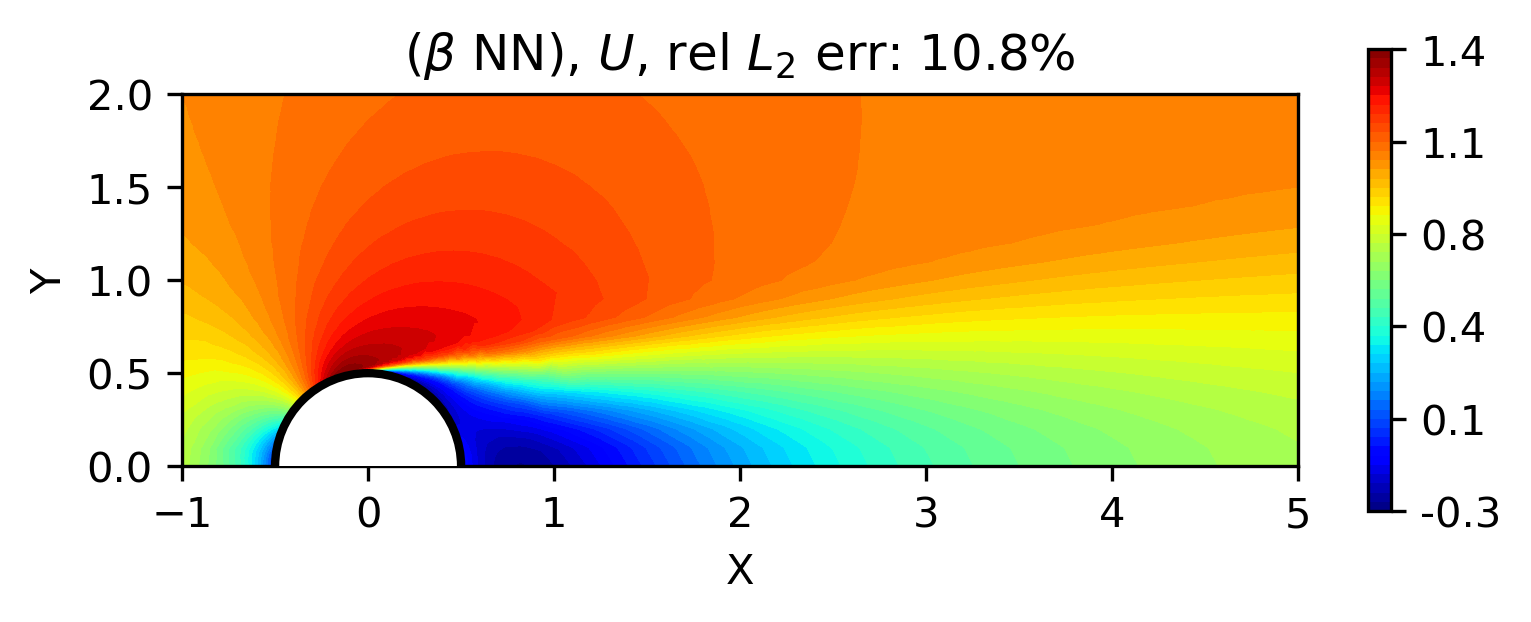}
    \end{subfigure}
      \begin{subfigure}{0.4\textwidth}
    \includegraphics[width=\textwidth]{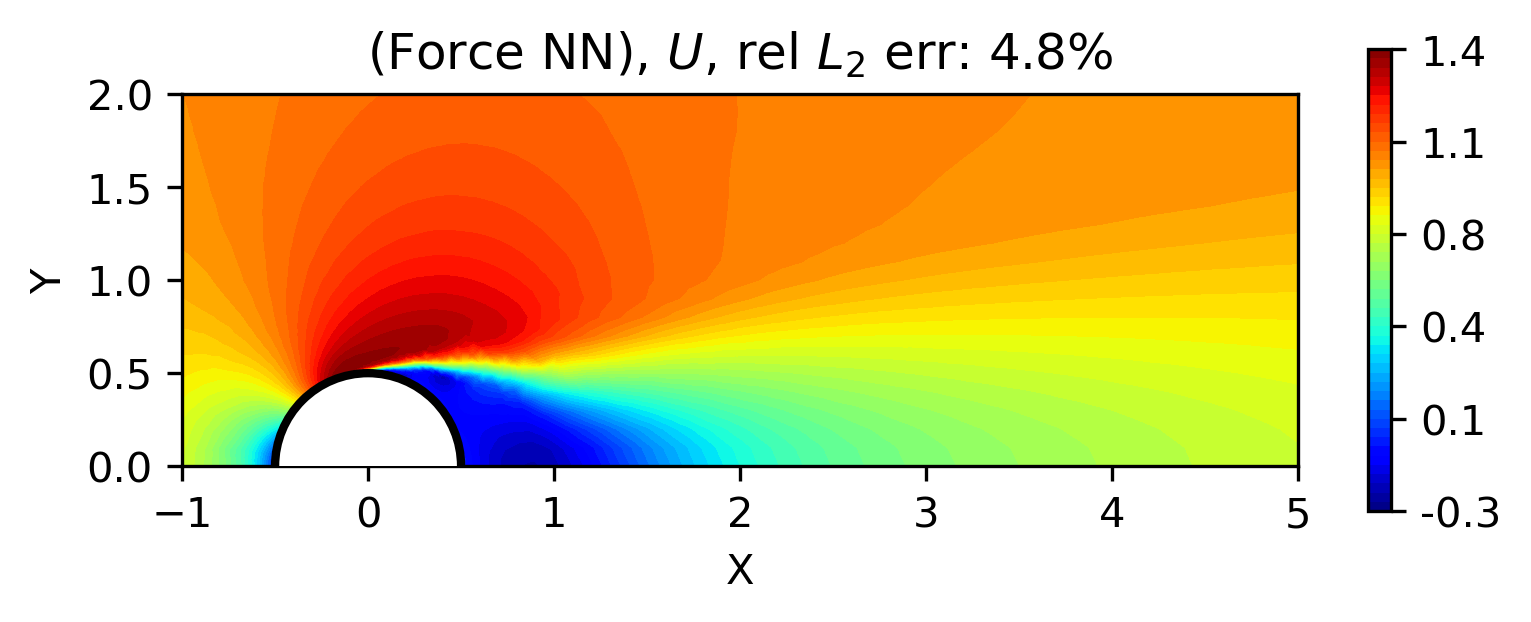}
    \end{subfigure}
    
      \begin{subfigure}{0.4\textwidth}
    \includegraphics[width=\textwidth]{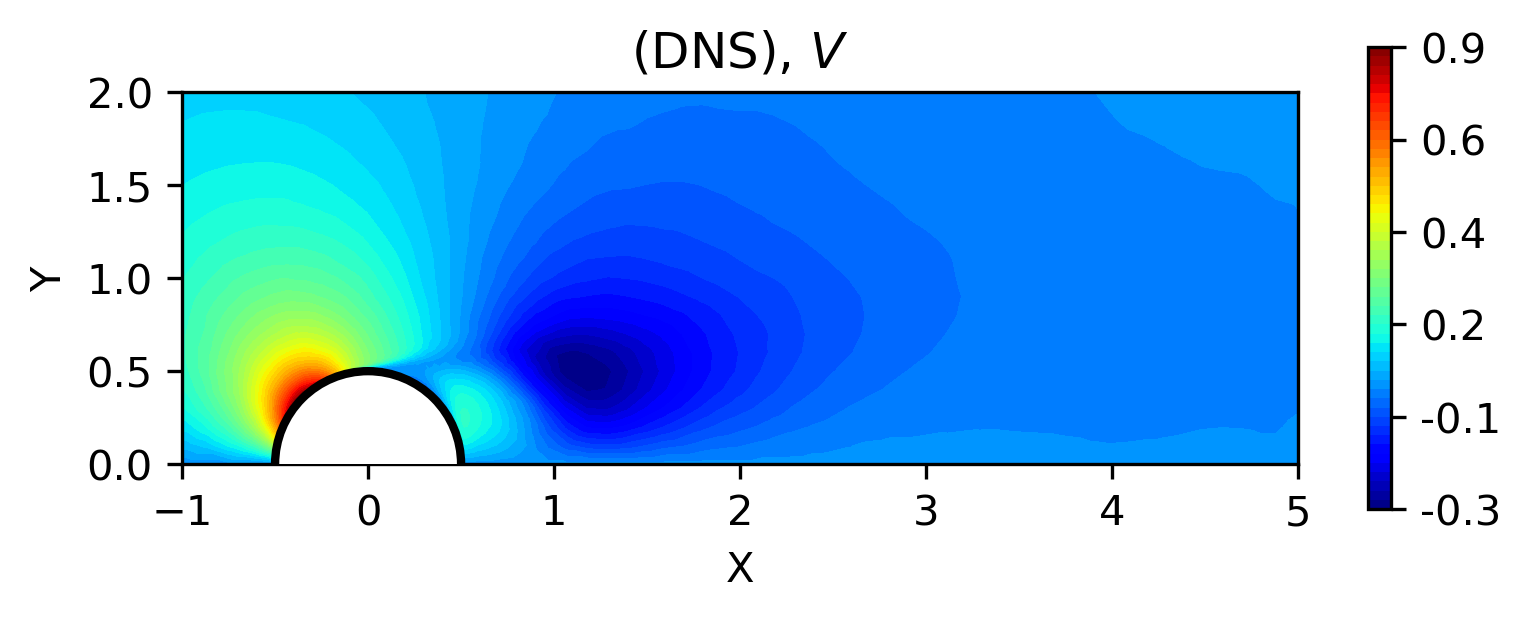}
    \end{subfigure}
      \begin{subfigure}{0.4\textwidth}
    \includegraphics[width=\textwidth]{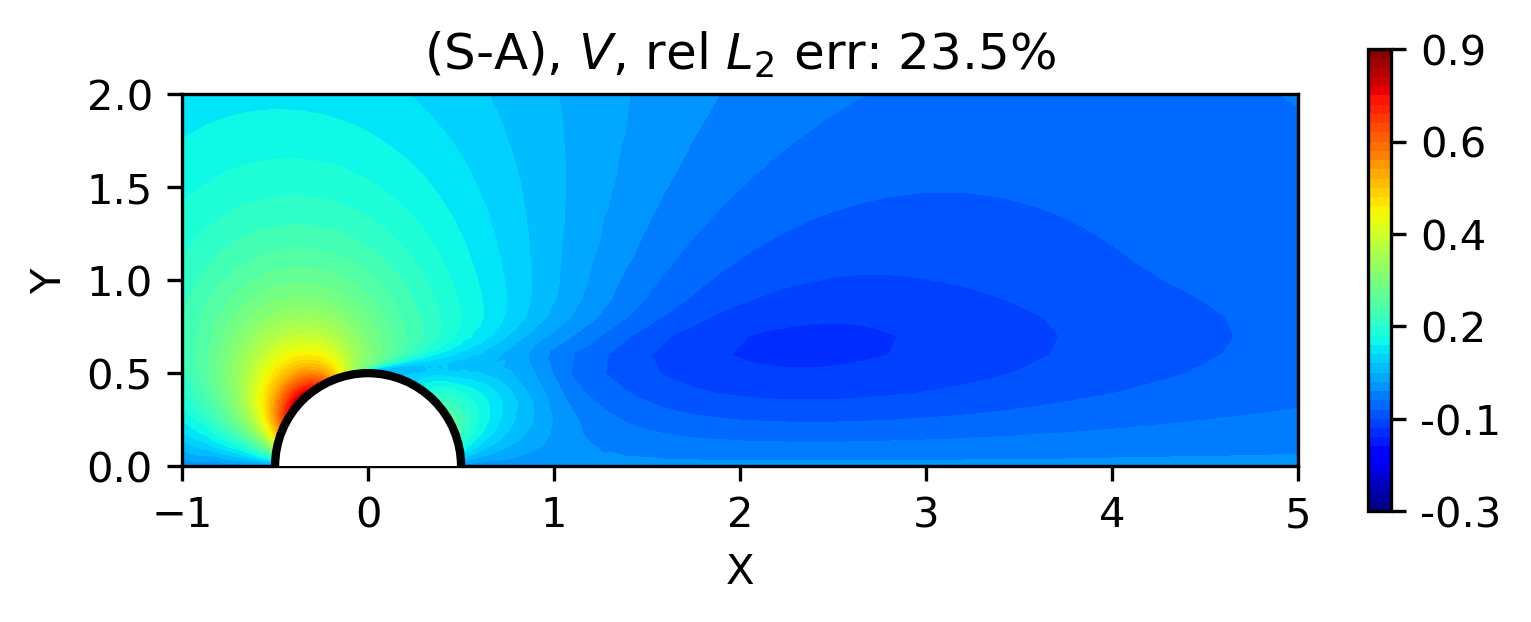}
    \end{subfigure}
      \begin{subfigure}{0.4\textwidth}
    \includegraphics[width=\textwidth]{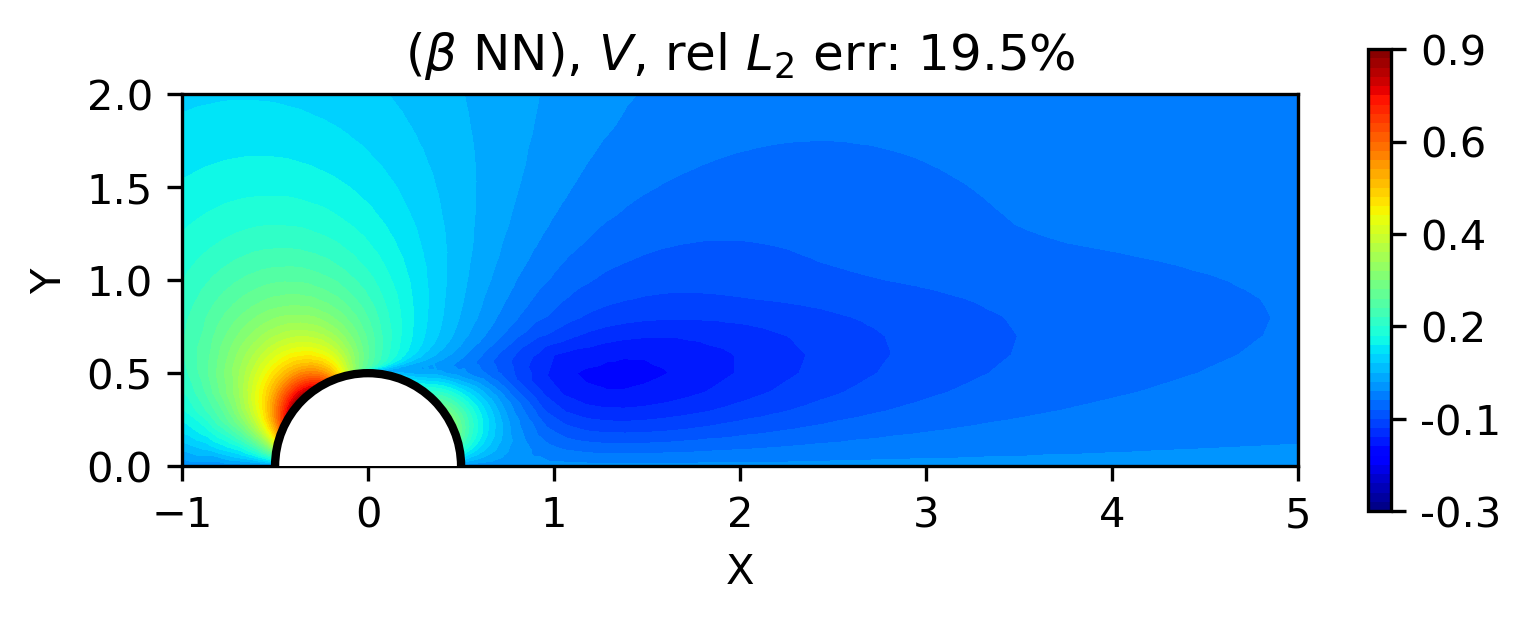}
    \end{subfigure}
      \begin{subfigure}{0.4\textwidth}
    \includegraphics[width=\textwidth]{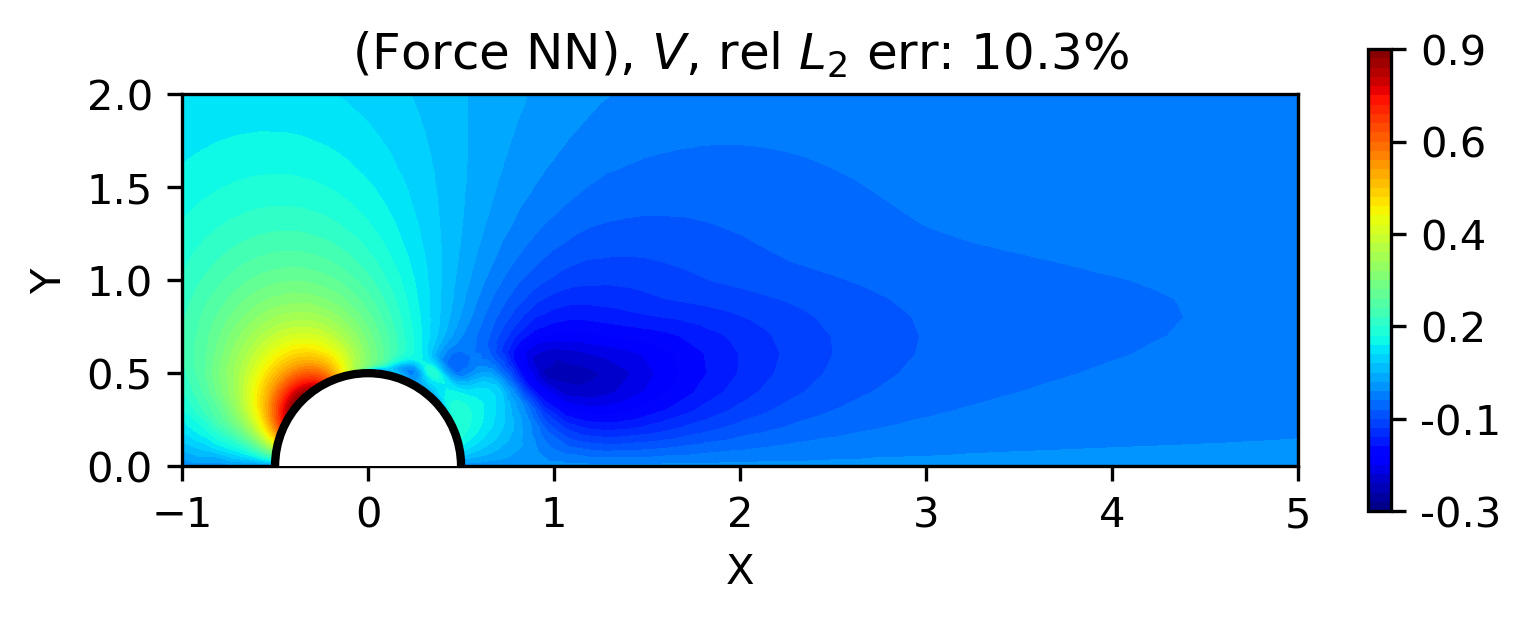}
    \end{subfigure}
     
    \caption{\textcolor{black}{Comparison of the velocity components among time-averaged DNS, standard S--A model, multiplicative S--A augmentation ($\beta$ NN in short), and the residual Reynolds force model (force NN in short). Relative $L_2$ errors calculated within the shown domain against DNS data are shown as well.}}
    \label{fig:solver_U}
\end{figure}

\textcolor{black}{\autoref{fig:solver_U} compares the time-averaged velocity components predicted by the baseline S--A model, the multiplicative S--A augmentation ($\beta$ NN), and the proposed Reynolds-force model against DNS. The baseline S--A model exhibits pronounced errors in the near-wake, particularly in the velocity deficit and cross-stream motion, reflecting the limitations of eddy-viscosity closures in separated flows. The multiplicative S--A augmentation improves the mean velocity prediction and reduces the relative $L_2$ errors for both velocity components; however, noticeable discrepancies persist in the recirculation region and along the shear layers. In contrast, the Reynolds-force model yields the closest agreement with DNS, accurately capturing both the streamwise recovery and the transverse velocity distribution throughout the wake. This result indicates that directly correcting the momentum equations through learned Reynolds-force residuals enables the solver to recover flow features that cannot be represented through eddy-viscosity-based modifications alone.
}

\begin{figure}
    \centering
      \begin{subfigure}{0.4\textwidth}
    \includegraphics[width=\textwidth]{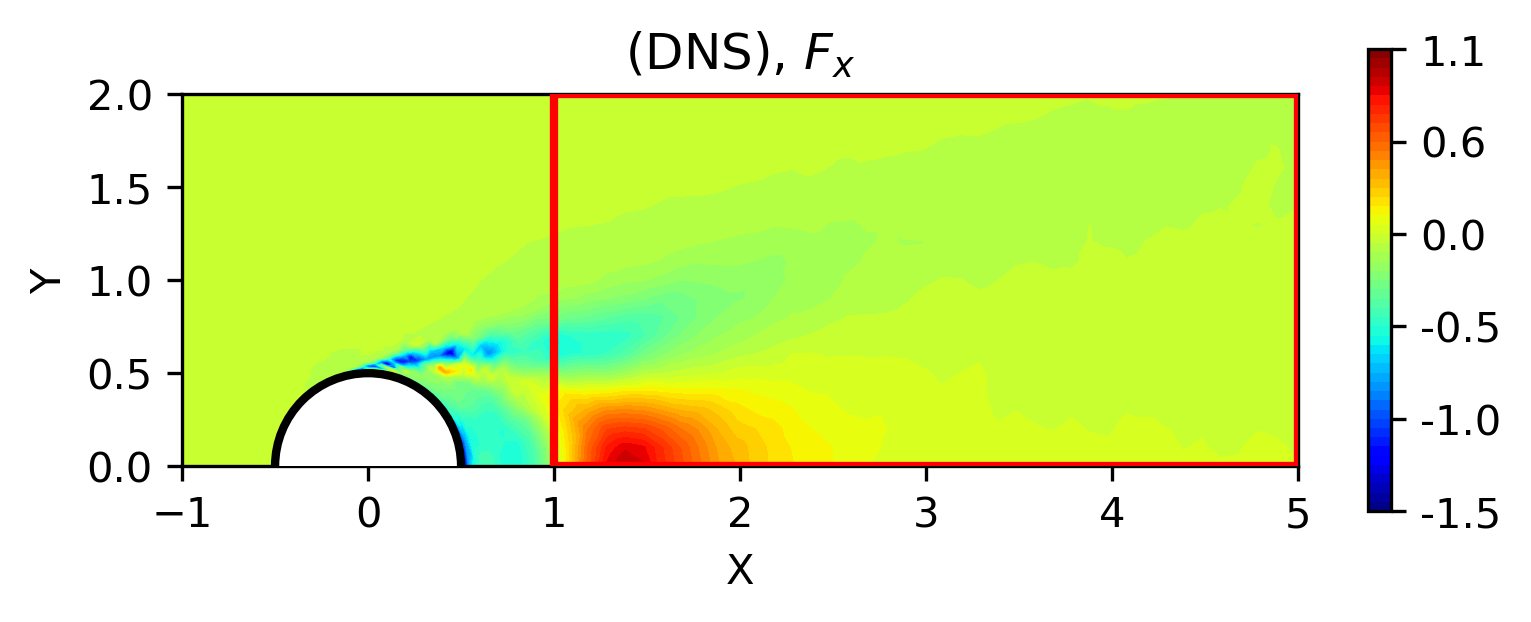}
    \end{subfigure}
      \begin{subfigure}{0.4\textwidth}
    \includegraphics[width=\textwidth]{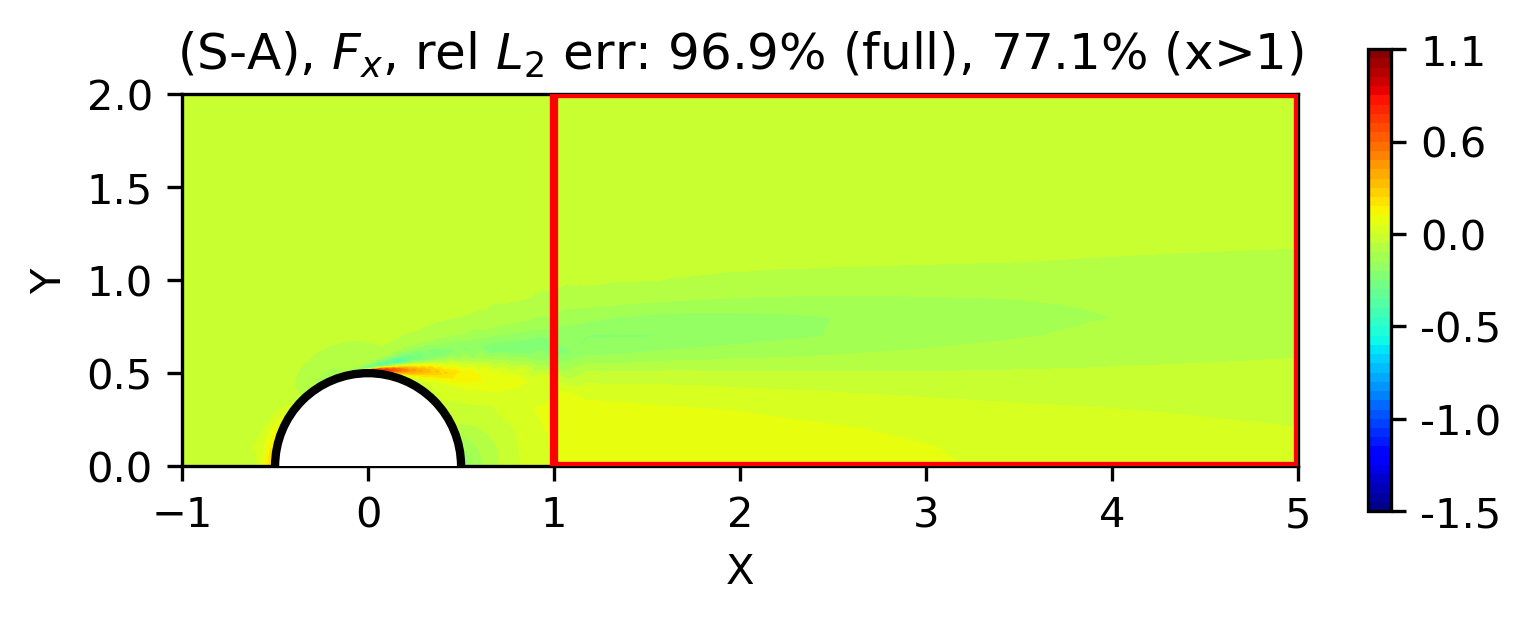}
    \end{subfigure}
      \begin{subfigure}{0.4\textwidth}
    \includegraphics[width=\textwidth]{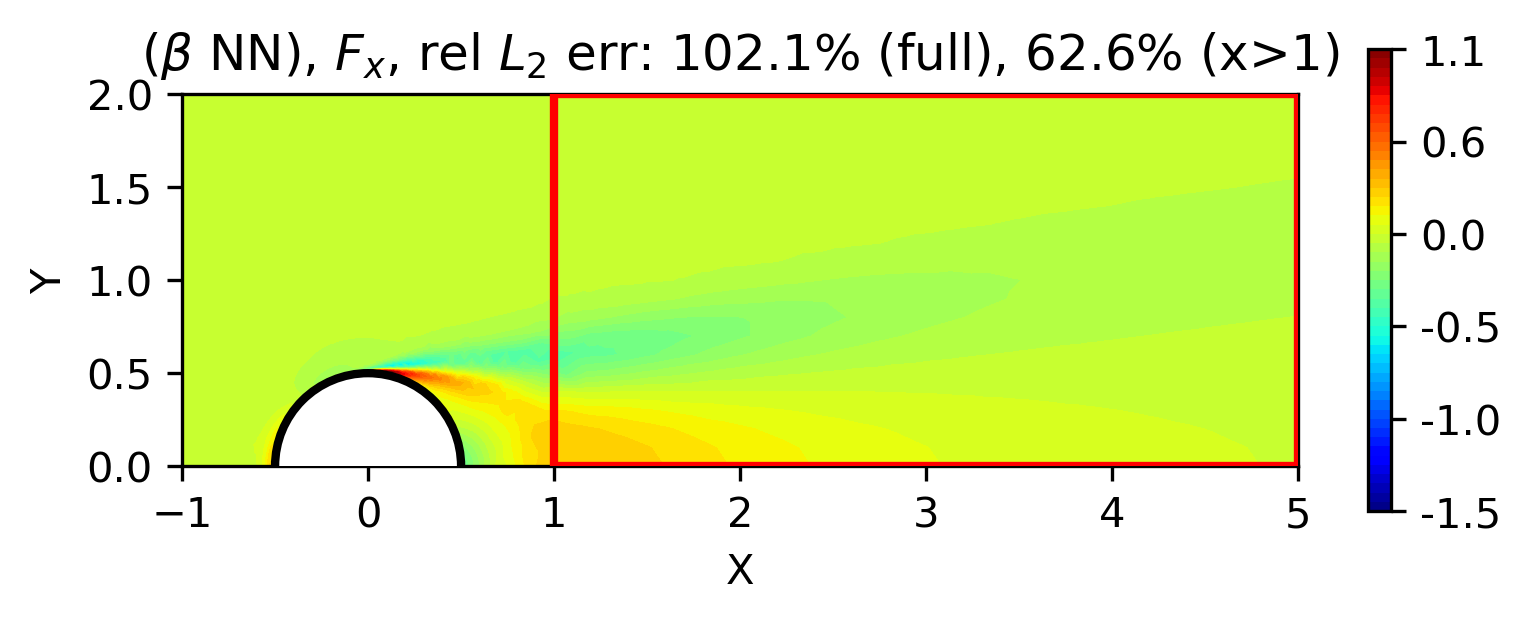}
    \end{subfigure}
      \begin{subfigure}{0.4\textwidth}
    \includegraphics[width=\textwidth]{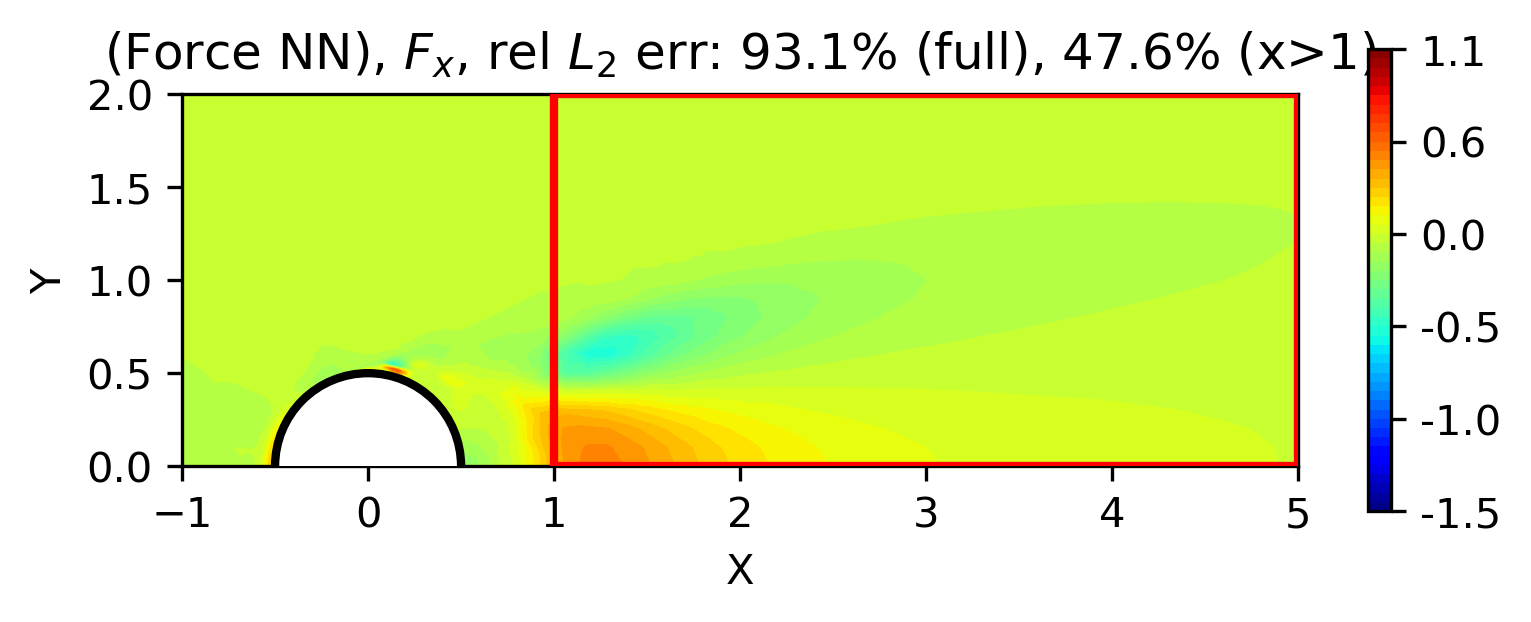}
    \end{subfigure}
     
      \begin{subfigure}{0.4\textwidth}
    \includegraphics[width=\textwidth]{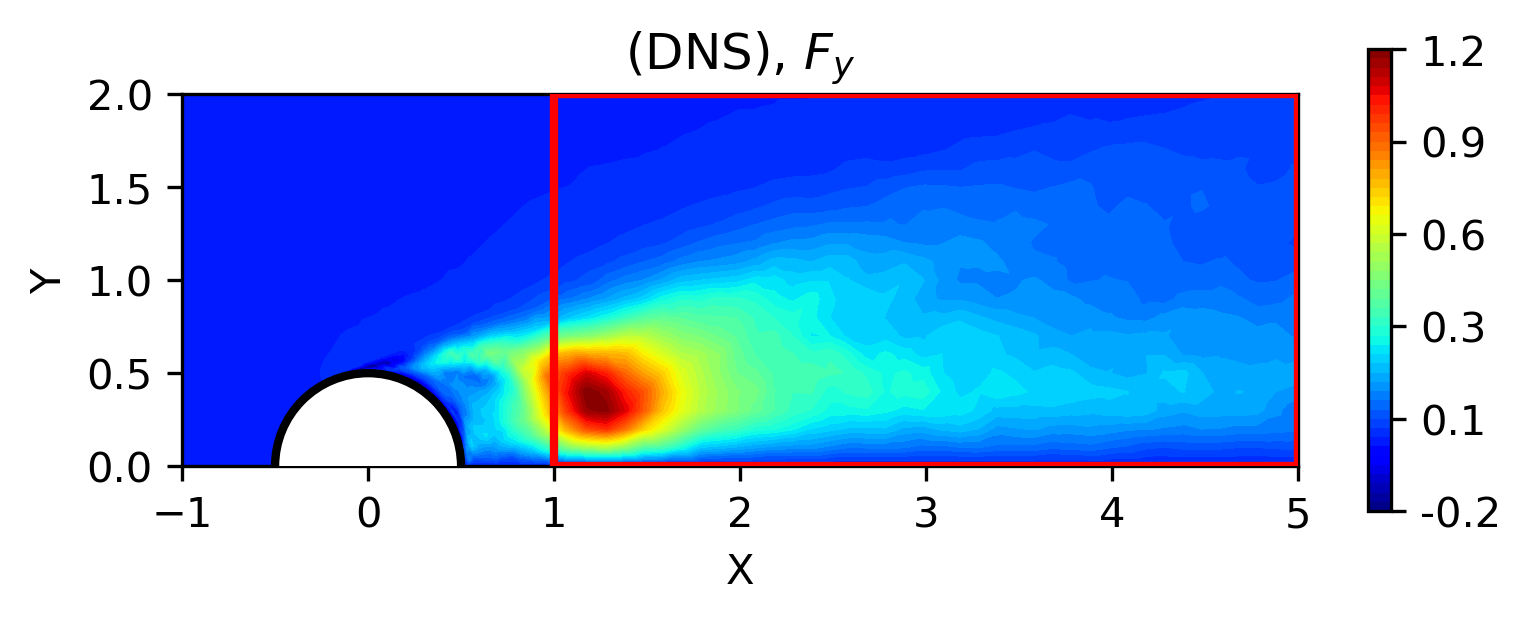}
    \end{subfigure}
      \begin{subfigure}{0.4\textwidth}
    \includegraphics[width=\textwidth]{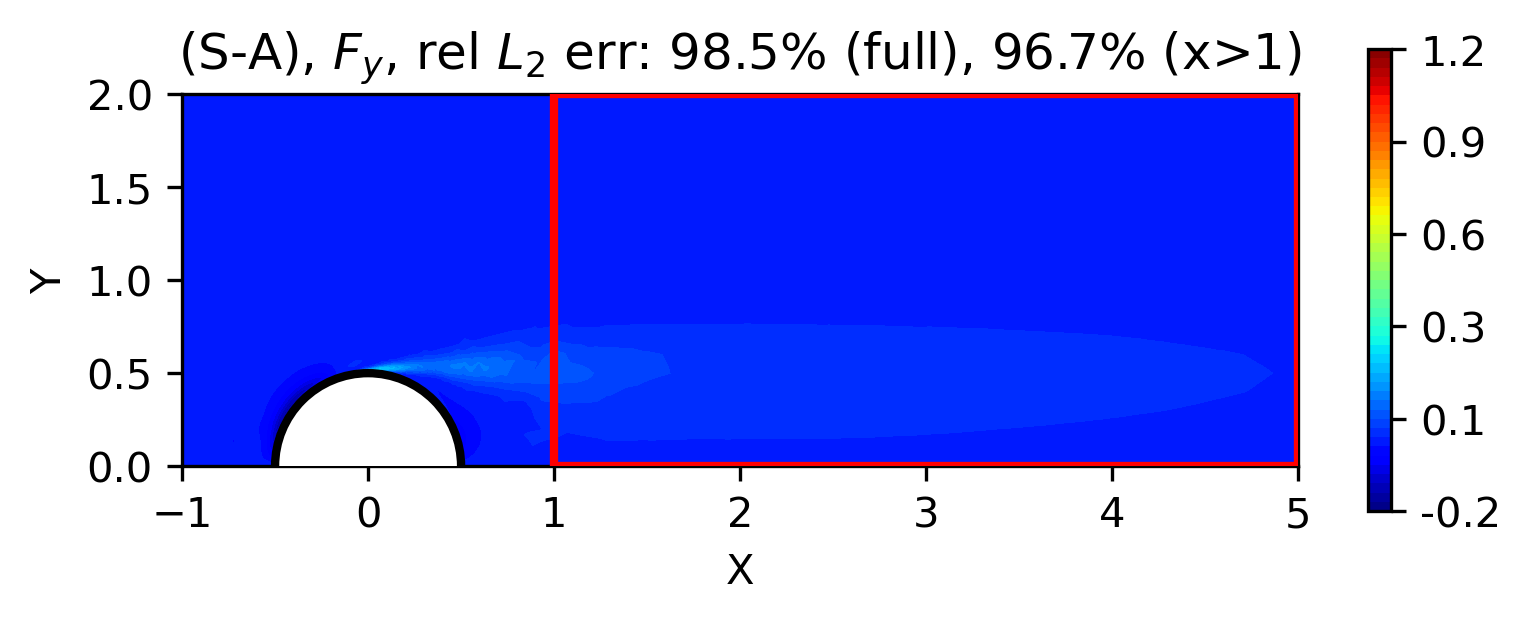}
    \end{subfigure}
      \begin{subfigure}{0.4\textwidth}
    \includegraphics[width=\textwidth]{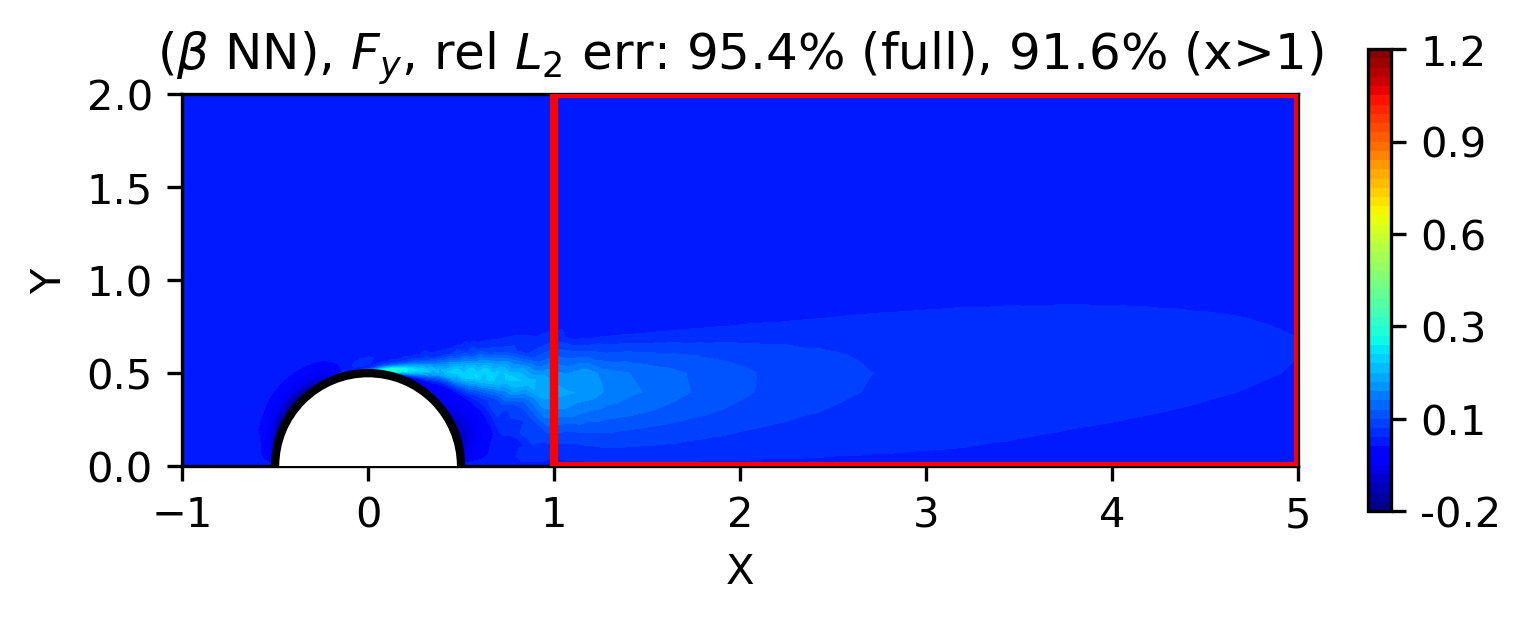}
    \end{subfigure}
      \begin{subfigure}{0.4\textwidth}
    \includegraphics[width=\textwidth]{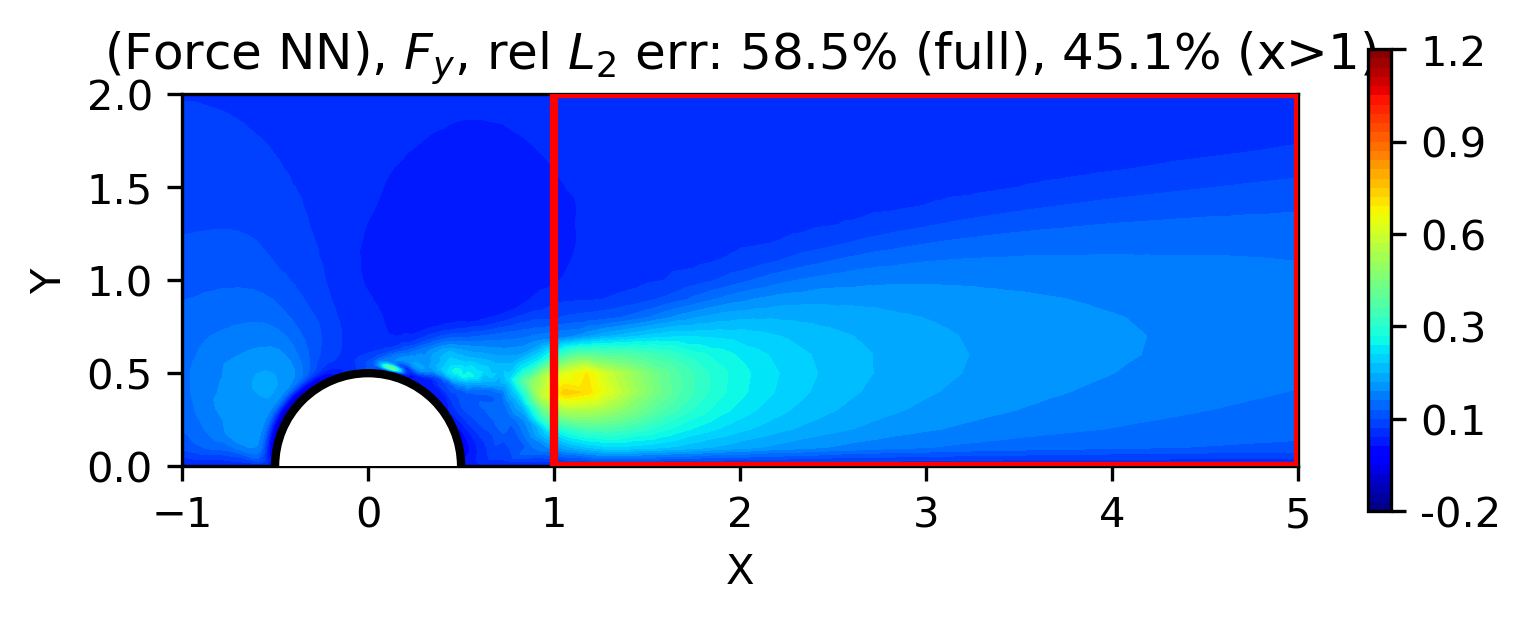}
    \end{subfigure}
    
    \caption{\textcolor{black}{Comparison of the Reynolds force components among time-averaged DNS, standard S--A model, multiplicative S--A augmentation ($\beta$ NN in short), and the residual Reynolds force model (force NN in short). For the force NN method, note that Reynolds forces are contributed by both the residual Reynolds force NN and the eddy viscosity predicted by the underlying S--A model. Relative $L_2$ errors calculated within the shown domain (full) and a subdomain in the wake ($x>1$) against DNS data are shown as well.}}
    \label{fig:solver_F}
\end{figure}

\textcolor{black}{\autoref{fig:solver_F} compares the Reynolds-force components obtained from DNS, the baseline S--A model, the multiplicative S--A augmentation, and the Reynolds-force model. The baseline S--A model severely underpredicts both force components, particularly in the near-wake region downstream of separation. The multiplicative S--A augmentation provides only limited improvement, as the resulting Reynolds forces remain constrained by the eddy-viscosity assumption. The Reynolds-force model substantially reduces the errors in both $F_x$ and $F_y$, especially in the wake region ($x>1$), where the dominant corrections to the mean flow arise. }

\textcolor{black}{Although the total Reynolds force contains contributions from both the baseline S--A eddy viscosity and the NN-predicted residual, the results demonstrate that the learned residual force is essential to recover the correct force balance. In the present configuration, the remaining error in the Reynolds-force prediction is on the order of $50\%$, which is primarily attributed to the use of a simple and compact set of input features that limits the expressive capacity of the Reynolds-force neural network. By enriching the input feature space following the same modeling strategy, the Reynolds-force prediction error can be further reduced. Such systematic improvement through feature design is not possible within eddy-viscosity-based turbulence models, highlighting the superior expressive capability of direct Reynolds-force modeling for cylinder flows.}

\section{Summary}\label{sec:conclusion}
We first built a comprehensive and cross-validated dataset for the flow past a cylinder using hydrodynamic PIV, aerodynamic PIV, and spectral-element based DNS/LES. The Reynolds number range is $Re=3,900-100,000$, and both incompressible and weakly compressible regimes are covered. The dataset includes mean velocity as well as the Reynolds stresses. A physics-based postprocessing method was proposed to ensure the dataset satisfies the continuity equation and the momentum equations. We found that there is a similarity in the Reynolds stresses along the Reynolds number, which provides a physical foundation for the search for a data-driven turbulence closure model. This dataset can also be used to validate the CFD code.

Subsequently, we formulated a flow inference problem for both incompressible and weakly compressible flows, where we made use of the unclosed form of the RANS equation and measurements at the domain boundary to infer the entire interior flow fields and Reynolds forcing terms by PINNs. This is an under-determined problem from the perspective of classical numerical analysis, because the governing RANS equation is not closed, and no data inside the domain is available. However, PINNs could  successfully infer the interior flow fields with satisfactory accuracy. We also reconstructed the flow field by leveraging boundary data of the mean flow velocities and employing a PINN, where the Reynolds stresses were decomposed using the Helmholtz decomposition and augmented with a turbulence model for wake region. These enhancements led to improved inference accuracy. 

Based on the similarity of the Reynolds stresses across different Reynolds numbers, we built a neural network as the turbulence closure model, which is a local algebraic model. We trained it at two Reynolds numbers and tested it on the other four. This model showed good generalization ability. We integrated the data-driven turbulence closure model into the forward PINN solver. Two models were tested, where one is explicit and the other is implicit. Results showed that the explicit model achieves higher accuracy than the implicit model and can substantially improve the accuracy of both the mean velocity and the Reynolds force vector.
\textcolor{black}{We further integrated the data-driven turbulence closure model into the numerical solver OpenFOAM. A hybrid implicit/explicit modeling strategy is adopted to guarantee numerical stability, and the discrete adjoint method is used to train the neural network coupled with the solver. Results show that the proposed data-driven turbulence closure method consistently outperforms the baseline Spalart--Allmaras model and a state-of-the-art data-driven turbulence closure model in predicting both mean-flow statistics and Reynolds-force distributions, demonstrating its effectiveness and robustness in practical RANS simulations of cylinder flows.}

\textcolor{black}{In summary and for reference, simulating the cylinder flow using existing models such as the Spalart--Allmaras model resulted in relative errors (compared to the time-averaged DNS) on the order of 100\%. The data-driven closures with sparse data developed herein reduced the errors in the Reynolds forces by approximately a factor of two.}

\section{Declaration of Interests}
The authors report no conflict of interest.

\section{Acknowledgments}
This research was primarily supported by the Defense Advanced Research Projects Agency (DARPA) under the Automated Prediction Aided by Quantized Simulators (APAQuS) program, Grant No. HR00112490526. Additional support was provided through the grant titled ``GPU Cluster for Neural PDEs and Neural Operators to Support MURI Research and Beyond,'' under Award No. FA9550-23-1-0671. KS acknowledges Prof. G. Rigas from Imperial College for valuable discussions on the turbulence-augmented model for RANS.


\newpage
\appendix
\section{Validation of NekRS with the Entropy Viscosity Method against Existing Experimental Data}\label{app:validation}

\begin{figure}
    \centering
    \begin{subfigure}{0.7\linewidth}
        \centering
        \includegraphics[width=\textwidth]{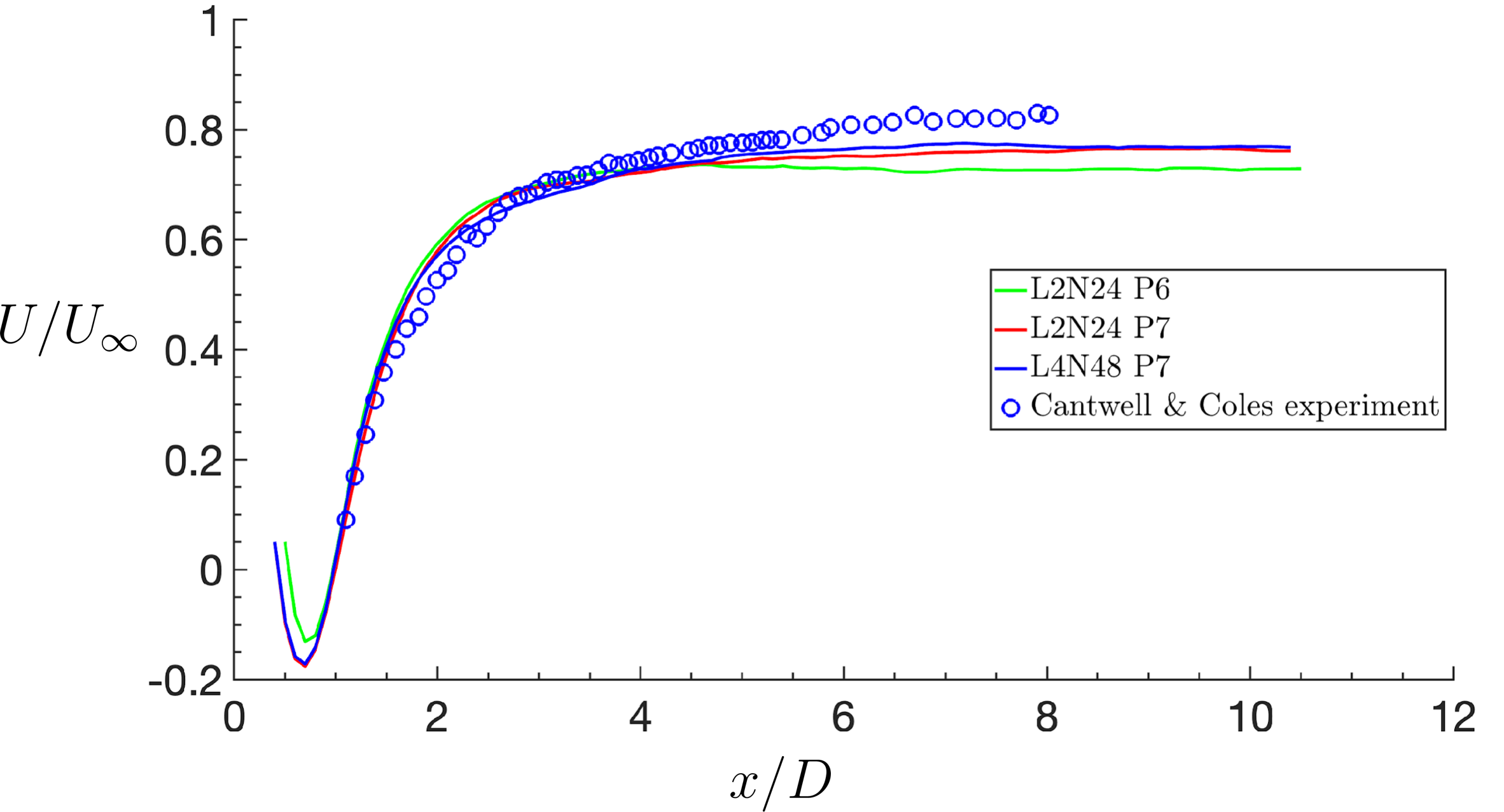}
        \caption{Mean streamwise velocity ($U$) along the center line ($y=0$) at different resolutions.}
        \label{fig:evm_vali1}
    \end{subfigure}
    \centering
    \begin{subfigure}{0.7\linewidth}
        \centering
        \includegraphics[width=\textwidth]{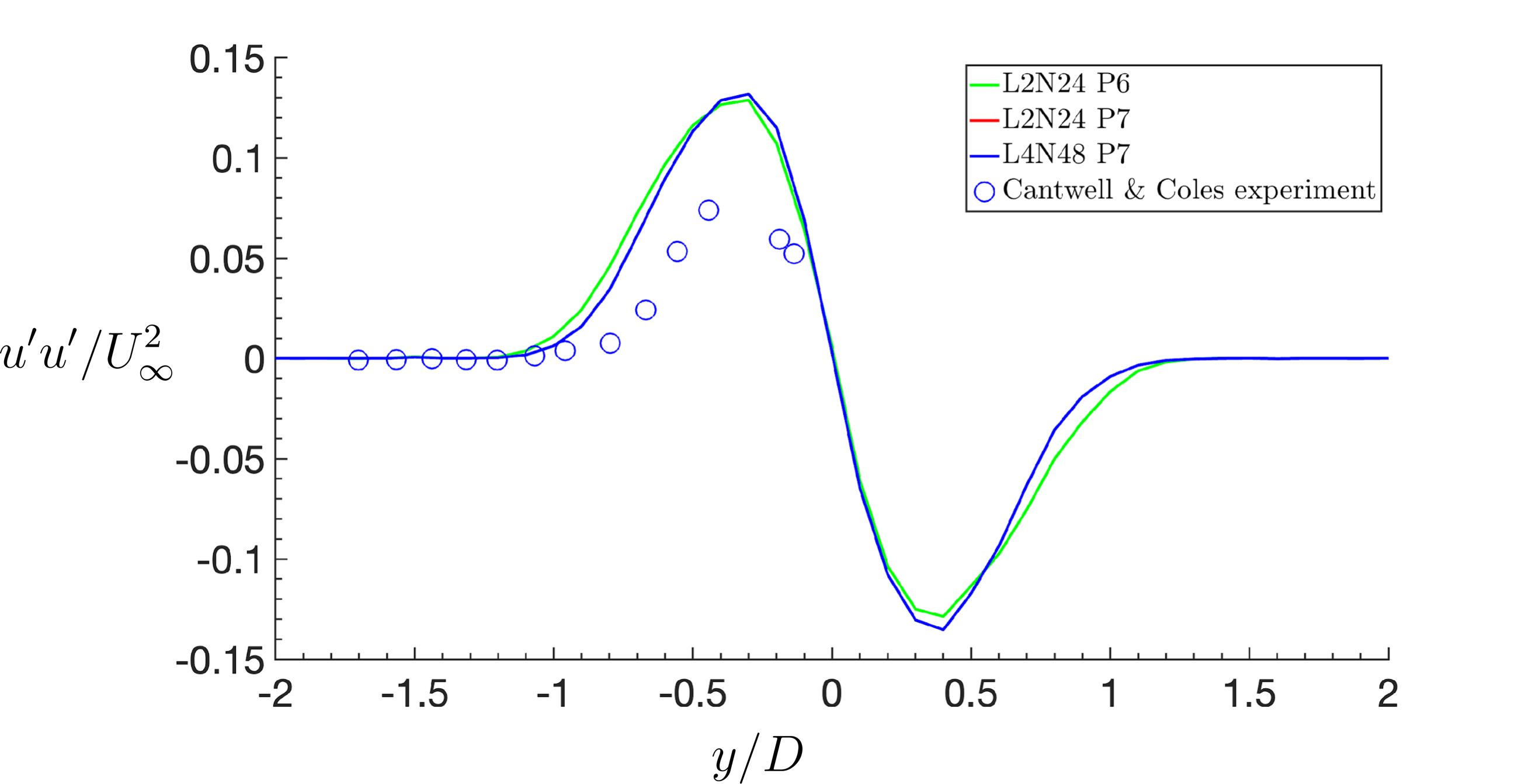}
        \caption{Reynolds stress ($u^{\prime}u^{\prime}$) along ($x/D=1$) at different resolutions.}
        \label{fig:evm_vali2}
    \end{subfigure}
    \centering
    \begin{subfigure}{0.7\linewidth}
        \centering
        \includegraphics[width=\textwidth]{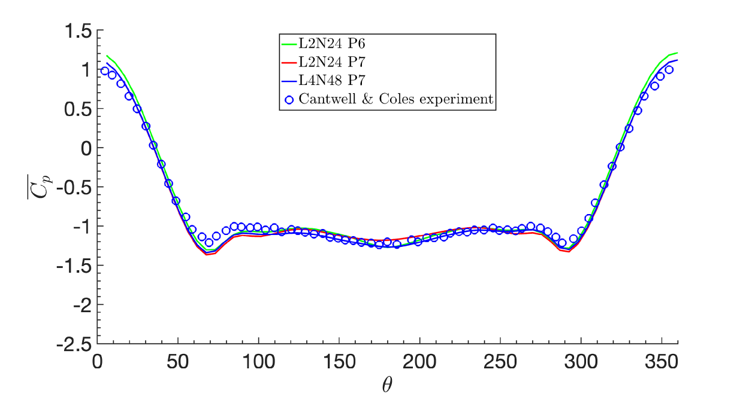}
        \caption{Local pressure coefficient ($C_P$) along the cylinder surface.}
        \label{fig:evm_vali3}
    \end{subfigure}
    \caption{Validation of the entropy viscosity method (EVM) implemented on nekRS by the simulation of flow past a stationary cylinder at $Re=140,000$. Note that $L$ denotes the length of the cylinder, N denotes the number of elements along the cylinder length and P denotes the spectral element polynomial order, specifically, L2N24 P6 means the cylinder length is $2D$, the number of elements along the cylinder is 24, and the SEM polynomial order is 6. The corresponding experimental measurements were performed by \cite{cantwell1983experimental}. \textcolor{black}{Note that, in the simulation, we have used $\alpha=0.1$ and $\beta=0.5$ of the EVM.}}
    \label{fig:evm_val}
\end{figure}

\autoref{fig:evm_val} compares the results of the cylinder flow at $Re=140,000$ obtained by nekRS with the experimental data. In these figures, different domain lengths in the cylinder span direction, different numbers of elements along the cylinder span, and different polynomial orders in SEM are compared. \autoref{fig:evm_vali1} shows the mean velocity component $U/U_\infty$ in the wake center line. \autoref{fig:evm_vali2} shows the Reynolds stress component $\overline{u'u'}/U_\infty^2$ in the steam wise position $x/D=1$. \autoref{fig:evm_vali3} shows the mean pressure coefficient distribution at the cylinder. The NekRS with the EVM LES closure is shown to accurately capture both the mean flow and turbulence statistics at high-Reynolds number, incompressible cylinder flows.

\section{Flow Inference of Incompressible RANS}\label{app:flow_inference}
\begin{figure}
    \centering
    \includegraphics[width=\linewidth]{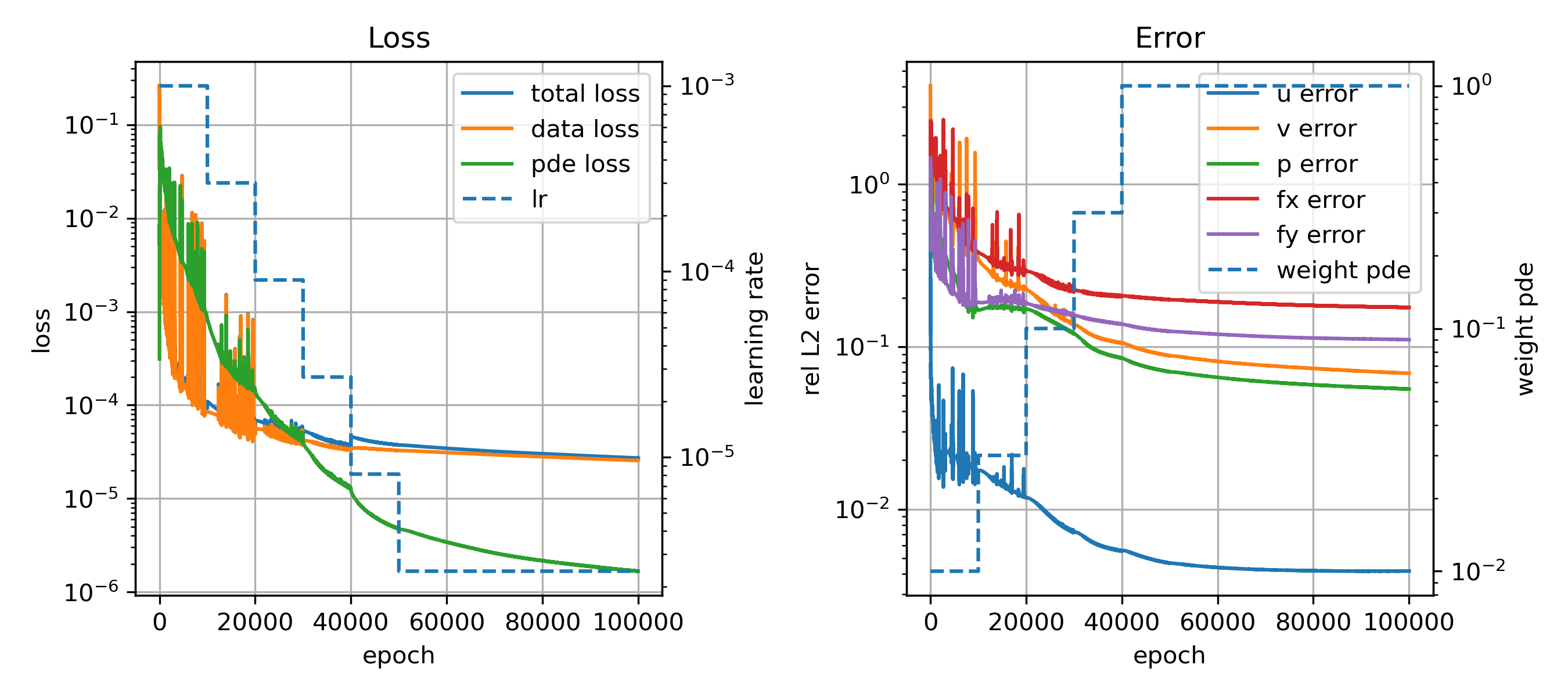}
    \caption{Training history for incompressible cylinder flow at $Re=11\,000$. The left panel shows training losses and the learning rate, while the right panel shows testing errors (relative $L_2$) and the weight of the PDE loss $\lambda_{PDE}$ ($L_{total}=L_{data}+\lambda_{PDE}L_{PDE}$).}
    \label{fig:flow_inference_DNS_Re11k_history}
\end{figure}

\autoref{fig:flow_inference_DNS_Re11k_history} shows the training history of the flow inference problem for incompressible cylinder flow at $Re=11\,000$. The training losses, testing errors, learning rate, and the varying weight $\lambda_{PDE}$ are shown. During training, the PDE weight is gradually increased from $0.01$ to $1$, and the PDE loss drops for roughly 5 orders. The data loss also drops by more than 3 orders. 

\begin{figure}
    \centering
    \includegraphics[width=\linewidth]{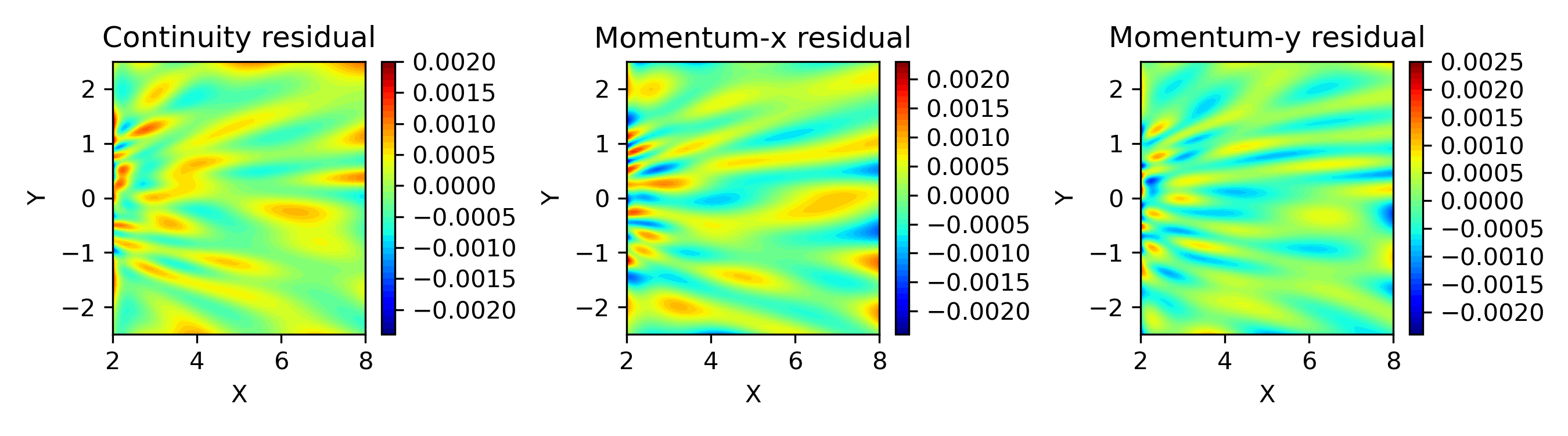}
    \caption{Final PDE residuals for incompressible cylinder flow at $Re=11\,000$. Residuals of the continuity equation, x-momentum equation, and y-momentum equation at the last epoch are shown.} 
    \label{fig:flow_inference_DNS_Re11k_residual}
\end{figure}

\autoref{fig:flow_inference_DNS_Re11k_residual} shows the final residuals for continuity, x-momentum, and y-momentum equations at the last training epoch. The magnitude of residuals is sufficiently reduced during the training stage, and the distribution is uniform and shows no large-scale structures, indicating the PDE is well satisfied during training. Note that the data is corrected to satisfy the governing equations, and thus there is no conflict between the data and the PDE.

\begin{figure}
    \centering
    \includegraphics[width=\linewidth]{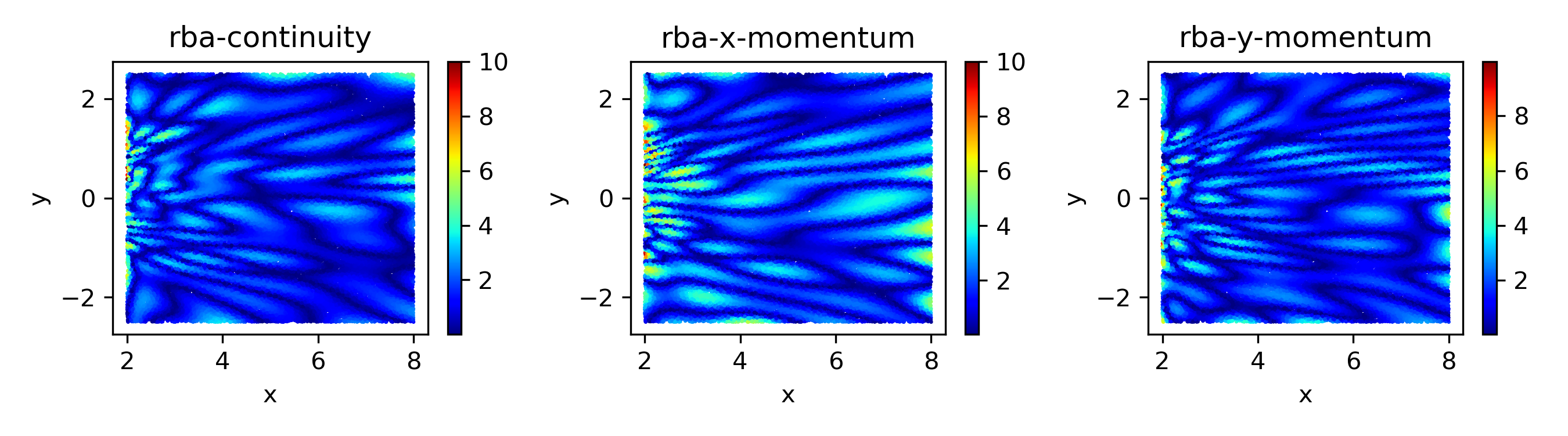}
    \caption{Final RBA weights for incompressible cylinder flow at $Re=11\,000$. RBA weights of the continuity equation, x-momentum equation, and y-momentum equation at the last epoch are shown. These weights are used to control the relative importance of different residual points in the PDE loss function during training based on the present and historical PDE residuals.} 
    \label{fig:flow_inference_DNS_Re11k_rba}
\end{figure}

\autoref{fig:flow_inference_DNS_Re11k_rba} shows the final RBA weights for continuity, x-momentum, and y-momentum equations at the last training epoch. These weights are calculated based on the present and historical residual distributions. The idea of RBA is to control the local weight of PINN's residual points in the PDE loss function during training based on the PDE residual values.

\begin{figure}
    \centering
    \begin{subfigure}{0.9\textwidth}
        \centering
        \includegraphics[width=\linewidth]{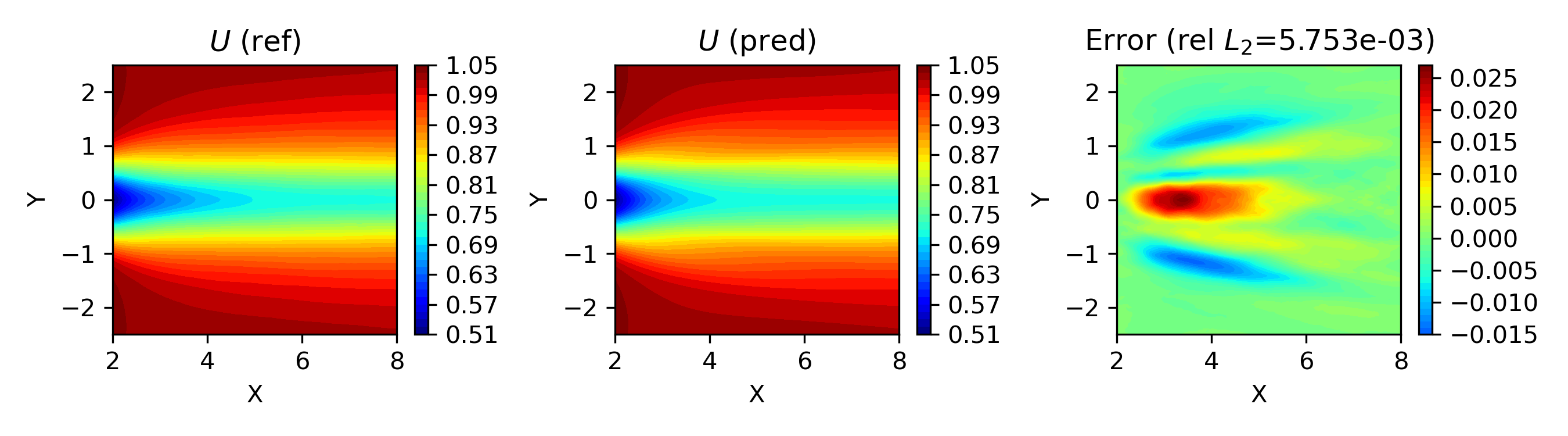}
    \end{subfigure}
    \begin{subfigure}{0.9\textwidth}
        \centering
        \includegraphics[width=\linewidth]{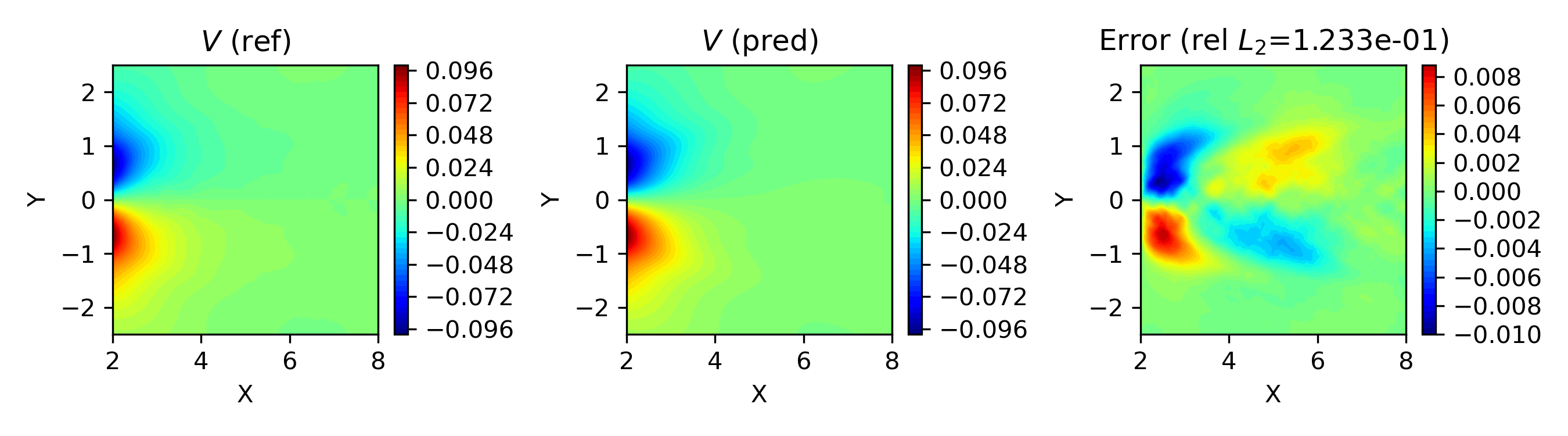}
    \end{subfigure}
    \begin{subfigure}{0.9\textwidth}
        \centering
        \includegraphics[width=\linewidth]{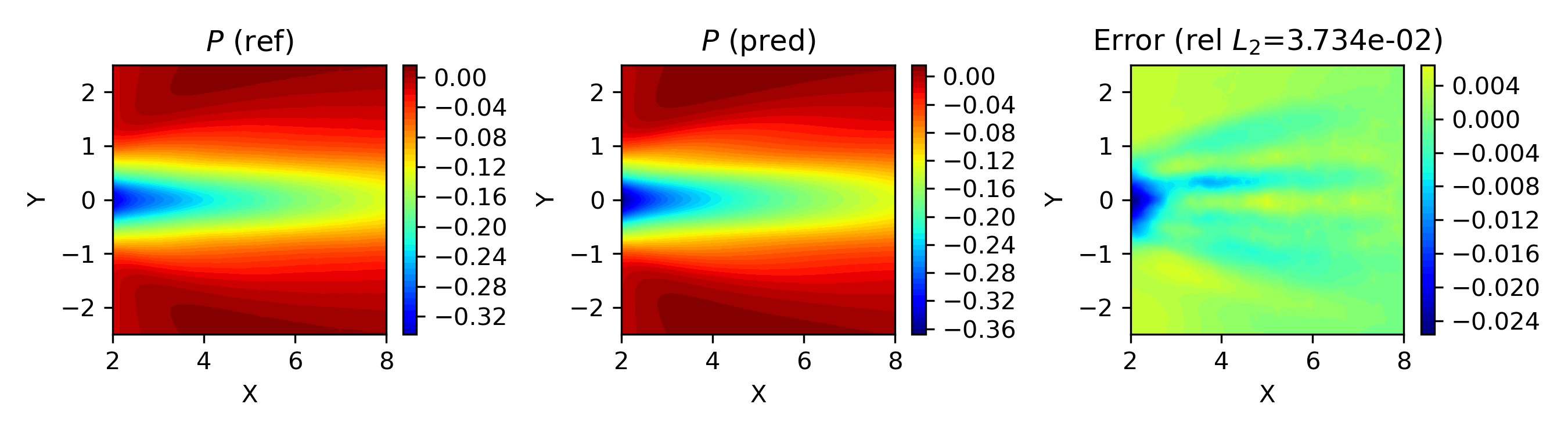}
    \end{subfigure}
    \begin{subfigure}{0.9\textwidth}
        \centering
        \includegraphics[width=\linewidth]{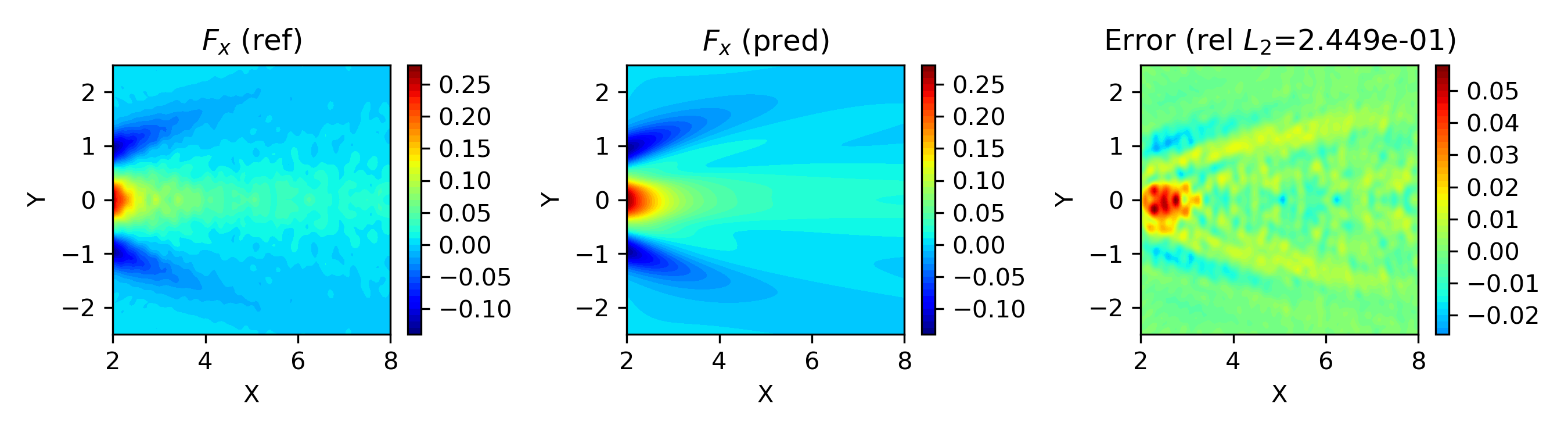}
    \end{subfigure}
    \begin{subfigure}{0.9\textwidth}
        \centering
        \includegraphics[width=\linewidth]{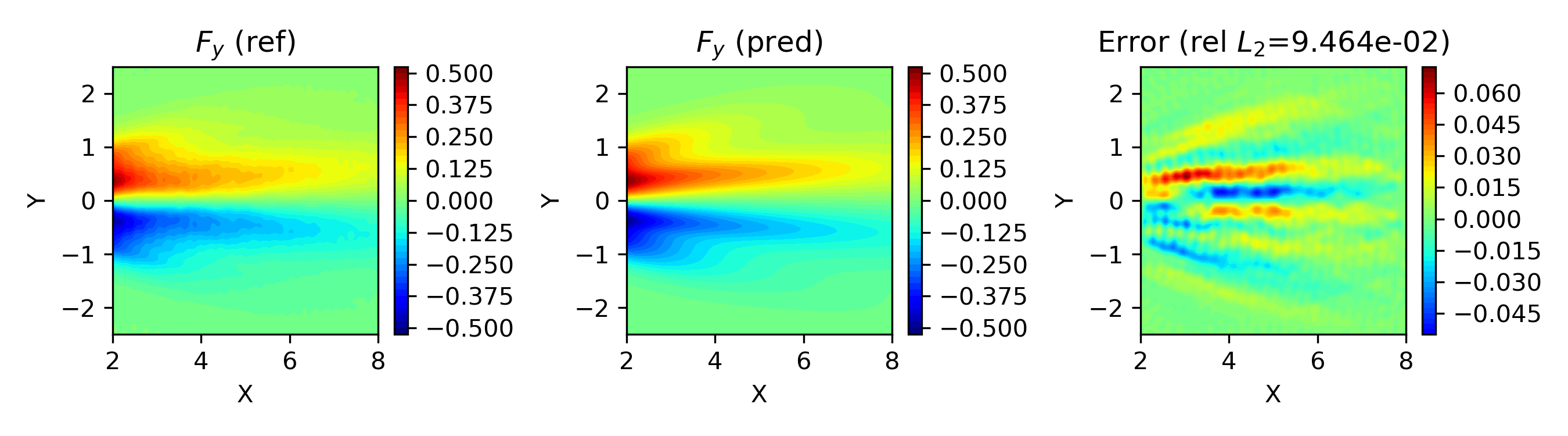}
    \end{subfigure}
    \caption{Flow inference for incompressible cylinder flow at $Re=140\,000$. The first column is reference values, the second column is PINN's reconstruction, and the last column is the pointwise error. Reference data is corrected time-averaged LES. Prediction is obtained by PINN only using the measurable data of $U,V,F_x,F_y$ at the domain boundary. Error distributions and the relative $L_2$ errors are shown in the third column.}
    \label{fig:flow_inference_DNS_Re140k_compare}
\end{figure}

\begin{figure}
    \centering
    \begin{subfigure}{0.9\textwidth}
        \centering
        \includegraphics[width=\linewidth]{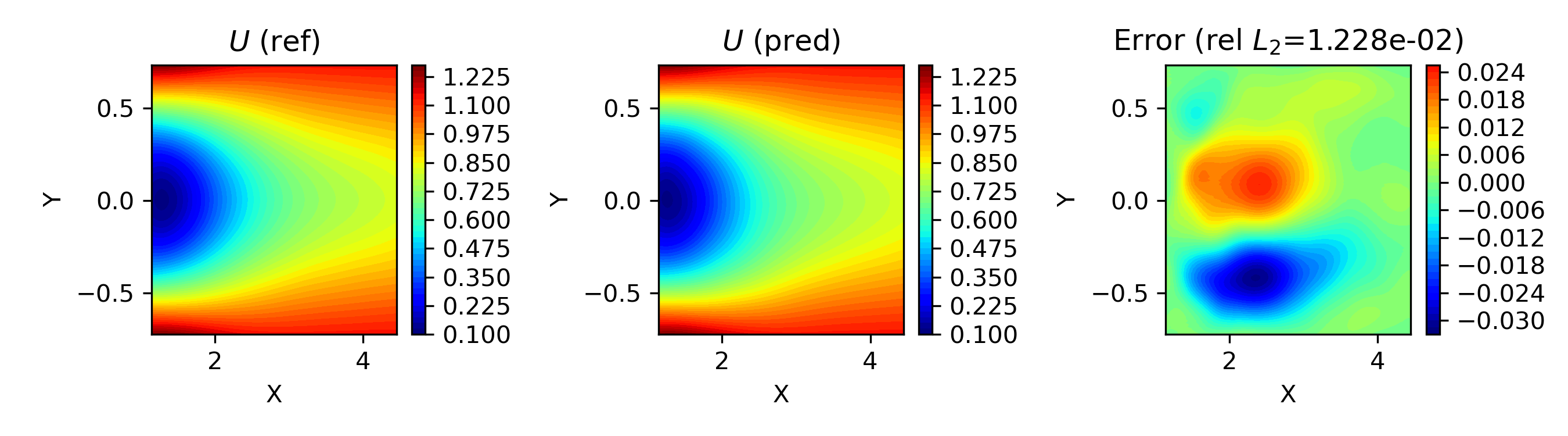}
    \end{subfigure}
    \begin{subfigure}{0.9\textwidth}
        \centering
        \includegraphics[width=\linewidth]{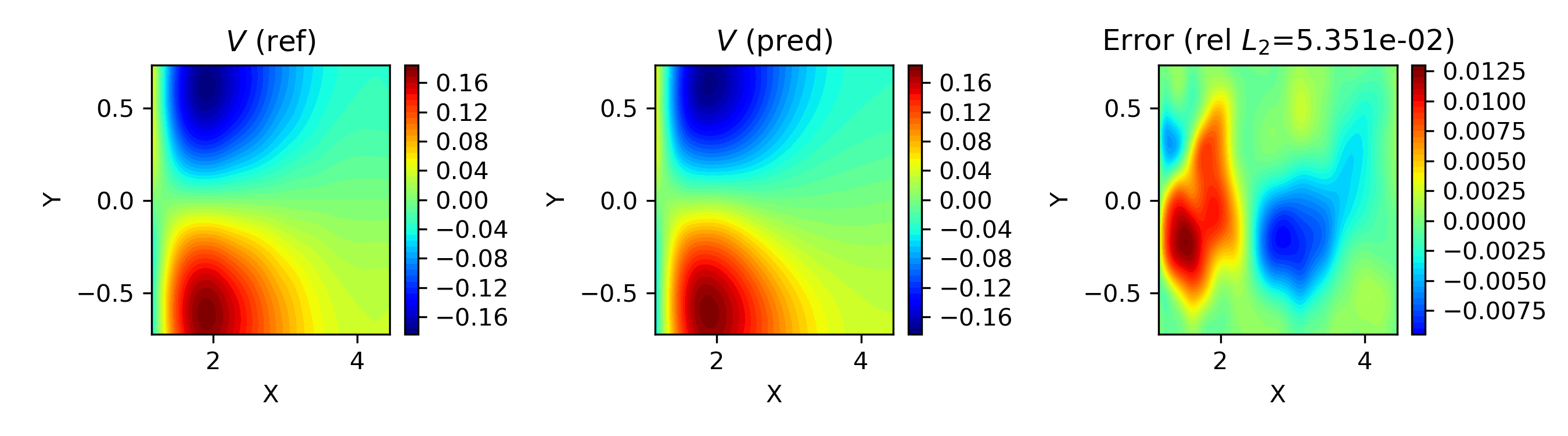}
    \end{subfigure}
    \begin{subfigure}{0.9\textwidth}
        \centering
        \includegraphics[width=\linewidth]{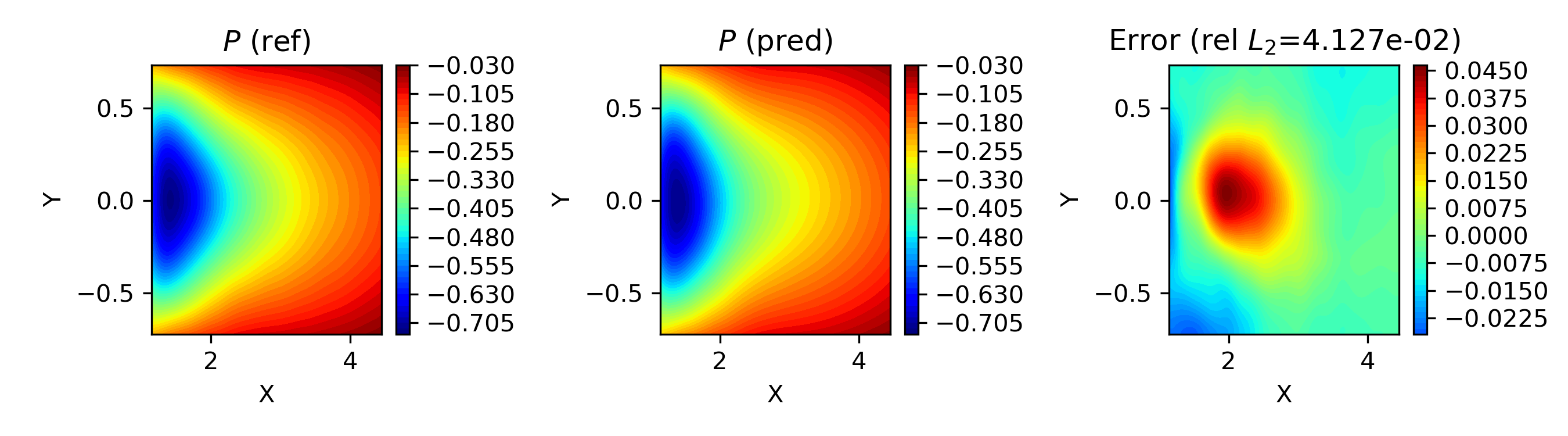}
    \end{subfigure}
    \begin{subfigure}{0.9\textwidth}
        \centering
        \includegraphics[width=\linewidth]{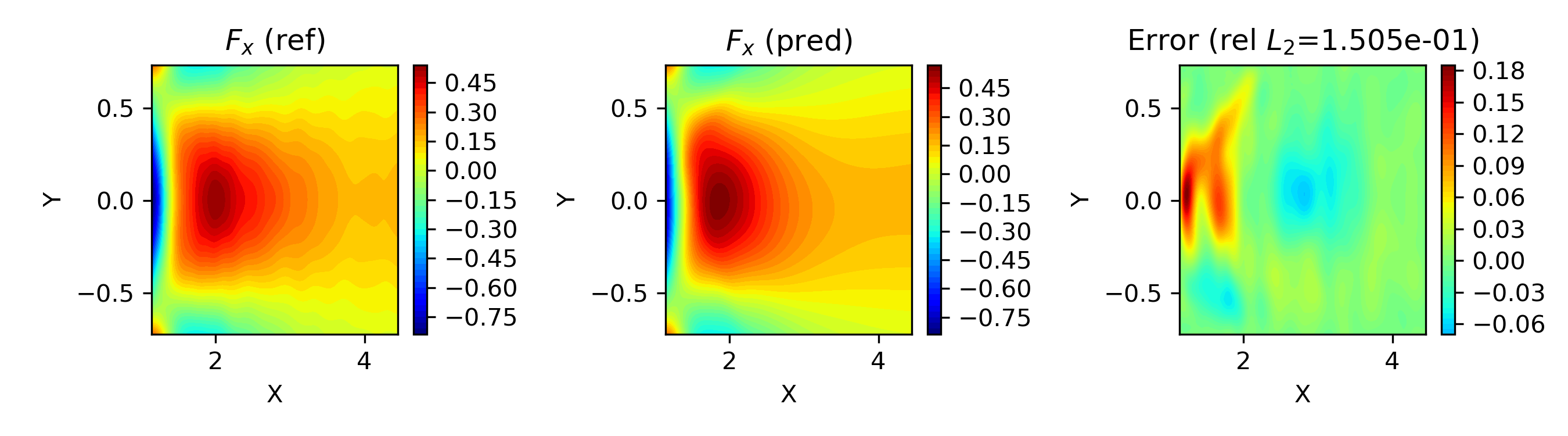}
    \end{subfigure}
    \begin{subfigure}{0.9\textwidth}
        \centering
        \includegraphics[width=\linewidth]{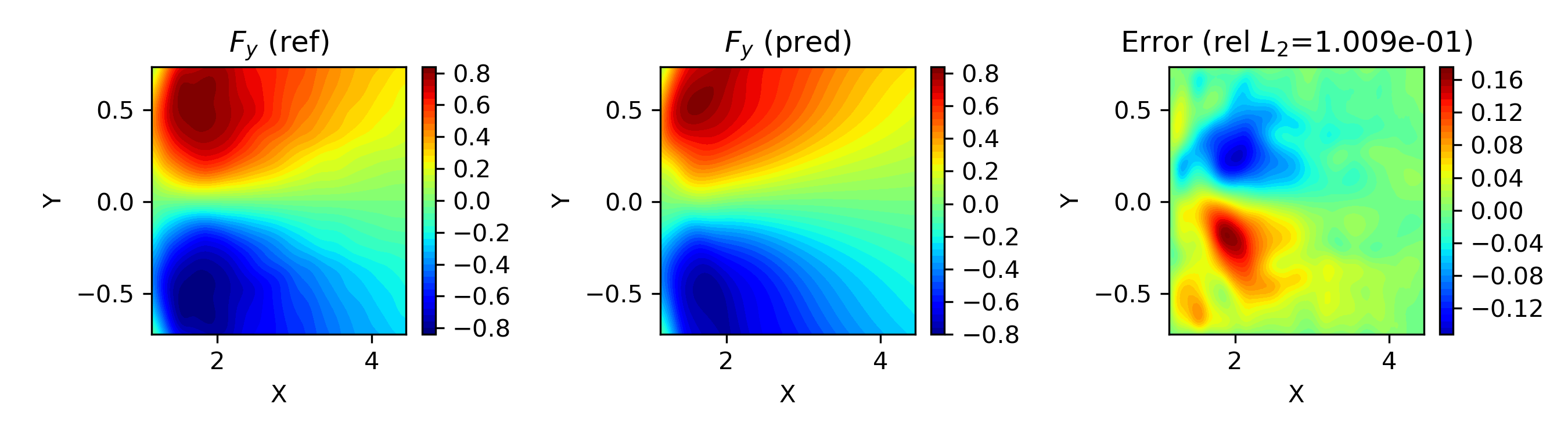}
    \end{subfigure}
    \caption{Flow inference for weakly compressible cylinder flow at $Re=50\,000$. The first column is reference values, the second column is PINN's reconstruction, and the last column is the pointwise error. Reference data is corrected time-averaged aerodynamic PIV. Prediction is obtained by PINN only using the measurable data of $U,V,F_x,F_y$ at the domain boundary. Error distributions and the relative $L_2$ errors are shown in the third column.}
    \label{fig:flow_inference_DNS_Re50k_UCF}
\end{figure}

\begin{figure}
    \centering
    \begin{subfigure}{0.9\textwidth}
        \centering
        \includegraphics[width=\linewidth]{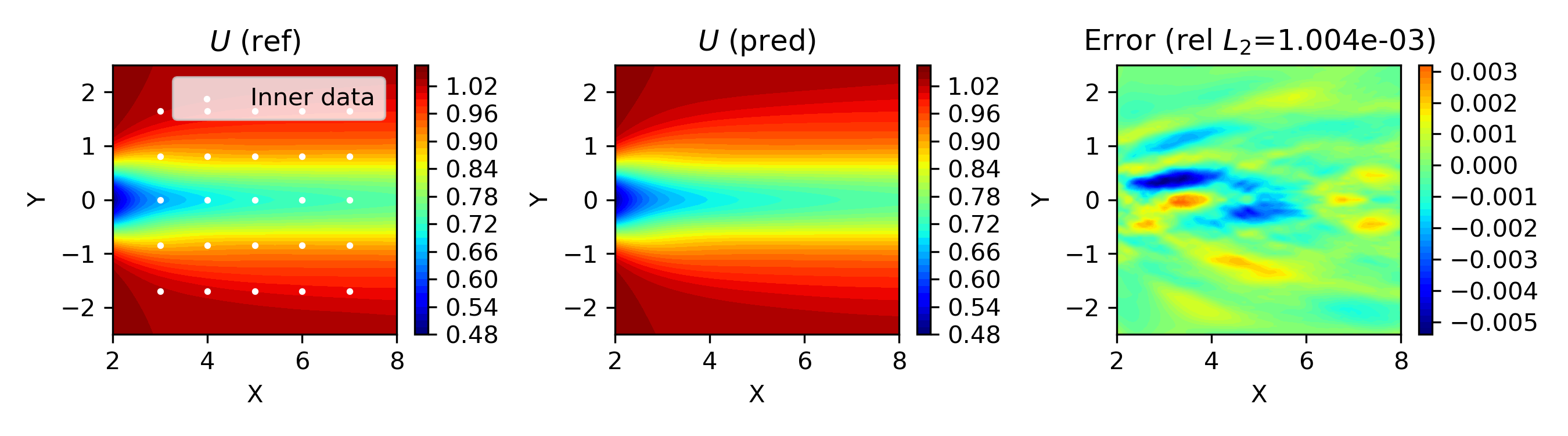}
    \end{subfigure}
    \begin{subfigure}{0.9\textwidth}
        \centering
        \includegraphics[width=\linewidth]{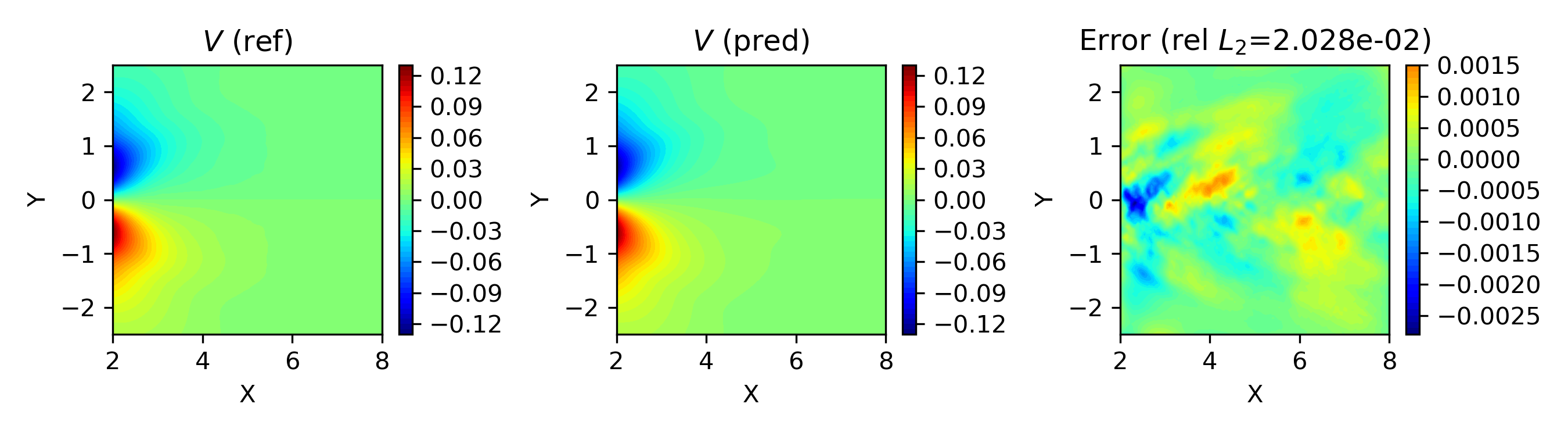}
    \end{subfigure}
    \begin{subfigure}{0.9\textwidth}
        \centering
        \includegraphics[width=\linewidth]{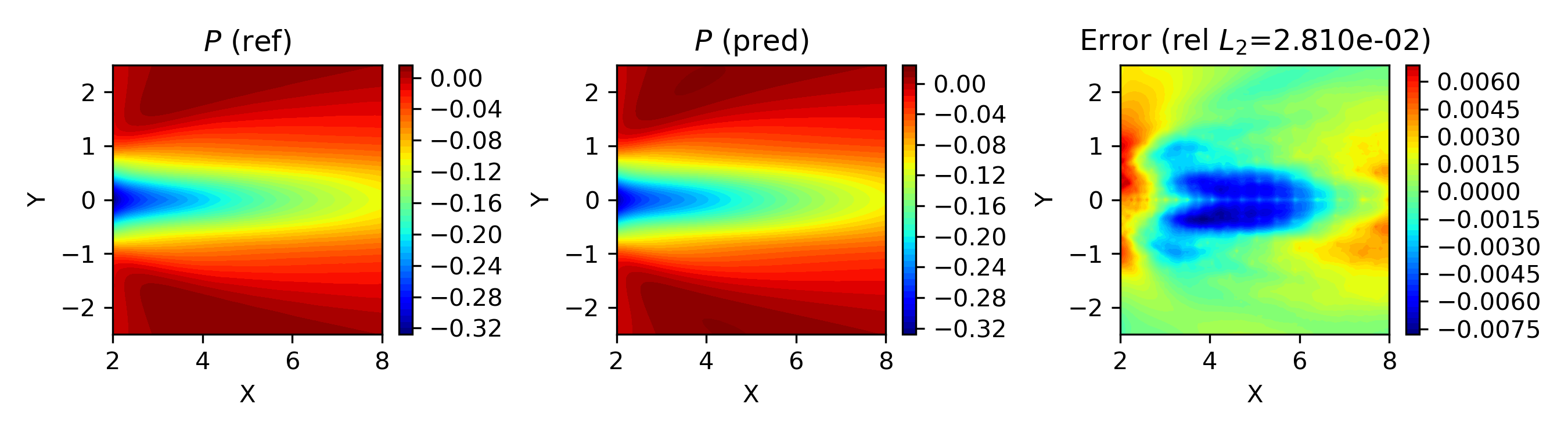}
    \end{subfigure}
    \begin{subfigure}{0.9\textwidth}
        \centering
        \includegraphics[width=\linewidth]{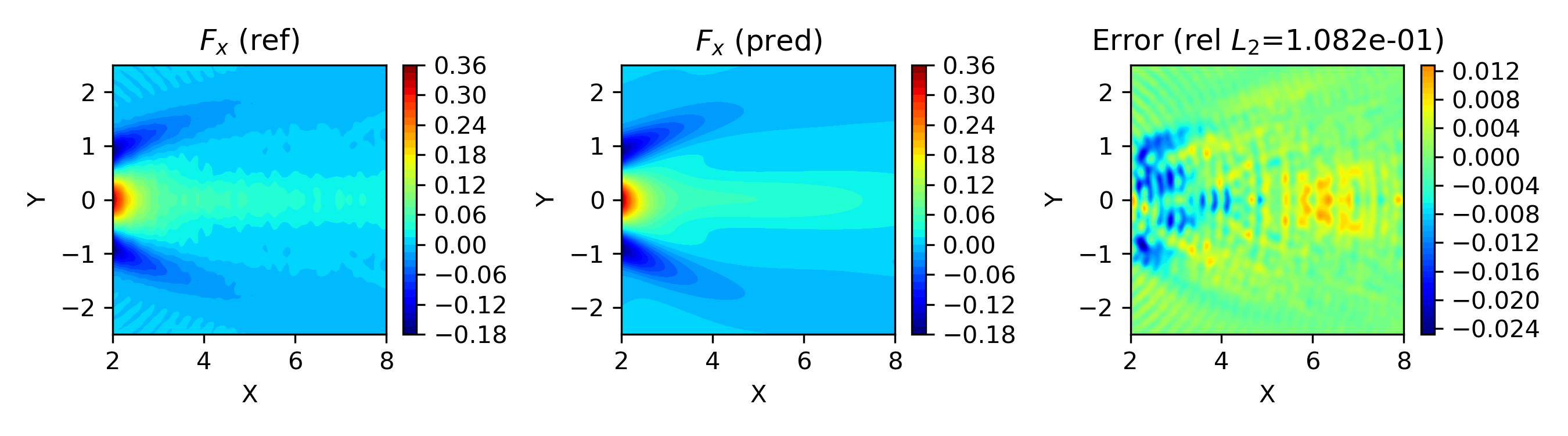}
    \end{subfigure}
    \begin{subfigure}{0.9\textwidth}
        \centering
        \includegraphics[width=\linewidth]{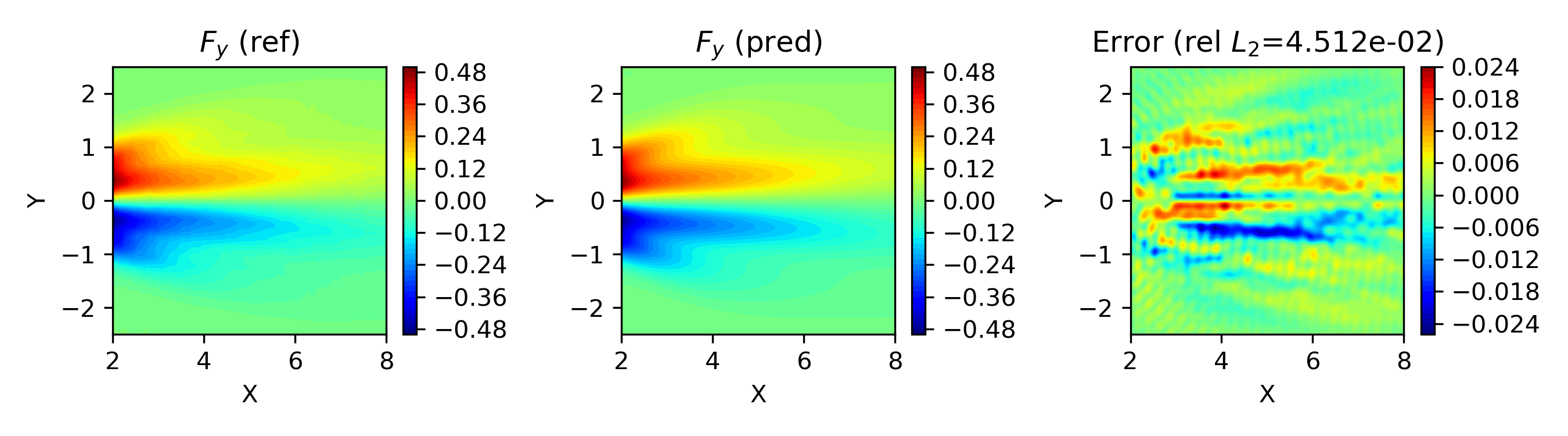}
    \end{subfigure}
    \caption{Flow inference for incompressible cylinder flow at $Re=11\,000$.  The first column is reference values, the second column is PINN's reconstruction, and the last column is the pointwise error. Reference data is corrected time-averaged DNS. Prediction is obtained by PINN using the measurable data of $U,V,F_x,F_y$ at the domain boundary and $5\times5$ inner points. Error distributions and the relative $L_2$ errors are shown in the third column.}
    \label{fig:flow_inference_DNS_Re11K_inner5}
\end{figure}

\autoref{fig:flow_inference_DNS_Re140k_compare}-\autoref{fig:flow_inference_DNS_Re11K_inner5} show some key plots of the flow inference problem listed in \autoref{tab:error_summary} and \autoref{tab:internal_data}.

\section{Helmholtz decomposition and turbulence model for flow inference}
\subsection{RANS equation with Helmholtz decomposition}
Substituting the Helmholtz decomposition of Reynolds forcing as shown in \autoref{eq:hz_decomp} in \autoref{eq:RANS_Euqtaion} yields \cite{patel2024turbulence}
\begin{align}\label{eq:RANS_HZ}
\begin{aligned}
U \frac{\partial U}{\partial x}+ V \frac{\partial U}{\partial y} +\frac{1}{\rho} \frac{\partial(P-\phi)}{\partial x}-\nu \left(\frac{\partial^2 U}{\partial x^2} + \frac{\partial^2 U}{\partial y^2}  \right)-F_{s, x} & =0 \\
U \frac{\partial V}{\partial x}+ V \frac{\partial V}{\partial y} +\frac{1}{\rho} \frac{\partial(P-\phi)}{\partial y}-\nu \left(\frac{\partial^2 V}{\partial x^2} + \frac{\partial^2 V}{\partial y^2}  \right)-F_{s, y} & =0 \\
\frac{\partial F_{s, x}}{\partial x} + \frac{\partial F_{s, y}}{\partial y} & =0\\
\frac{\partial U}{\partial x} + \frac{\partial V}{\partial y} & =0.
\end{aligned}
\end{align}

\begin{figure}
    \centering
   \includegraphics[width=\textwidth]{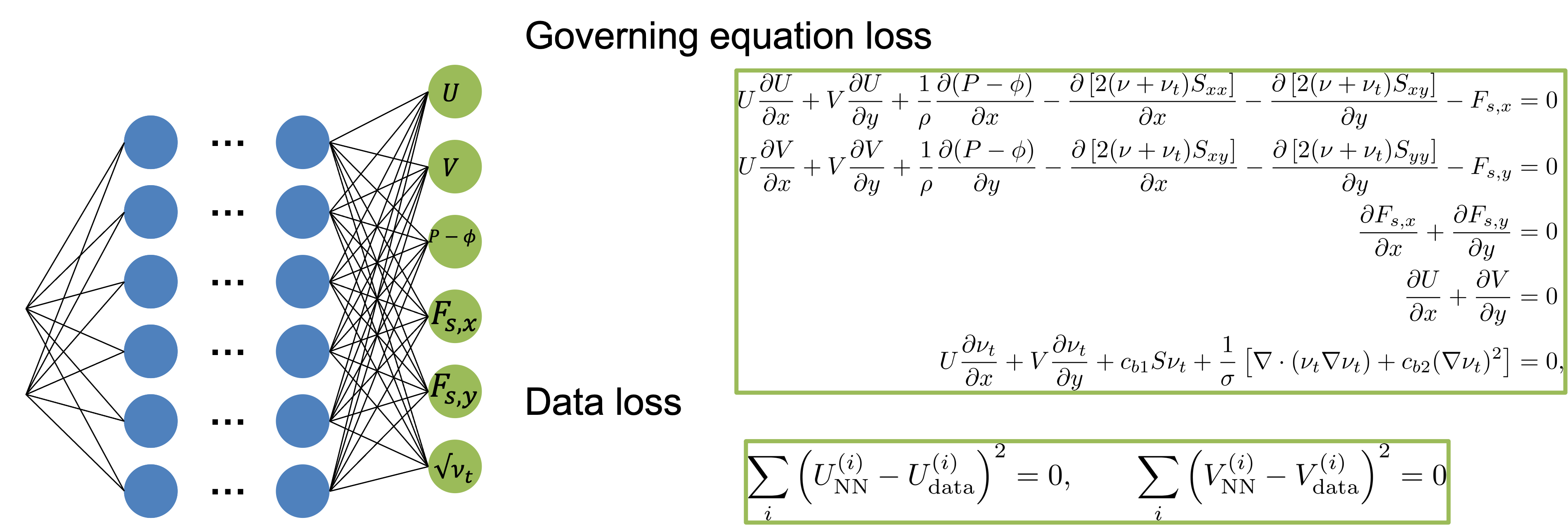}
   \caption{Architecture of a PINN leveraging the RANS equations with Reynolds forcing represented through Helmholtz decomposition and turbulent model for wake region as proposed by \cite{spalart1992one}.}
    \label{fig:pinn_hd_sa}
  \end{figure}

\subsection{RANS equation with Helmholtz decomposition and augmented with turbulence model} \label{app:RANS_SA}
 We augment the governing PINN formulation with a turbulence model tailored for the wake region, as proposed by \cite{spalart1992one}. The resulting augmented system of equations is expressed in \autoref{eq:RANS_HZ_SA},

\begin{align}\label{eq:RANS_HZ_SA}
\begin{aligned}
U \frac{\partial U}{\partial x}+ V \frac{\partial U}{\partial y} +\frac{1}{\rho} \frac{\partial(P-\phi)}{\partial x}-\frac{\partial \left[ 2(\nu+ \nu_t) S_{xx} \right]}{\partial x} - \frac{\partial \left[ 2(\nu + \nu_t) S_{xy} \right]}{\partial y} -F_{s, x} & =0 \\
U \frac{\partial V}{\partial x}+ V \frac{\partial V}{\partial y} +\frac{1}{\rho} \frac{\partial(P-\phi)}{\partial y}-\frac{\partial \left[ 2(\nu+ \nu_t) S_{xy} \right]}{\partial x} - \frac{\partial \left[ 2(\nu + \nu_t) S_{yy} \right]}{\partial y} -F_{s, y} & =0 \\
\frac{\partial F_{s, x}}{\partial x} + \frac{\partial F_{s, y}}{\partial y} & =0\\
\frac{\partial U}{\partial x} + \frac{\partial V}{\partial y} & =0 \\
U\frac{\partial \nu_t} {\partial x} + V\frac{\partial \nu_t}{\partial y} +  c_{b1} S \nu_t + \frac{1}{\sigma}\left[ \nabla \cdot (\nu_t \nabla \nu_t) + c_{b2} (\nabla \nu_t)^2 \right] &=0.
\end{aligned}
\end{align}

where $S_{xx} = \frac{\partial U}{\partial x},~S_{yy} = \frac{\partial V}{\partial y}$ and \textcolor{black}{$S_{xy} = \frac{1}{2}\left( \frac{\partial U}{\partial y} + \frac{\partial V} {\partial x}\right)$} are mean strain rate tensor. $S = || \omega||$, where $\omega$ is vorticity and expressed as {\textcolor{black}{$\omega=\frac{\partial V}{\partial x} - \frac{\partial U}{\partial y}$} and $c_{b1}$, $c_{b2}$ and $\sigma$ aree empirical turbulence constants and taken from \cite{spalart1992one} as $c_{b1}=0.1355,~c_{b2}=0.622$ and $\sigma=2/3.$ 
\begin{figure}
  \centering
  \begin{subfigure}[b]{\textwidth}
    \centering
     \includegraphics[width=\textwidth, trim=100 80 100 100, clip]{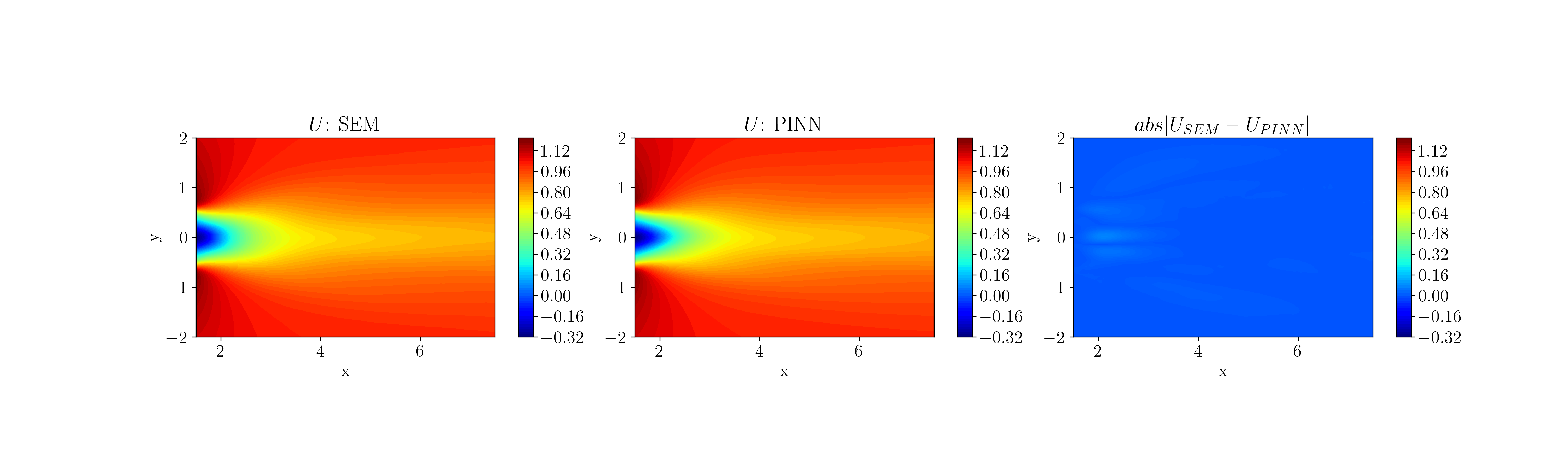}
    \caption{Reference (SEM) vs Inferred (PINN) $(U)$, Rel. $L_2$ err: 0.83\%}
    \label{fig:pinn_sa_u}
  \end{subfigure}
  \hfill
  \\
  \begin{subfigure}[b]{\textwidth}
    \centering
   \includegraphics[width=\textwidth, trim=100 80 100 100, clip]{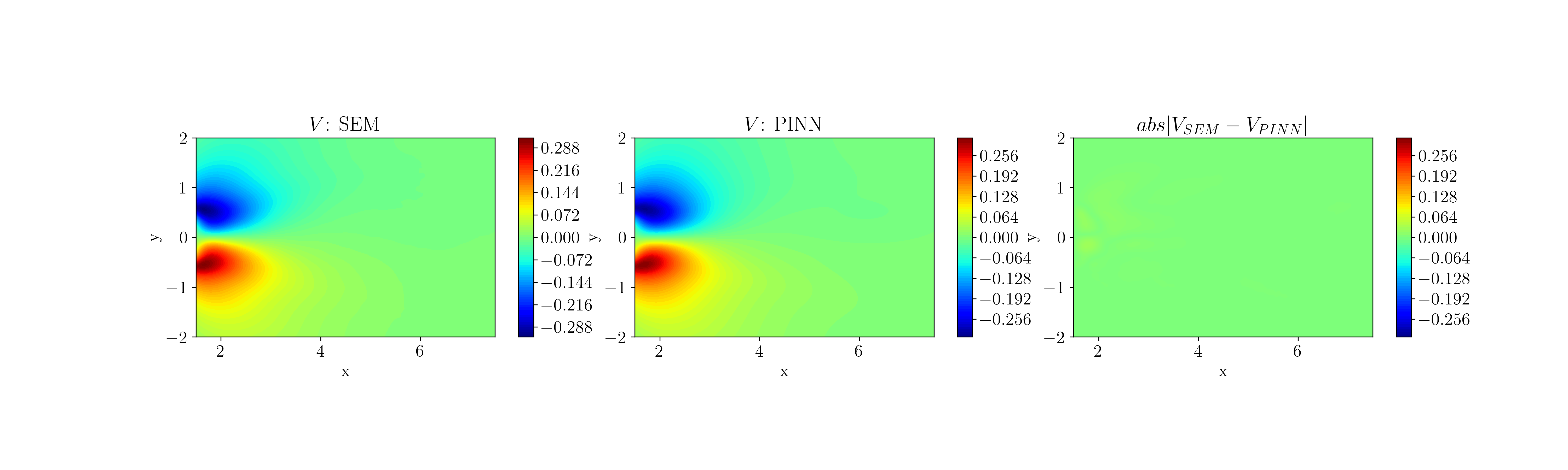}
    \caption{Reference (SEM) vs Inferred (PINN) $(V)$, Rel. $L_2$ err: 4.00\%}
    \label{fig:pinn_sa_v}
  \end{subfigure}
     \begin{subfigure}[b]{\textwidth}
    \centering
   \includegraphics[width=\textwidth, trim=100 80 100 100, clip]{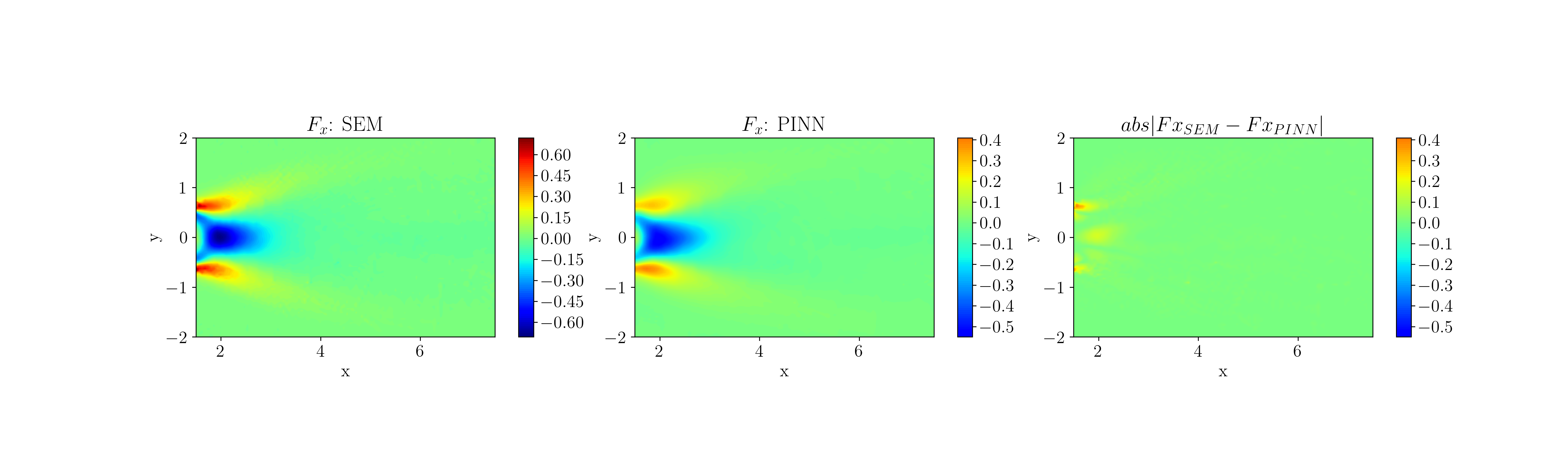}
    \caption{Reference (SEM) vs Inferred (PINN) $(F_x)$, Rel. $L_2$ err: 22.66\%}
    \label{fig:pinn_sa_fx}
  \end{subfigure}
     \begin{subfigure}[b]{\textwidth}
    \centering
   \includegraphics[width=\textwidth, trim=100 80 100 100, clip]{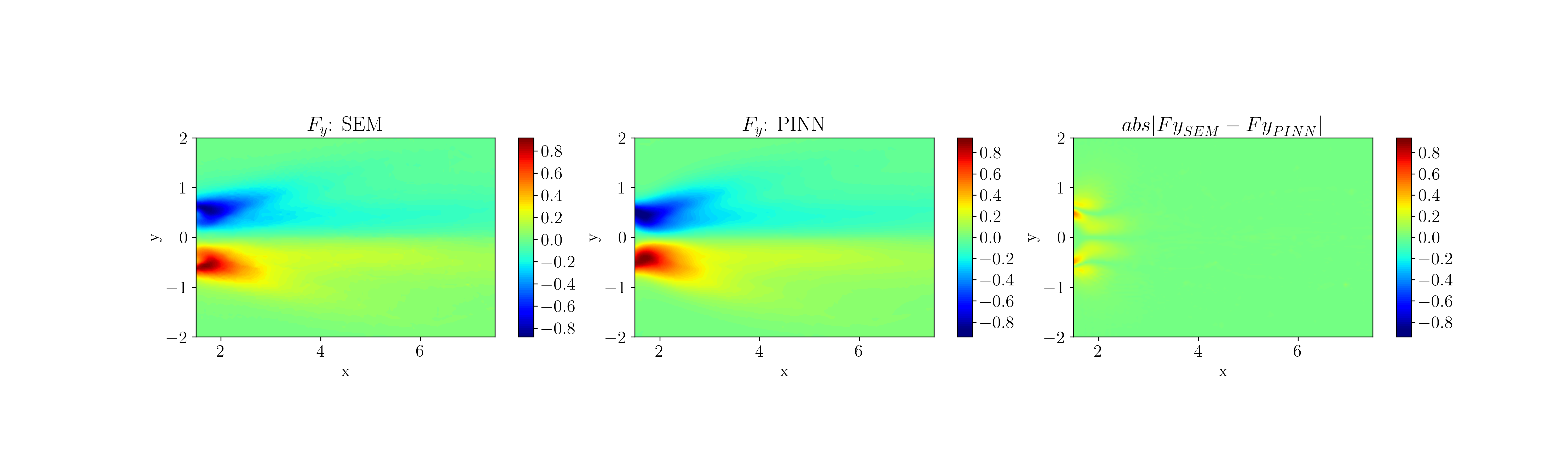}
    \caption{Reference (SEM) vs Inferred (PINN) $(F_y)$, Rel. $L_2$ err: 23.85\%}
    \label{fig:pinn_sa_fy}
  \end{subfigure}
\caption{\textcolor{black}{(a)~$U$, (b)~$V$, (c)~$F_{x}$, and (d)~$F_{y}$, obtained using boundary data of DNS at $Re=3900$ together with the Helmholtz decomposition and the turbulence-augmented model described by equation \autoref{eq:RANS_HZ_SA}. The left, middle, and right panels in each subfigure display the reference flow field (computed using the spectral element method), the PINN-inferred flow field, and the absolute pointwise error, respectively.
It is observed that the flow fields reconstructed using the turbulence-augmented model \autoref{eq:RANS_HZ_SA} achieve higher accuracy compared to those predicted by both the standard RANS equation and the RANS equation with Helmholtz-decomposed forcing—except for $F_x$.}}
  \label{fig:overall_sa}
\end{figure}

  Architecture of PINN and loss functions using \autoref{eq:RANS_HZ_SA}  is shown in \autoref{fig:pinn_hd_sa}. The flow inference results are presented in \autoref{fig:overall_sa}, where subfigures (a), (b), (c), and (d) correspond to $U$, $V$, $F_x$, and $F_y$, respectively. In each subfigure, the left, middle, and right panels represent the reference flow field (computed via the spectral element method), the PINN-inferred flow field, and the absolute pointwise error, respectively. A relative $L_2$ error metric is provided in \autoref{tab:RANS_models}. It is to be noted that the flow fields reconstructed using the turbulence-augmented model \autoref{eq:RANS_HZ_SA} exhibit higher accuracy than those predicted by both the standard RANS equation and the RANS equation with Helmholtz-decomposed forcing—except for $F_y$. This discrepancy may arise from the turbulence model’s limited accuracy in the wake region.

\section{\textcolor{black}{Unsteady flow Inference from phase-averaged PIV data}}\label{app:flow_inference_unsteady}

\textcolor{black}{
The flow inference framework described above can be naturally extended to unsteady flows by incorporating time as an additional input to the neural network. In this section, we demonstrate this extension using phase-averaged unsteady data, which is commonly available in experimental measurements of periodically shedding wakes.}

\textcolor{black}{
We consider the unsteady cylinder wake at $Re=11,000$. The reference data are corrected phase-averaged aerodynamic PIV measurements over one shedding period. Prior to training, the data are smoothed using a combined spatial--temporal filtering procedure and corrected using the methodology described earlier in this work. }

\textcolor{black}{
The neural network input is extended to $(x,y,t)$, and the network outputs the time-dependent fields $(U,V,P,F_x,F_y)$. The governing equations are the unsteady incompressible RANS equations with an unclosed Reynolds force vector, enforced through the PDE loss over the full space--time domain. The training strategy closely follows that of the steady case, including adaptive weighting between data and PDE losses and the use of residual-based attention to accelerate convergence.}

\textcolor{black}{
Data loss is imposed using only measurable quantities. Specifically, the phase-averaged velocity components $(U,V)$ and Reynolds force components $(F_x,F_y)$ are provided at the domain boundary for all times, together with the initial condition at the beginning of the shedding period. In addition, one interior point is included in the data loss to further constrain the solution. No pressure data are supplied to the network.}

\textcolor{black}{
~\autoref{fig:flow_inference_unsteady} presents a representative snapshot of the inferred unsteady flow field at the midpoint of one Strouhal period. The PINN successfully reconstructs the spatially and temporally varying velocity, pressure, and forcing fields throughout the domain using only sparse measurements and the governing equations as physical regularization. The results demonstrate that the proposed framework is capable of inferring unsteady flow dynamics from limited phase-averaged experimental data.}

\begin{figure}
    \centering
    \includegraphics[width=\linewidth]{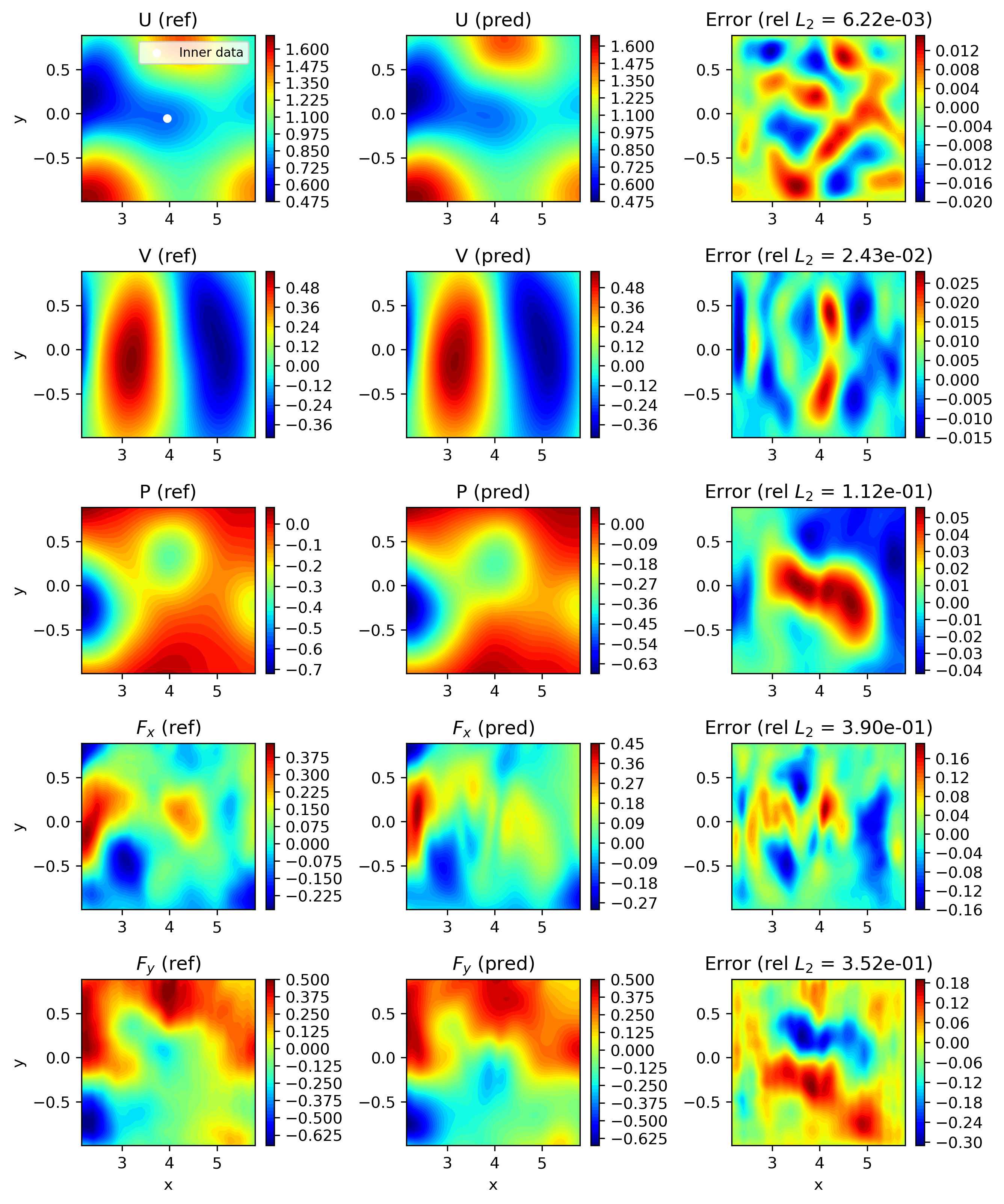}
    \caption{\textcolor{black}{Flow inference for weakly compressible unsteady flow at $Re=11,000$.  The first column is reference values, the second column is PINN's reconstruction, and the last column is the pointwise error. Reference data is corrected phase-averaged aerodynamic PIV data. Prediction is obtained by PINN using the measurable data of $U,V,F_x,F_y$ at the initial time, the domain boundary, and one inner point. The spatial-temporal varying solution is inferred, and the snapshot at the middle of one Strouhal period is shown. Error distributions and the relative $L_2$ errors are shown in the third column.}}
    \label{fig:flow_inference_unsteady} 
\end{figure}

\end{document}